\documentclass{ws-jai}

\usepackage{subfig}
\usepackage{jnl_macros}
\usepackage[flushleft]{threeparttable}
\usepackage{multirow}

\begin{document}

\catchline{}{}{}{}{} 

\newcommand{\mata}[1]{\mathbf{#1}}
\newcommand{\mats}[1]{\boldsymbol{#1}}
\newcommand{\ift}[1]{\mathcal{FT}^{-1}\left\{#1\right\}}
\newcommand{\ftf}[1]{\mathcal{FT}\left\{#1\right\}}
\newcommand{\fts}[1]{\hat{#1}}
\newcommand{\imgdir}{.}
\newcommand{\irelandpavoatsusi}{(Ireland \emph{et al.}, 2013, in prep.)}

\markboth{Y.~Kok et al.}{Phase-referenced Interferometry and Narrow-angle Astrometry
with SUSI}

\title{\uppercase{Phase-referenced Interferometry and Narrow-angle Astrometry with
SUSI}}

\author{Y.~Kok$^1$, M.~J.~Ireland$^{2,3}$, P.~G.~Tuthill$^1$,
J.~G.~Robertson$^1$, B.~A.~Warrington$^2$, A.~C.~Rizzuto$^2$ and W.~J.~Tango$^1$}

\address{
$^1$Sydney Institute for Astronomy, School of Physics, The University of Sydney,
NSW 2006, Australia\\
$^2$Department of Physics and Astronomy, Macquarie University, NSW 2109,
Australia\\
$^3$Australian Astronomical Observatory, PO Box 915, North Ryde, NSW 1670,
Australia
}

\maketitle

\footnotetext[1]{Email: y.kok@physics.usyd.edu.au}

\begin{history}
\received{(to be inserted by publisher)};
\revised{(to be inserted by publisher)};
\accepted{(to be inserted by publisher)};
\end{history}

\begin{abstract}
The Sydney University Stellar Interferometer (SUSI) now incorporates a new beam
combiner, called the Microarcsecond University of Sydney Companion Astrometry
instrument (MUSCA), for the purpose of high precision differential astrometry of
bright binary stars. Operating in the visible wavelength regime where
photon-counting and post-processing fringe tracking is possible, MUSCA will be
used in tandem with SUSI's primary beam combiner, Precision Astronomical Visible
Observations (PAVO), to record high spatial resolution fringes and thereby
measure the separation of fringe packets of binary stars. In its current phase
of development, the dual beam combiner configuration has successfully
demonstrated for the first time a dual-star phase-referencing operation in
visible wavelengths. This paper describes the beam combiner optics and hardware,
the network of metrology systems employed to measure every non-common path
between the two beam combiners and also reports on a recent narrow-angle
astrometric observation of $\delta$ Orionis A (HR 1852) as the project enters
its on-sky testing phase.
\end{abstract}


\section{Introduction}
Phase-referencing in interferometry is a technique which uses the phase
information in the interference signal of a reference object to correct the
phase of the signal from a target object for path length fluctuations in the
atmosphere. The target can either be the same or a nearby object. This
technique has been widely used in radio
interferometry but its potential in optical long baseline interferometry was
only realized in the early 1990s \citep{Shao:1992,Colavita:1992}. This
technique is mainly used (1) to extend the effective atmospheric coherence
time beyond the limit imposed by atmospheric turbulence in order to increase the
sensitivity of an interferometer
\citep{Quirrenbach:1994,Hummel:2003,Lane:2003} and (2) to measure the phase
difference between the target and the reference signal in order to determine the
relative position of the target object in the sky
\citep{Lane:2004,Delplancke:2008,Woillez:2010}.

Unlike radio interferometry where phase-referencing can be achieved by rapidly
slewing the antenna back and forth between the target object and the reference,
in optical interferometry both objects must be observed simultaneously because
the atmospheric coherence time scale is measured in milliseconds. Therefore,
phase-referencing optical interferometers usually have two beam combiners to
record interference signals from two objects at the same time. A dual beam
combiner for an optical interferometer is analogous to an adaptive optics (AO)
instrument for a single mirror imaging telescope. The first dual beam combiner
setup for an optical long baseline interferometer was installed at the Palomar
Testbed Interferometer (PTI). The beam combiners operated in the near IR
wavelengths (K~band) and were designed to perform phase-referenced narrow-angle
astrometry \citep{Colavita:1999}.

With similar scientific motivation, a second beam combiner was recently
installed at the Sydney University Stellar Interferometer (SUSI)
\citep{Davis:1999}. The new setup at
SUSI aims to perform phase-referenced narrow-angle astrometry as a mean to
search for extra-solar planets around bright close binary stars. Despite the
similar goal, the implementation of the dual beam combiner configuration at SUSI
as compared to PTI is vastly different. The beam combiners at SUSI operate at
visible wavelengths (0.77--0.9$\mu$m) and are unconventional in that they perform dual-star
phase-referencing in post-processing. This offers the beam combiners at SUSI the
potential of measuring the relative position of a target object about three
times more precisely at the same baseline at a cost of tighter tolerances, and a
method to perform real time group-delay tracking at much lower bandwidth. The
achievable astrometric precision, which is limited by the phase uncertainty of
the interference signal, is proportional to the mean wavelength of the signal
obtained with the beam combiners. A shorter wavelength translates to a higher
astrometric precision. The mean wavelength in which the beam combiners in SUSI
are operating at is $\sim$3$\times$ of K~band (i.e.\ $\sim$2.2$\mu$m).
For this same reason, the phase information of the reference signal must be
accurately known to a fraction of the mean wavelength (i.e.\ a tighter tolerance
with shorter wavelength) so that the phase uncertainty of interference signal of
the target object as a result of the phase-referencing process is small and the
visibility of the fringe packet is not lost due to incoherent integration.

The optical setup of the beam combiners at SUSI is described in the following
section. Then, the approach to phase-referencing via a network of metrology
systems and the details of the data processing required are discussed in
Sec.~\ref{sec:metrology} and Sec.~\ref{sec:pipeline} respectively. Sample
results from recent test observations are then presented and discussed in
Sec.~\ref{sec:results} and Sec.~\ref{sec:discussion}. Finally, a summary of future work to be performed and
scientific goals of the instruments are discussed in Sec.~\ref{sec:summary}.

\section{Instrument configuration} \label{sec:optics}
\begin{figure}
\centering
\includegraphics[width=0.4\textwidth]{\imgdir/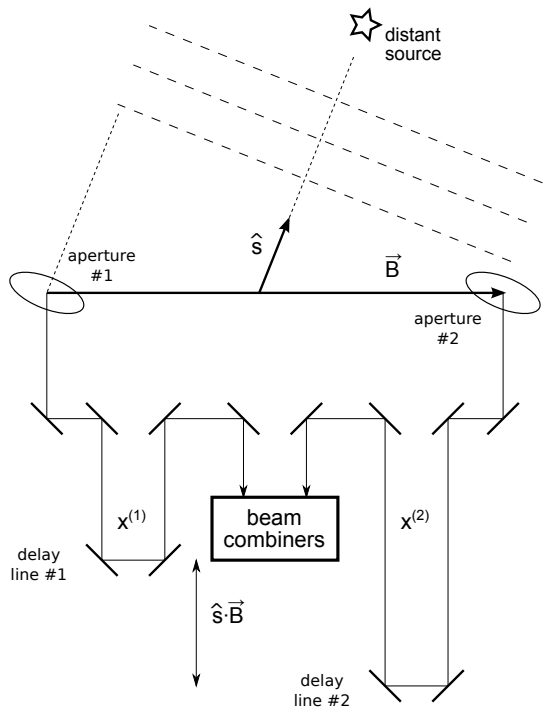}
\caption{A minimalist 2-element optical long baseline interferometer. The
baseline, $\vec{B}$, defines the separation between the 2 light collecting
elements of the interferometer while the unit vector $\hat{s}$ (commonly known
as the pointing vector) describes the angle between a star of interest and the
baseline. The optical path length of each delay line is denoted by the symbol
$X$. Interference fringes are formed when the OPD between the delay lines is
$\approx\hat{s}\cdot\vec{B}$.}
\label{fig:opt_olbi}
\end{figure}

SUSI is a 2-element optical long baseline interferometer located near Narrabri,
New South Wales, Australia. Fig.~\ref{fig:opt_olbi} shows a schematic diagram of
such an interferometer. Starlight collected from 2 siderostats (labeled as
apertures in Fig.~\ref{fig:opt_olbi}) are reflected by a series of mirrors into one or more light
combining instruments. SUSI has a configurable (5--160m) North-South oriented
baseline (distance between 2 siderostats, commonly denoted as
$\vec{B}$ or $B$). It was completed in 1991 and for over two decades, the
subsystem instruments within the
interferometer, especially the
beam combiners, have been constantly upgraded to increase its sensitivity and
capability \citep{ten-Brummelaar:1994b,Lawson:1994,Davis:2007,Robertson:2012}.
The current main beam combiner in SUSI is the Precision Astronomical Visible
Observation (PAVO) instrument \citep{Ireland:2008} which has been operating
since 2009 and has been used mainly for stellar diameter measurements and
astrometry of binary stars \citep{Antoci:2013,Rizzuto:2013}. Joining it is the
recently installed Microarcsecond University of Sydney Companion Astrometry
(MUSCA) beam combiner instrument. MUSCA is designed to operate alongside PAVO
for high precision narrow-angle astrometry.

\begin{figure}
\centering
\includegraphics[width=0.5\textwidth,trim=0 -3em 0 0]{\imgdir/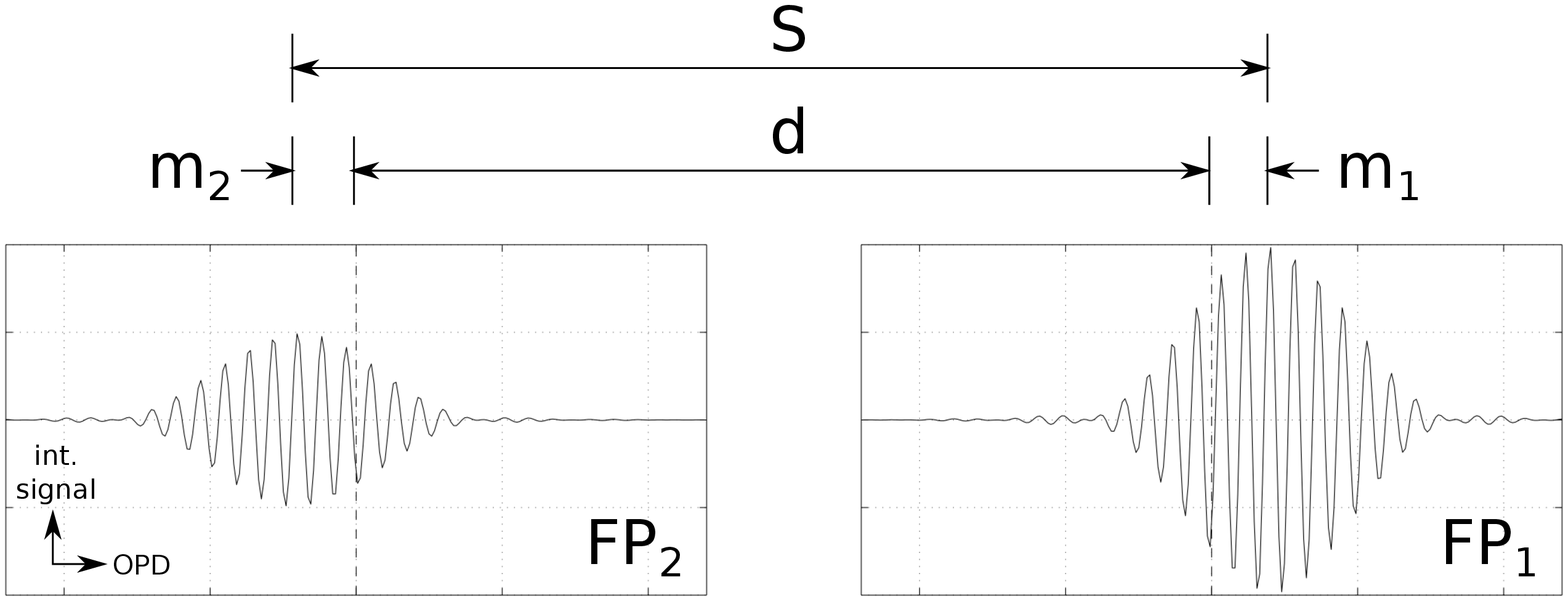}
\caption{The interference signal of two nearby stars. Due to the range of
optical modulation, represented by the width of the rectangular window, often
only one separated fringe packet (FP) is visible to MUSCA at a given time. The
selection of a fringe packet to be observed is controlled by the position of the
differential delay line (DDL) in MUSCA. The separation of the fringe packets can
be estimated from the measurement of the displacement of the DDL, $d$, and the
phase delay difference of the fringe packets, $m_2 - m_1$.}
\label{fig:opt_strategy}
\end{figure}

The two beam combiners are fed with a pair of starlight beams from the two arms
of the interferometer but they each use a slightly different part of the visible
light spectrum. They share the same main delay line for the tracking of
their interference signals in the form of interference fringes which are only
visible when the optical path difference (OPD) between the two
arms of the interferometer is near zero (within the coherence length of the
starlight whose range of wavelengths are usually defined by a bandpass filter).
Such interference signal is usually known as a fringe packet (see
Fig.~\ref{fig:opt_strategy} for examples).
The main delay line is made up of 2 folded optical paths, one for each arm.
Both paths, represented by the structures labeled `delay line \#1' and `delay line
\#2' in Fig.~\ref{fig:opt_olbi}, are simultaneously controlled to differentially
add optical path to one arm and remove the same amount from the other. Detail
description of the main delay line is given by \citet{Davis:1999}. 
In addition, MUSCA
has a differential delay line (DDL) for switching between fringe packets of two
stars in a binary system. The DDL is needed because the optical path modulation used to sweep
through a fringe packet in MUSCA is often smaller than the typical optical
path length between peaks of the envelope of the fringe packets of two stars in
a binary system. The limit of the optical path modulation is illustrated by the
width of the rectangular window around each fringe packet in
Fig.~\ref{fig:opt_strategy}
while the fringe packet separation, which is the main observable of the
PAVO+MUSCA dual beam combiner configuration, is represented by the dimension $S$
in the figure. Suppose the phase delay of a fringe packet, denoted by the
symbol $m_1$ or $m_2$, is defined as the OPD between two arms of the
interferometer computed at the mid-range of the optical path modulation in
MUSCA, then a measurement of the displacement of the DDL, denoted by the symbol
$d$, and the difference in phase delay between the pair of stellar fringe
packets yield the fringe packet separation, $S$. The relation between $S$ and
the on-sky separation of the two
stars is given as,
\begin{equation} \label{eq:opt_S}
S = (\hat{s}_2 - \hat{s}_1) \cdot \vec{B},
\end{equation}
where $\hat{s}_1$ or $\hat{s}_2$ is a unit vector that describes the angle
between a star and the baseline of the interferometer (see
Fig.~\ref{fig:opt_olbi}).

During observations, PAVO is used as a group-delay tracker to keep the fringe
packet of a star within the modulation range of optical path in MUSCA. Then in
post-processing, PAVO and a network of interferometer-based metrology systems
are used to estimate the OPD at each optical path modulation step in the MUSCA.
PAVO observes only one reference fringe packet (usually the one of the primary
star in a binary system) throughout an observation session but MUSCA alternately
observes either one of the two fringe packets by toggling the position of the
DDL every 5--15 minutes depending on the atmospheric seeing conditions. The
relative position of the fringe packets is measured simultaneously but
indirectly because as PAVO tracks the reference (primary) fringe packet MUSCA
observes the secondary. The reference phase of the reference fringe packet is
assumed to be stable over time. The stability of the reference fringe packet is
evaluated and discussed in Sec.~\ref{sec:obs_calibrators}. With a typical
integration time of 10--15 minutes on the fringe packets (during dual-star
phase-referencing mode), the astrometric precision due to anisoplanatism of the
atmospheric turbulence \citep{Shao:1992} for star separation of $<$1$''$ is
negligible ($<$10$\mu$as) as compared to the reference phase stability (refer to
Sec.~\ref{sec:results} for detail discussion). This assumes a baseline of $>$60m
is used and the atmospheric scaling factor at SUSI is
$\sim$780m$^{2/3}$s$^{1/2}$arcsec.

\subsection{PAVO} \label{sec:pavo}
\begin{figure}
\centering
\includegraphics[width=0.8\textwidth]{\imgdir/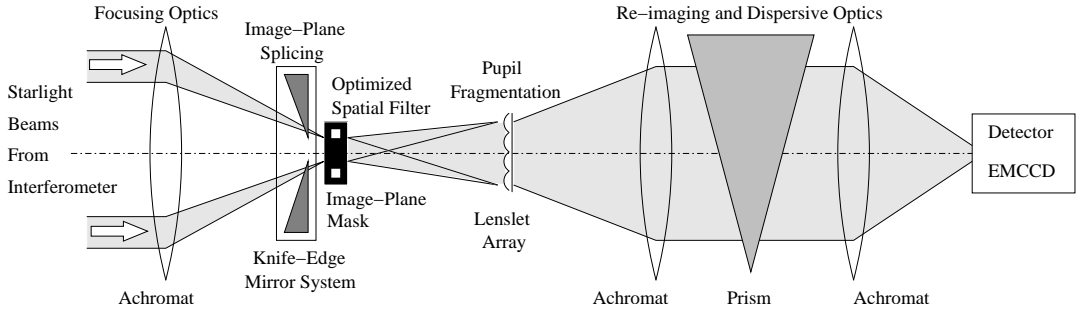}
\caption{Schematic diagram of the PAVO beam combiner at SUSI.}
\label{fig:opt_pavo_sch}
\end{figure}

PAVO is a multi-axially aligned Fizeau-type interferometer, but unlike a typical
Fizeau interferometer, PAVO forms spatially modulated interference fringes in
a pupil plane and then spectrally disperses the fringes.
It also employs spatial filtering in its image plane and an array of cylindrical
lenslets to utilize the full multi-r$_0$ aperture of the siderostats at SUSI. A
similar setup \citep{Ireland:2008} is operational at the Center for High
Angular Resolution Astronomy (CHARA) array. Fig.~\ref{fig:opt_pavo_sch} shows
the schematic diagram of the PAVO beam combiner. 

\begin{figure}
\begin{center}
\subfloat[]{\includegraphics[width=0.45\columnwidth]{\imgdir/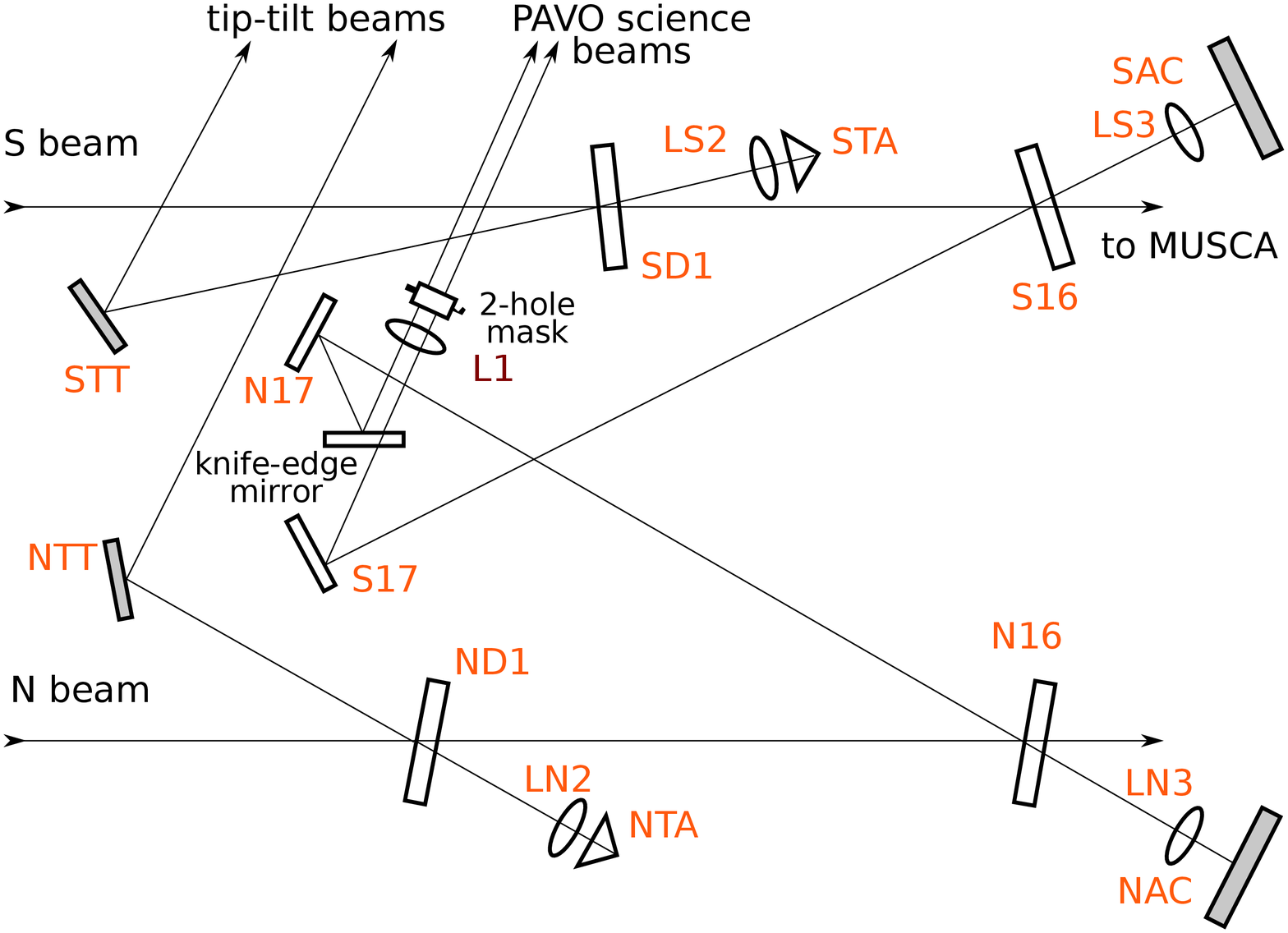}}\,\,
\subfloat[]{\includegraphics[width=0.45\columnwidth]{\imgdir/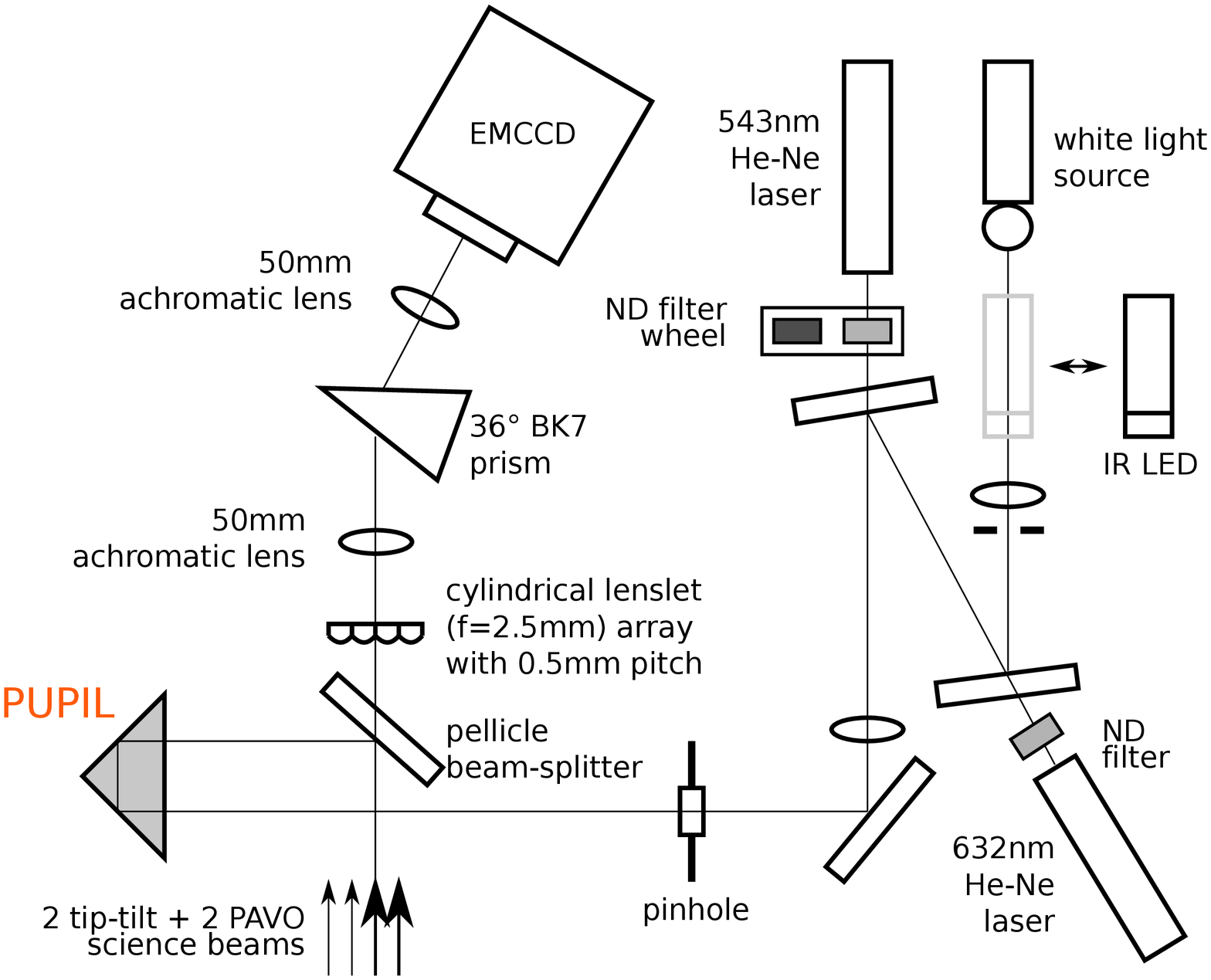}}
\caption{The (a) front- and (b) back-end optical setup of PAVO in the
laboratory.}
\label{fig:opt_pavo_current}
\end{center}
\end{figure}

The actual optical setup is divided into two parts: a front-end and a back-end
arrangement. The front-end optics splits the incoming beams into various
components while the back-end optics (Fig.~\ref{fig:opt_pavo_sch}) recombines
the beams. Fig.~\ref{fig:opt_pavo_current} shows the overall arrangement of the
PAVO optics. Fig.~\ref{fig:opt_pavo_current}(b) also shows the location of the
metrology sources which are discussed in Sec.~\ref{sec:metrology}.

The North (N) and South (S) beams from the siderostats propagate from the left
to right in Fig.~\ref{fig:opt_pavo_current}(a). A pair of dichroic filters, ND1
and SD1, separate the tip-tilt from the science beams. The filters reflect light
of wavelength shorter than 0.54$\mu$m and transmit light of longer wavelengths.
The reflected blue end of the visible spectrum is fed into a tip-tilt system
which acts as a first order AO system that removes the tip and tilt components
(or the phase ramp) from the wavefront. The science beams are then
separated by wavelength into 2 parts, one for PAVO and another for MUSCA, by
another pair of dichroic filters, N16 and S16.  These filters reflect light of
wavelength shorter than 0.79$\mu$m and transmit
light of longer wavelengths. As a consequence, the middle part of the visible
spectrum (0.54--0.79$\mu$m) is reflected into PAVO while the red end of the
spectrum is transmitted into MUSCA.

The beams are actually converging ($\mathtt{\sim}f$/700) because there is a
focusing lens in each arm of the interferometer (not shown in
Fig.~\ref{fig:opt_pavo_current}) which focuses to an image plane located at the
PAVO two-hole mask. Each of the $\sim$1.2mm apertures in the mask corresponds to
a FOV of $\sim$2$''$ on the sky. The holes are separated by twice their width
and are used as low-pass spatial filters to clean aberrated images of stars. A
knife-edge mirror is used to align the beams through the apertures of the mask.

The beams are aligned at the pupil plane to form spatially modulated fringes.
Fringes are formed in the pupil plane so that they are less
sensitive to seeing and misalignment. At the pupil plane, different parts of the
combined pupil are sampled by an array of lenslets that act as virtual slits
across the image plane for a spectrograph. The spectrograph, which consists of a
BK7 prism and a pair of six-element 50mm lenses, spectrally disperses the
fringes before they are re-imaged onto an EMCCD camera.

More in-depth description of PAVO at SUSI can be found in several other papers
\citep{Tuthill:2008,Robertson:2010,Robertson:2012}.

\subsection{MUSCA} \label{sec:musca}
\begin{figure}
\begin{center}
\begin{tabular}{c}
\includegraphics[width=0.45\columnwidth]{\imgdir/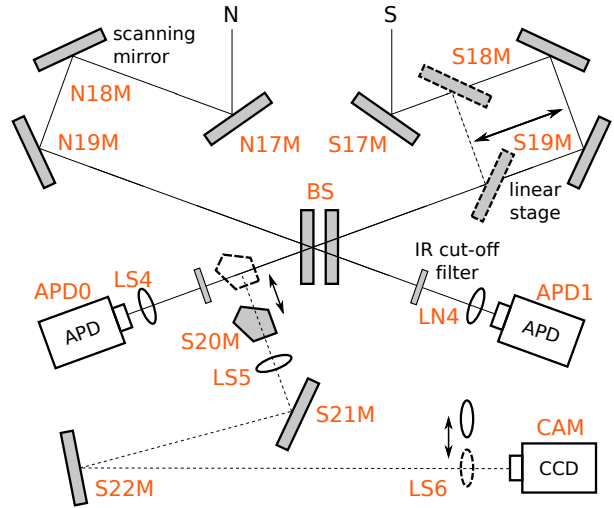}
\end{tabular}
\end{center}
\caption{Optical setup of MUSCA.}
\label{fig:opt_musca}
\end{figure}

MUSCA is a pupil-plane Michelson interferometer which operates at wavelengths
between 0.77$\mu$m and 0.91$\mu$m. It forms temporal fringes by modulating the
optical path length of one arm with time. Fig.~\ref{fig:opt_musca} shows the
optical path and components of MUSCA. Incoming converging light beams from the
top of the figure continue from Fig.~\ref{fig:opt_pavo_current}(a).

The beams are combined co-axially at the beam-splitter, BS. A scanning mirror
N18M is used to modulate the optical path length of the North arm of MUSCA. It
is attached to a piezoelectric transducer (PZT) which converts voltage to
physical displacement. Voltage is applied in small steps and at an interval of
about 0.2--0.3ms per step.  In order to avoid excitation of unnecessary
mechanical resonance the modulation signal is passed through a 1ms low-pass RC
filter before it is fed to the transducer. The intensity variation of the
combined beams as a result of the optical path length modulation is recorded as
temporal fringes using a pair of SPCM-AQR-14-FC self-contained single photon
counting avalanche photodiode (APD) modules from PerkinElmer.

The maximum path length modulation attainable from the transducer is
$\sim$140$\mu$m (maximum throw of the PZT is $\sim$100$\mu$m in a folded optical
path) but the typical value used for MUSCA is $\sim$25$\mu$m. This is chosen for
MUSCA because (1) it matches with the PAVO's group delay tracking window and (2)
it is approximately 2$\times$ larger than the expected width of the MUSCA fringe
packet. Making the range larger is undesirable because the PAVO group delay
estimates are unreliable beyond its tracking window and therefore
phase-referencing is futile at that range. 

A linear translational stage (T-LS28 from Zaber Technologies) and two mirrors on
the South arm of MUSCA form the DDL for the beam combiner. It is used for fringe
finding by equalizing the optical path length between the North and South arms.
The fiducial location where OPD between the arms is defined to be zero is chosen
at the middle of the scan range of the scanning mirror. As fringe packets from
two different stars are separated in optical path length the DDL must alternate
between two positions in order to place either one of the fringe packets into
the scan window. Effectively the DDL is used to `select' one fringe packet for
observation at any one time.

The stage is stepper-motor driven and has a built-in open loop position control
system. The stepper motor converts rotary motion into linear motion via a
leadscrew. The total travel distance of the stage is more than 8mm, which is the
required span to observe binary star separations as wide as 10$''$ using a 160m
baseline. At an average slew speed of $\sim$2mm/s, it takes the stage less than
5 seconds to cover that distance. The leadscrew-based open loop position control
system of the stage, which has a nominal accuracy of 12$\mu$m and a
repeatability of $<$0.4$\mu$m, is not the only mechanism put in place to measure
the position of the stage for MUSCA's astrometric application. It is also being
measured by a dual laser metrology system which has a precision better than 5nm.
The metrology system works hand in hand with the open loop position control
system and is discussed in detail in a separate paper \citep{Kok:2013}.

The role of each field lens, LN4 and LS4 ($f$=35mm) is to ensure that the pupil
of any off-axis star (in MUSCA's case, the secondary of a binary system) is
formed within the active (photo-sensitive) region of an APD. Otherwise MUSCA
will not be able to detect fringes from stars further than 0.2$''$ from the
on-axis reference star.

The IR cut-off filters in front of each APD are intended to reduce the number of
background photon counts originating from an IR metrology laser ($\lambda = $
1.152591$\mu$m) which is used with the main delay line at SUSI. The filter cut-off
wavelength is about 1$\mu$m.

\subsection{Optical alignment} \label{sec:alignment}
Alignment light sources in SUSI can be divided into two sets according to their
function. The first set consists of a laser and a bright halogen lamp and is
used for general optical alignment and internal fringe searching. They are
injected into the optical path in front of PAVO (see
Fig.~\ref{fig:opt_pavo_current}(b)) through a pin hole and a pellicle. The
second set consists of LEDs and is used to align the narrow-angle baseline of
MUSCA with the wide-angle baseline of SUSI. The LEDs are attached precisely in
the center of the back of the mirror mount at every siderostat. The alignment
process takes advantage of the imperfection in the dichroic filter fabrication,
e.g.\ the finite slope of the filter profile at the cut-off wavelengths, to
reflect light back into the beam combiners. PAVO uses its EMCCD camera for
optical alignment but MUSCA uses a separate CCD camera specifically for the same
purpose. In order to use the latter, a prism is slid into the optical path
during alignment and is removed during observation.

The wide-angle baseline of an interferometer is defined as a vector between
pivot points of two siderostats. The pivot point of a siderostat is a virtual
point where the two rotational (elevation and azimuth) axes of the siderostat
intersect and typically coincides with the center of the siderostat mirror. The
wide-angle baseline vector is obtained by observing stars separated by large
angles and located in various parts of the sky, noting the differences between
the expected and the actual optical delays (offsets) required to produce stellar
fringes based on a set of estimated dimensions and finally solving the
simultaneous equations of offsets to obtain a correction to the estimated
baseline vector initially used. The uncertainty of the wide-angle baseline
obtained with this method is in the regime of several micrometers in principle,
and $\sim$100$\mu$m for SUSI once all errors are taken into account. The
baseline vector only needs to be updated on a time-scale of the order of a few
weeks if the accuracy of its value is to remain better than 50$\mu$m
\citep{Davis:1999a}. On the other hand, the narrow-angle baseline of the same
interferometer (also known as the imaging baseline according to a definition by
\citet{Woillez:2013}) is defined as a vector between the center of two entrance
pupils of a beam combiner projected onto the siderostats. The following is an
example to easily understand the relation between the wide- and narrow-angle
baselines. Suppose a two-hole aperture-mask is put over each siderostat mirror,
fringes from each pair of apertures (one from each siderostat) can be used to
make astrometric measurements of a binary star on a slightly different
narrow-angle baseline simultaneously, despite the aperture pairs between the two
masks share the same pivot points, which defines the wide-angle baseline of the
two interferometers.

If the wide- and narrow-angle baselines are not co-aligned, then using the
wide-angle baseline solution for narrow-angle astrometry of a close binary star
system is erroneous which then systematically limits the precision of the
astrometric measurement.
In order to achieve a precision of 10$\mu$as, the co-alignment between the two
baselines must be better than 1mm per 100m baseline for a binary separation of
$>$1$''$.
The effect of
baseline uncertainty on the astrometric precision of the measurement has been
extensively discussed elsewhere \citep{Muterspaugh:2010a,Woillez:2013}.

The image of an LED at a siderostat formed on the MUSCA camera has a diameter of
about 12 pixels, where $\sim$1 camera pixel in the image plane corresponds to a
dimension of 1mm at the siderostats. Therefore, in order to meet the baseline
uncertainty requirement, the centroid of the LED images from the North and South
siderostats must be aligned to within 1 pixel. The centroid of an LED image is
computed by fitting a two-dimensional Gaussian model to the image in real time
while mirrors are being adjusted for alignment. At the same time, the LED images
are also aligned to the fiducial center of the tip-tilt system so that the
tip-tilt beams have the same optical axis as the MUSCA science beams. This
alignment procedure is carried out every night before an astrometric observation
with PAVO and MUSCA commences.

\section{Metrology systems} \label{sec:metrology}
There are 4 metrology systems in SUSI which are designed to measure the OPD of
different parts of the interferometer. The main metrology is a double laser
homodyne interferometer and operates at an IR wavelength of 1.152591$\mu$m with
a single acousto-optic modulator shifting the reference beam by 40 MHz. It is
used to measure the OPD introduced by the main delay line in SUSI. The metrology feeds
PAVO with measurements of the displacement of the main delay line so that PAVO can track
stellar fringes in a closed servo loop by driving the main delay line to compensate any
fringe motion during observation. Since a detailed description of this metrology
system in its original design was given by \citet{Davis:1999} and later by
\citet{Robertson:2012} after the system underwent a recent upgrade, it will not
be discussed further in this paper.

The remaining three metrology systems are newly installed and used for
phase-referenced narrow-angle astrometric observations. Other than additional
light sources, which consist partly of those shown in
Fig.~\ref{fig:opt_pavo_current}(b), the metrology systems do not require any
extra hardware or optical components. The systems are made up of a white-light
(WL) metrology, a single laser (SL) metrology and a dual laser (DL) metrology
system. Together, they form a two-prong approach to measuring the optical path
length between fringe packets of two stars in a binary system. First, the WL and the SL
metrology systems are used to measure the OPD of every optical path non-common
to PAVO and MUSCA. The objective is to ensure that any change in the position of
a MUSCA phase-referenced fringe packet is caused by the DDL within the beam
combiner. Second, the DL metrology system is used to measure the displacement of
the DDL.

\begin{figure}
\begin{center}
\includegraphics[width=0.45\columnwidth]{\imgdir/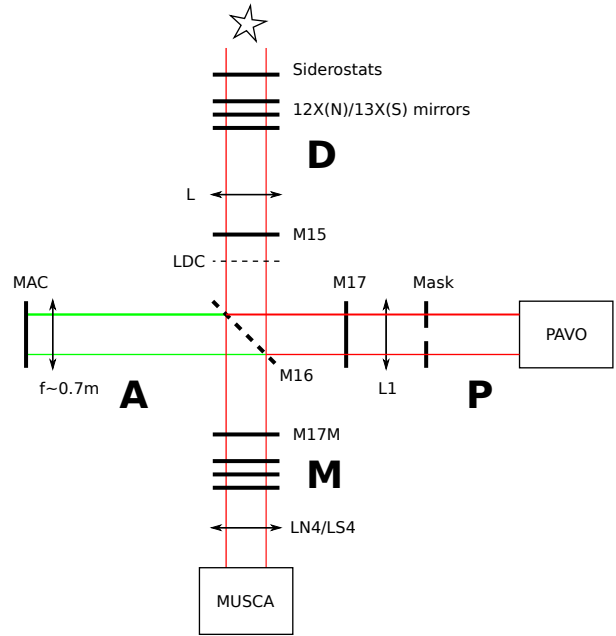}
\end{center}
\caption{Schematic of beam combiners at SUSI. Part D includes all the optical
components from the siderostat until M16 (a collective term for N16 and S16).
Part P consists of all optical components from M16 to the PAVO camera. Part A
includes MAC (collective term for NAC and SAC), LS3 and LN3. Part M consists of
all the optical components from M16 to the MUSCA APDs.}
\label{fig:met_susisch}
\end{figure}

A schematic diagram showing the optical paths of SUSI is given in
Fig.~\ref{fig:met_susisch}. The labels in bold in the figure represent the
different parts of SUSI. The labels in regular type font are optical components
which can be cross-referenced back to Fig.~\ref{fig:opt_pavo_current} and
\ref{fig:opt_musca}. The non-common paths between PAVO and MUSCA are labeled as
\textbf{P} and \textbf{M} in Fig.~\ref{fig:met_susisch}. In the first prong of
the approach, the phase delays of interference signals in MUSCA, $m$, are
estimated using the phase delays of interference signals measured by PAVO, $z$,
and the OPDs of the non-common paths measured by the metrology systems. This is
given as,
\begin{equation} \label{eq:met_m}
\begin{split}
m &= \text{OPD}_{\rm{D}} + \text{OPD}_{\rm{M}} \\
  &= z + \left(\text{OPD}_{\rm{M}} - \text{OPD}_{\rm{P}}\right),
\end{split}\end{equation}
where,
\begin{equation} \label{eq:met_z}
z = \text{OPD}_{\rm{D}} + \text{OPD}_{\rm{P}},
\end{equation}
and the subscripts in the equations represent different parts of the SUSI
optical paths as illustrated in Fig.~\ref{fig:met_susisch}. Then in the second
prong of the approach, the true phase delay difference between two MUSCA fringe
packets, $S$, is estimated by correcting the apparent phase delay difference
with the change in the OPD within MUSCA, $\Delta$OPD$_{\rm{M}}$. This is given
as,
\begin{equation} \label{eq:met_S}
S = (m_2 - m_1) + \Delta\text{OPD}_{\rm{M}},
\end{equation}
where $m_i$ is the phase delay of either fringe packet.

In this paper, the phase delay of an interference signal is defined as the OPD
between two arms of an interferometer computed at a fiducial location and is
usually quoted in units of $\mu$m. The fiducial location for MUSCA is in the
middle of the scan range of the scanning mirror. Similarly, the fiducial
location for PAVO is in the center of the FOV of the EMCCD camera. The fiducial
locations for computing the remaining OPD terms in Eq.~\eqref{eq:met_m} and
\eqref{eq:met_z} are not important because they are not measured independently.

\subsection{White-light metrology} \label{sec:met_wl}

The light source for this metrology system is an LED which emits light at a
peak wavelength of $\lambda_{\rm{I}} = \sigma_{\rm{I}}^{-1} = 0.940\mu$m. It is
injected into the SUSI optical path in front of PAVO with a pellicle
beam-splitter (see Fig.~\ref{fig:opt_pavo_current}(b)). It then propagates to
a pair of retro-reflecting mirrors, NAC and SAC, and back to PAVO. This optical
setup constitutes a Fizeau interferometer. Less than 5\% of the total light from
the LED is reflected into MUSCA because the dichroic mirrors (N16 and S16) are
only highly reflective at wavelengths shorter than 0.79$\mu$m. This has
negligible effect on the phase-referencing performance of PAVO and MUSCA.

The optical path of the metrology is coaxially aligned with the incoming
starlight path in order to minimize the non-common path error in the
measurement. The interference fringes formed by the LED are recorded
simultaneously with the stellar fringes during observation with the
same PAVO camera for the same purpose. The lower quantum efficiency of the
camera at the metrology wavelength ($\sim$25-30\%) enables PAVO to observe
fringes of a star and the metrology simultaneously. Suppose $w$ represents the
phase delay of the fringes formed by this metrology, then $w$ is the sum of OPDs
along the propagation path of the metrology, i.e.,
\begin{equation} \label{eq:met_w}
w = 2\text{OPD}_{\rm{P}} + 2\text{OPD}_{\rm{A}}.
\end{equation}
However, the phase measurement is modulo of 2$\pi$. Therefore the non-ambiguity
range (NAR) of the metrology is equal to one wavelength of the LED light.
Despite the short NAR, the range is adequate because fringes are recorded by the
PAVO camera at a rate of one frame per $\mathtt{\sim}$5ms and the OPD measured
by $w$ is not expected to change by more than one wavelength within this time
interval. 

Over longer timescales, the range of OPD variation measured by $w$ can exceed
one wavelength of light over a timescale of several hours depending on the
laboratory seeing. This is not a concern because the PAVO spectrograph disperses
the metrology fringes into narrow bands so that the coherence length of the
fringe packet is $\mathtt{\sim}$30$\mu$m. The sum of OPD$_{\rm{P}}$ and
OPD$_{\rm{A}}$ is not expected to vary beyond 30$\mu$m in one night as there are
no moving parts along the optical path. 

The WL metrology measures only the relative OPD. The reference OPD is taken to
be the OPD at the start of an observation session. The value is near to zero
(within $\pm$10$\mu$m approximately). The absolute value of the reference OPD is
not important nor does it need to be precisely at zero because it will be
cancelled out in the data analysis.

The major source of noise in the phase delay measurement is photon noise because
the LED is relatively bright as seen with the PAVO's low read noise EMCCD
camera. The estimated measurement uncertainty is 80nm or less, but this value
does not translate directly into astrometric error because the error in
determining the
relative position of the phase-referenced fringe packet is averaged over many
MUSCA scans, which reduces the error by a factor of $\sqrt{N_{\rm{SC}}}$, where
$N_{\rm{SC}}$ is the number of good scans used in the integration. It does
however affect the overall integration time because more scans are required to
average out larger uncertainties.

\subsection{Single-laser metrology} \label{sec:met_sl}

The light source for this metrology system is a green He-Ne laser which emits
light at a peak wavelength of $\lambda_{\rm{G}} = \sigma_{\rm{G}}^{-1}$ =
0.5435161$\mu$m. This laser also serves a dual purpose for optical alignment.
The metrology laser is injected into the SUSI optical path the same way as the
LED but it propagates via the pair of retro-reflecting NAC and SAC mirrors
and into the MUSCA beam combiner. The intensity of the metrology laser is
controlled by the selection of neutral density filters at the source (see
Fig.~\ref{fig:opt_pavo_current}(b)) so that it does not `blind' the APDs when
observing faint stars. This optical setup constitutes a Mach-Zehnder
interferometer.

Like the WL metrology, the optical path of the SL metrology is also coaxially
aligned with the starlight path in order to minimize the non-common path error
in the measurement. For the same reason, the interference fringes formed by the
laser at the MUSCA beam-splitter are recorded by the MUSCA APDs simultaneously
with the stellar fringes during observation. Suppose $x$ represents the phase
delay of the fringes formed by this metrology, then the OPD probed by $x$ is,
\begin{equation} \label{eq:met_x}
x = \text{OPD}_{\rm{P}} + 2\text{OPD}_{\rm{A}} + \text{OPD}_{\rm{M}}.
\end{equation}
Although the phase measurement is modulo of 2$\pi$, the NAR is adequate because
the scanning mirror makes an up and down scan (one period) per 150ms and the OPD
measured by $x$ is not expected to change by more than one laser wavelength
within this time interval. Over longer timescales, when laboratory conditions
are unstable, the OPD variation measured by $x$ can exceed one laser wavelength
over a timescale of several hours, but is still negligible compared to the
coherence length of the laser.

The SL metrology measures only the relative OPD between the two arms of the
interferometer. The OPD measured at the start of an observation session is set
as the reference OPD. It is also not necessary for the reference OPD to be
precisely zero although it is usually set to near zero (within $\pm$5$\mu$m
approximately) before the start of an observation. The absolute value of the
reference OPD is not important because it will be cancelled out in the data
analysis.

Similar to the WL metrology, the measurement uncertainty of $x$ does not
translate directly into astrometric error.

\subsection{Dual-laser metrology} \label{sec:met_dl}

The light sources for this metrology system are green and red He-Ne lasers. The
former is the same green laser used for the SL metrology. The latter emits light
at a peak wavelength of $\lambda_{\rm{R}} = \sigma_{\rm{R}}^{-1}$ =
0.6329915$\mu$m. The intention of using two lasers of different wavelengths is
to extend the NAR of a single laser metrology. Conventionally, this is achieved
using a long synthetic optical wavelength by means of heterodyne interferometry
\citep{Daendliker:1988,Schuhler:2006} but the dual wavelength metrology in MUSCA
employs only a simple homodyne fringe counting detection scheme, thereby
avoiding the need for specialized optical elements and higher development cost.
Neither laser is frequency stabilized, but based on the full width half maximum
(FWHM) of their theoretical gain profiles the relative uncertainties of the
laser wavelengths are better than 3$\times$10$^{-7}$. This level of uncertainty
is adequate because the displacement of the DDL is typically less than 0.5mm.
Suppose $d$ represents the physical quantity measured by this metrology, then
the OPD probed by $d$ is,
\begin{equation} \label{eq:met_d}
d = \Delta\text{OPD}_{\rm{M}},
\end{equation}
where $\Delta$OPD$_{\rm{M}}$ is the change in optical path within MUSCA brought
about by the displacement of the DDL. This metrology can achieve a measurement
uncertainty of $\lesssim$5nm and is discussed in detail in a separate paper
\citep{Kok:2013}.

\section{Data reduction pipeline} \label{sec:pipeline}
Raw data from a PAVO-MUSCA observation, in the form of interferograms, are
reduced in two stages to produce its primary observable, which is the
differential phase delay of two fringe packets. The first stage of the
data reduction pipeline computes $m$ in Eq.~\eqref{eq:met_m} using estimates
of $z$, $w$ and $x$ while the second stage computes $S$ in Eq.~\eqref{eq:met_S}
by using estimates of $d$.

\begin{figure}
\begin{center}
\includegraphics[width=0.5\columnwidth]{\imgdir/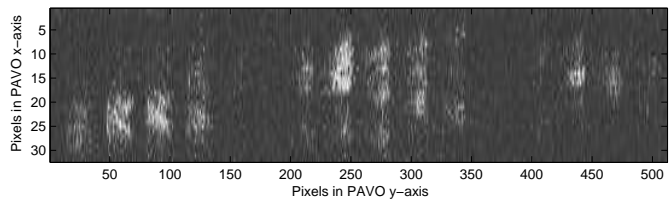}
\end{center}
\caption{An interferogram of $\alpha$~Gru recorded by PAVO. It contains images
of spectrally dispersed (horizontally) pupils as sampled by an array of four
lenslets. The science pupil is on the left of the frame. The pupils in the
middle and right of the frame are used for tip-tilt control. They are ignored by
the data reduction pipeline.}
\label{fig:drp_pavogram}
\end{figure}

The PAVO interferograms are 32 by 512 pixels images recorded in 4.2ms of
exposure time. Fig.~\ref{fig:drp_pavogram} shows an example of a PAVO
interferogram. They contain images of three pupils, one to be used for fringe
tracking (left in the figure) and two to be used for tip-tilt correction (middle
and right in the figure). Each pupil is imaged by 4 lenslets and spectrally
dispersed by a prism. Each dispersed segment of a pupil is 33-pixel wide, which
represents individual spectral channels of the PAVO spectrograph. The PAVO
interferograms are stored as FITS files with headers that contain various
information such as the status of the fringe lock, timestamp, etc.

\begin{figure}
\begin{center}
\includegraphics[width=0.45\columnwidth]{\imgdir/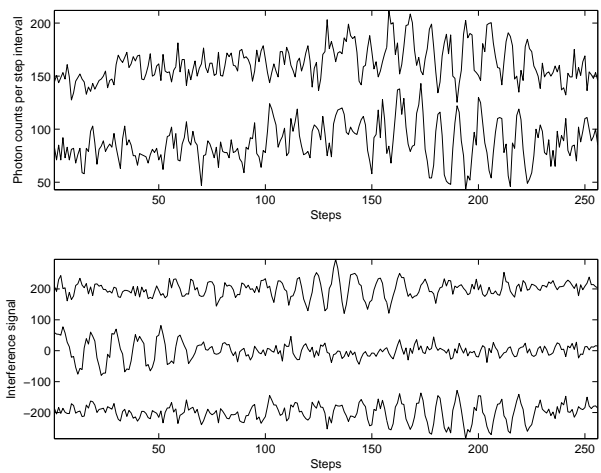}
\end{center}
\caption{Interferograms of $\beta$~Cru recorded by MUSCA. The top plots are
photon counts recorded by the pair of APDs, one on each output port of the
beam-splitter, hence the $\pi$ radian phase difference for the fringe signals.
The bottom plots show the difference of photon counts recorded by the APDs.
Three separate sweeps through the fringe envelope are depicted, illustrating the
varying position of the fringes at different times.}
\label{fig:drp_muscagram}
\end{figure}

The MUSCA interferograms are time series of photon counts recorded by the pair
of APDs as the scanning mirror scans through a predetermined range periodically.
The photon counts are stored as ASCII text files and each count has a timestamp
for synchronization with post-processed data from PAVO in the pipeline. The
temporal fringes recorded by the two APDs have a phase difference of $\pi$
radian due to the beam-splitter. Fig.~\ref{fig:drp_muscagram} shows several
scans of MUSCA interferograms with the upper panel illustrating the out-of-phase
relationship between signals from the two APDs. Let the \emph{interference}
signal of the time series be defined as the difference between the photon counts
recorded by APD0 and APD1 and the \emph{photometry} signal of the series be
defined as the sum of the photon counts recorded by the APDs. The interference
signal of the same scan in the upper panel of Fig.~\ref{fig:drp_muscagram} is
the lowest line in the lower panel in the same figure. Also plotted in the
lower panel are interference signals of two other scans at different times which
show the position of stellar fringe packets vary with time due to atmospheric
turbulence.

\subsection{Stage I} \label{sec:drp_pre}

\begin{figure}
\begin{center}
\includegraphics[width=0.7\textwidth]{\imgdir/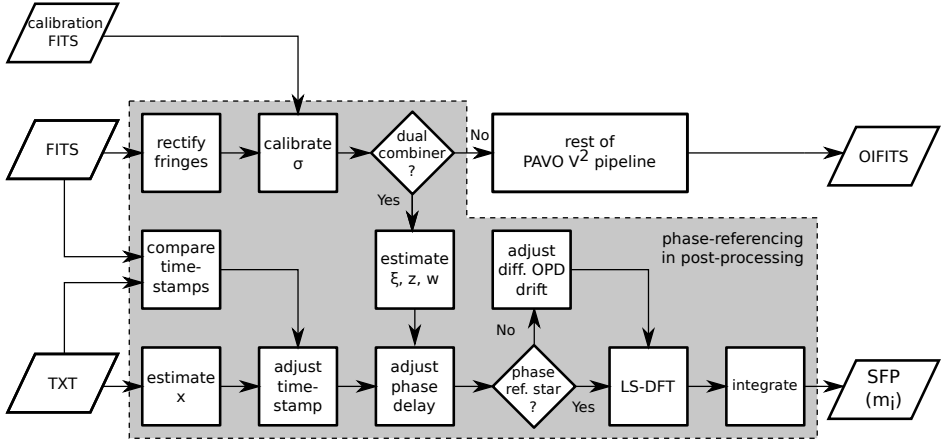}
\end{center}
\caption{The logical flow of Stage I of the PAVO+MUSCA pipeline. Subprocesses
within the shaded region are the phase-referencing part of the pipeline. The
main inputs to the pipeline are the PAVO (FITS) and MUSCA (TXT) interferograms.
If necessary, calibration FITS files which contain PAVO interferograms of
metrology light sources are also included to calibrate the PAVO spectrograph.
The output is a phase-referenced fringe packet (SFP).}
\label{fig:drp_phsref}
\end{figure}

The main goal of the first stage of the pipeline is to produce a
phase-referenced fringe packet that has an average phase error better than 5nm.
This stage is implemented in two parts, one for each type of interferogram. The
PAVO part of the pipeline is modified from an existing $V^2$ pipeline developed
for the PAVO $V^2$ science observation \irelandpavoatsusi. It is written in the
IDL programming language. On the other hand, the MUSCA part of the pipeline was
developed from scratch by the first author and is written in MATLAB. A diagram
showing the interdependency between the PAVO and the MUSCA parts in this stage
of the pipeline and the summary of their various subprocesses is given in
Fig.~\ref{fig:drp_phsref}. Only subprocesses within the shaded region in the
figure are related to the phase-referencing part of the PAVO+MUSCA pipeline. The
following subsections describe some of the more important subprocesses in the
figure in the context of phase-referencing.

\subsubsection{Estimating $z$} \label{sec:drp_opdz}

Only PAVO interferograms with stellar fringes are selected for further
processing. First, the interferograms are pre-processed to retain only the
relevant parts of the image (i.e.\ the spectrum of the science pupil) and to
correct for a small distortion in the spectrum (along the $x$-axis of the image,
the spatial fringes direction) which is caused by the glass prism. Then the
spatially modulated fringes are rectified by treating the fringes in each
spectral channel as an analytic signal and removing the carrier frequency
through a series of forward and inverse Fourier transform operations and data
manipulation in the Fourier domain. The outcome of the rectification is a
complex function that describes the envelope of the fringes in each spectral
channel, which is given as,
\begin{equation} \label{eq:drp_rhoj}
\begin{split}
\rho_{j,k} = \Omega_{j,k}\,e^{-i\phi_{j,k}},
\end{split}\end{equation}
where $j$ and $k$ are indices of a pixel in the $y$- and $x$-axis respectively
(see Fig.~\ref{fig:drp_pavogram}) while $\Omega_{j,k}$ is the function of the
fringe envelope and $\phi_{j,k}$ is a phase term attributed to the OPD of the
combining beams. The index $j$ refers to one of the 33 spectral channels while
$k$ refers to one of the 4 lenslets in PAVO. Only the first 21 channels are
within the PAVO science bandwidth and used for estimating $z$. Suppose
$\sigma_j$ represents the wavenumber of a spectral channel and $z(\sigma_j)$ or
$z_j$ is the OPD between two arms of SUSI at a particular wavenumber, then,
according to \citet{Tango:1990},
\begin{equation} \label{eq:drp_opdz}
\begin{split}
\sigma_jz_j = \bar{\sigma}z(\bar{\sigma}) + \xi(\sigma_j-\bar{\sigma}) + \psi_j,
\end{split}\end{equation}
where $\bar{\sigma}$ is the mean of the wavenumbers of all the spectral
channels, $\xi$ is the group delay of the fringes at $\bar{\sigma}$ and $\psi_j$
represents the higher order terms of the Taylor's expansion of $\sigma_jz_j$.
With an additional noise term, $\varepsilon_{j,k}$, to represent atmospheric
phase noise and/or optical aberration, then the argument of $\rho_{j,k}$ in
Eq.~\eqref{eq:drp_rhoj} is,
\begin{equation} \label{eq:drp_phi}
\begin{split}
\phi_{j,k} = 2\pi\sigma_jz_j + \varepsilon_{j,k}.
\end{split}\end{equation}

The first step in estimating $z_j$ is to estimate the group delay of the
fringes. This is done as follows. First, the autocorrelation of the phasor $p_j$
is calculated. The phasor, $p_j$, is defined as,
\begin{equation} \label{eq:drp_pj}
\begin{split}
p_j = \sum_k \rho_{j,k} 
    = \bar{\Omega}_j\,e^{-i\bar{\phi}_j}
    = \bar{\Omega}_j\,e^{-i\left(2\pi\sigma_jz_j+\bar{\varepsilon}_j\right)},
\end{split}\end{equation}
and its autocorrelation is,
\begin{equation} \label{eq:drp_qa}
\begin{split}
q_a &= \sum_j p_jp_{j+a}^* \\
    &= e^{-i2\pi a\Delta\sigma\xi}
       \sum_j\,\bar{\Omega}_j\bar{\Omega}_{j+a}\,e^{-i2\pi\Delta\psi_{j,a}}, \\
\end{split}\end{equation}
where $a\Delta\sigma = \sigma_j-\sigma_{j+a}$ and $\Delta\psi_{j,a} =
\psi_j+\bar{\varepsilon}_j - (\psi_{j,a}+\bar{\varepsilon}_{j+a})$. The
wavenumber difference between two adjacent spectral channels, $\Delta\sigma$, is
sufficiently regular across the entire spectral bandwidth to be considered
constant and to have negligible effect on the following analysis step.

Next, the Fourier transform of the vector $\mata{q}$ from Eq.~\eqref{eq:drp_qa}
is calculated using a Fast Fourier Transform (FFT) routine. Using the notation
$\hat{\mata{q}}$ to denote the Fourier transform of $\mata{q}$, then,
\begin{equation} \label{eq:drp_ftq}
\begin{split}
\fts{q}_b
  = \delta_{b-\check{b}} \ast \fts{\bar{\Omega}}_b = \fts{\bar{\Omega}}_{b-\check{b}},
\end{split}\end{equation}
where $\delta_{b-\check{b}}$ is a delta function at index $\check{b}$ and
$\check{b} = \Delta\sigma \xi N_{\text{FFT}}$. The second term of the
convolution in Eq.~(\ref{eq:drp_ftq}) is the power spectral density (PSD) of
$\bar{\Omega}_je^{-i(\psi_j+\bar{\varepsilon}_j)}$ and has a global maximum at
the origin. As a result the group delay of the fringes can be estimated from the
position of the global maximum of $\fts{q}_b$. The estimated group delay is,
\begin{equation} \label{eq:drp_xi}
\begin{split}
\tilde{\xi} &= \Delta\sigma^{-1}\tilde{b}/N_{\rm{FFT}},
\end{split}\end{equation}
where $\tilde{b}$ is the location of the global maximum of $\fts{q}_b$, which
may not necessarily be at $\check{b}$ as discussed in the next paragraph, and
$N_{\rm{FFT}}$ is the size of (or number of elements in) the vector as input to
the FFT routine. The tilde notation in Eq.~\eqref{eq:drp_xi} and throughout this
paper denotes estimation of a parameter. Since the coherence length of each
spectral channel, $\Delta\sigma^{-1}$, is $\mathtt{\sim}$30$\mu$m, this estimate
is reliable if the true group delay is within $\pm$15$\mu$m.

The assumption that $\psi_j$ is approximately constant across the bandwidth of
the PAVO spectrograph is ensured by the usage of a longitudinal dispersion
compensation system, or LDC \citep{Davis:1999}. However the similar assumption
regarding $\bar{\varepsilon}_j$ is only true when the atmospheric phase noise
across the pupil is small (i.e.\ seeing is good). If the phase noise variation
is large then the location of the global maximum of $\hat{q}_b$ may not be at
zero. In addition, a large variation also gives rise to spurious peaks in the
PSD.

In order to gauge the phase noise variation, a signal-to-noise (SNR) ratio,
$\eta$, is introduced. $\eta$ is defined as the ratio of the value of the
dominant peak to the noise floor of the PSD where the noise floor is defined as
the standard deviation of the PSD with the dominant peak removed. The value of
$\eta$, which is inversely proportional to the phase noise variation, is also
used to gauge the accuracy of the OPD estimate, $\tilde{z}_j$, because the
requirement is that $\tilde{\xi}$ must be accurate to within one wavelength.
From computer simulation, an $\eta$ of 5 is acceptable while an $\eta$ of 6 is
good \citep{Kok:2013a}.

If the optical media in which the starlight propagates to reach the beam
combiners are non-dispersive then the group delay is also the OPD affecting the
fringes at all wavenumbers. However, except for the optical path in the vacuum
pipe near the siderostats, all optical media along the optical path of PAVO and
MUSCA are dispersive.

The last step in estimating $z_j$ is to adjust the group delay estimate by a
fraction of a wavelength according to the dispersion factor, $\delta_{\ell}$,
and length of the optical path propagated by the light source, $L_{\ell}$.
Suppose $\alpha$ is the value to be adjusted and its estimate is defined as,
\begin{equation}
\begin{split}
\alpha
  &= \sum_{\ell=1}^{N_{\rm{MED}}} \delta_{\ell}L_{\ell}, \quad \text{or}, \\
e^{i\tilde{\alpha}}
  &= \left\langle e^{i\bar{\phi}_j}e^{-i2\pi\sigma_j\tilde{\xi}}\right\rangle,
\end{split}\end{equation}
where $N_{\rm{MED}}$ is the number of different optical media along the
propagation path, then,
\begin{equation} \label{eq:z}
\tilde{z}_j = \tilde{\xi} + \frac{\tilde{\alpha}}{2\pi\sigma_j}.
\end{equation}
Finally, suppose $\sigma_{\rm{M}}$ is the mean wavenumber of the MUSCA's
operating bandwidth, which is $\mathtt{\sim}$1.2$\mu$m$^{-1}$, then the OPD
estimate at that wavelength is obtained by evaluating Eq.~\eqref{eq:z} at
$\sigma_{\rm{M}}$.

\subsubsection{Estimating $w$} \label{sec:drp_opdw}

\begin{figure}
\begin{center}
\includegraphics[width=0.45\columnwidth]{\imgdir/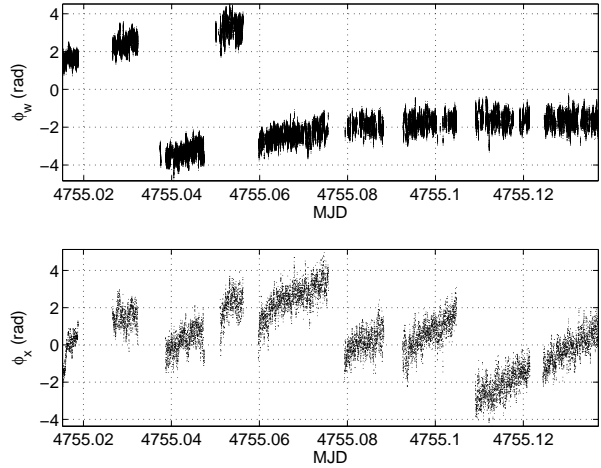}
\end{center}
\caption{Wrapped phases of the WL (top) and the SL (bottom) metrology fringes
used to estimate $w$ and $x$ respectively.}
\label{fig:drp_opdwx_wrapped}
\end{figure}

The WL metrology fringes are recorded at the same time as the stellar fringes in
PAVO. After undergoing the same pre-processing, the metrology fringes are used
to estimate $w$ in a slightly different way than the stellar fringes to estimate
$z$. Since the optical paths probed by $w$ are contained within an enclosed
climate controlled environment it is not expected to vary beyond one wavelength
of the metrology fringes within the one unit of a PAVO interferogram exposure
time. Therefore the OPD probed by the metrology can measured by converting the
phase of the metrology fringes, which is the argument of $p_j$ in
Eq.~\eqref{eq:drp_pj}, into distance. The estimate of $w$ is given as,
\begin{equation} \label{eq:drp_w}
\begin{split}
\tilde{w}
  = \frac{2\pi N_{\sigma\rm{I}} + \arg(p_{\rm{I}})}{2\pi\sigma_{\rm{I}}},
\end{split}\end{equation}
where the subscript I is the spectral channel at $\sigma_{\rm{I}}$ and
$N_{\sigma\rm{I}}$ represents an integer number of cycles over which the fringes
may have wrapped around. $N_{\sigma\rm{I}}$ is obtained by unwrapping the phase
measurement collected throughout several sets of observations (or the entire
night) because the phase measurement obtained directly from a `single'
wavelength fringes is modulo of $2\pi$. The top plot in
Fig.~\ref{fig:drp_opdwx_wrapped} shows the argument of $p_{\rm{I}}$ without
taking into account phase wrapping.

\subsubsection{Estimating $x$} \label{sec:drp_opdx}

\begin{figure}
\begin{center}
\subfloat[]{\includegraphics[width=0.35\textwidth]{\imgdir/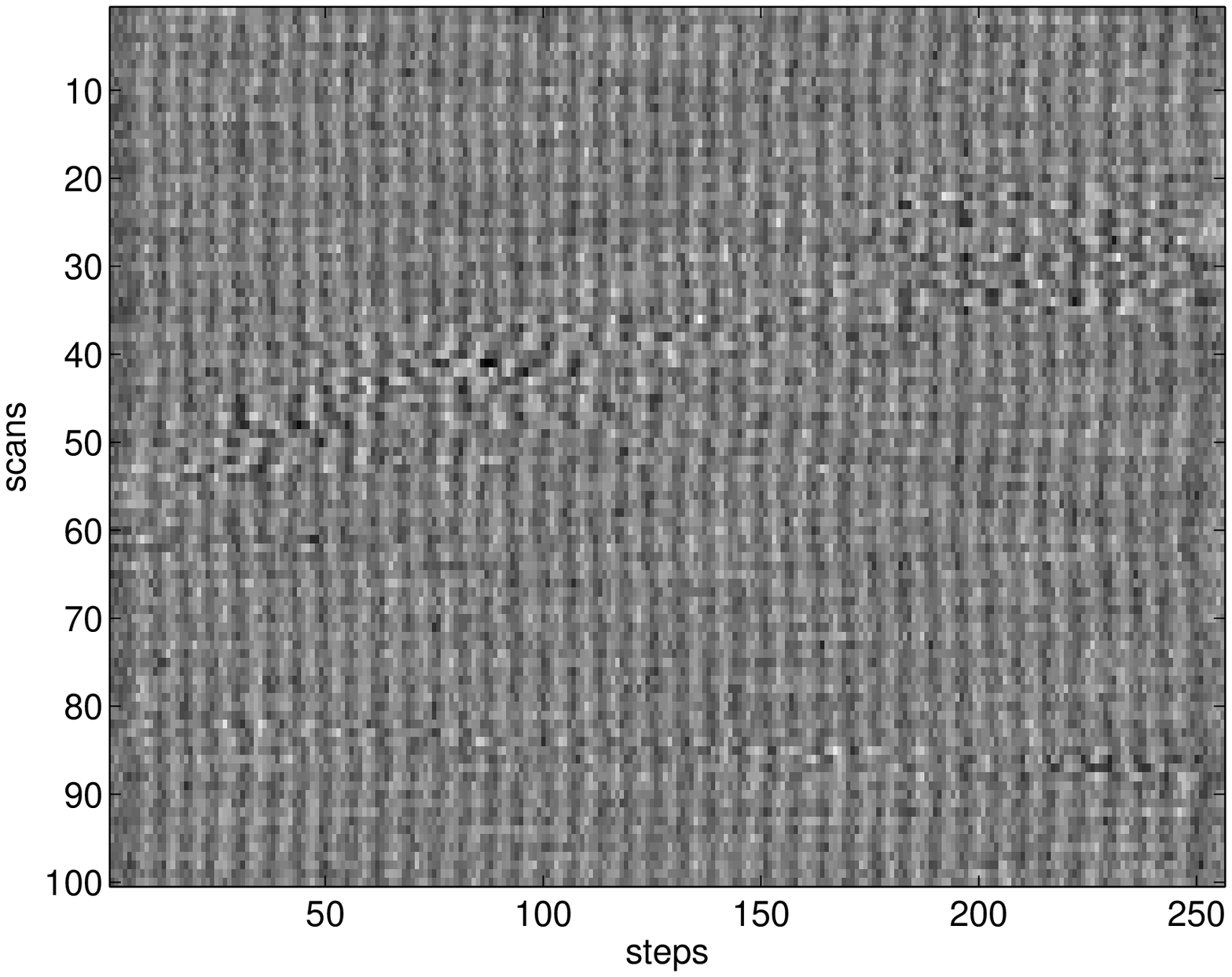}}
\subfloat[]{\includegraphics[width=0.35\textwidth]{\imgdir/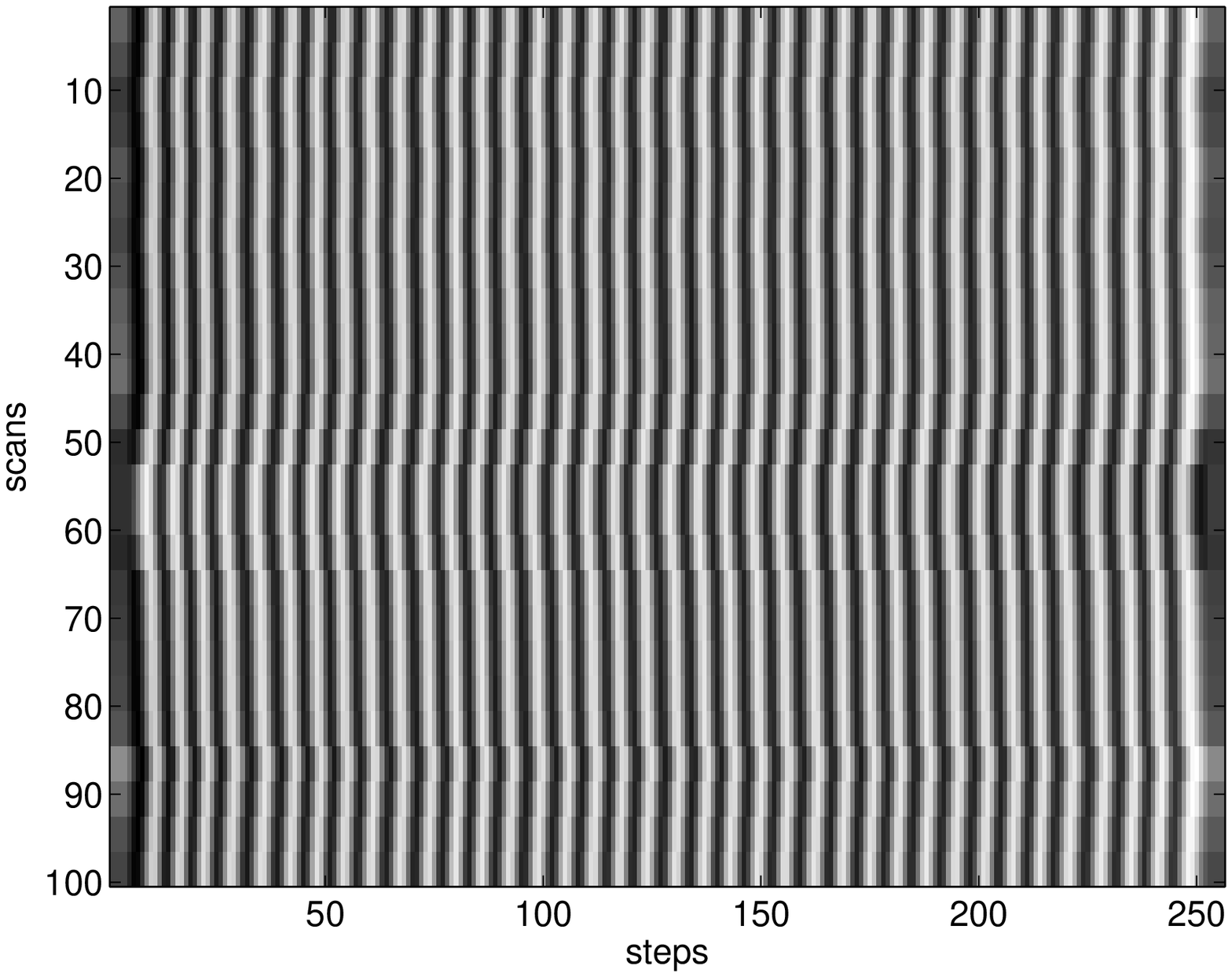}}
\end{center}
\caption{Waterfall plot of (a) SL metrology and MUSCA science fringes as
recorded and (b) SL metrology fringes after SNR enhancement and removal of the
science fringes.}
\label{fig:drp_slmet}
\end{figure}

The SL metrology fringes are also recorded by MUSCA together with science
fringes. As a result of that, the interference signal of the metrology fringes
imposes a measurement grid across the scan range of the scanning mirror. The
spacing of the lines, which is defined by the wavelength of the metrology laser,
is used to measure the change in OPD induced by the mirror. A visualization
of this grid is shown in Fig.~\ref{fig:drp_slmet}(a). Instead of using a Fourier
filtering method, the faint metrology fringes are extracted by shifting a
pre-recorded scan of high SNR reference fringes in the spatial domain and the
amount of shift to be applied is measured from the phase of the metrology
fringes recorded during the observation. This method assumes that the metrology
fringes produced by each scan of the scanning mirror in OPD, denoted by
$\ell(x)$, is a spatially shifted version of fringes produced by a reference
scan, $\ell_{\rm{R}}(x)$. Physically, the spatial shift is caused by internal
laboratory seeing which has a timescale\footnote{duration over which the average
phase variation is 1 radian, computed from the spectral density of the measured
phase variation} of several tens of seconds to a minute. This is very slow
compared to the time taken by the scanning mirror to complete an up and down
scan cycle, which is about 150ms. Therefore the assumption that path length
variations are much less than one wavelength between successive scans is
reasonable. Suppose $\fts{\ell}_{\rm{R}}$ represents the Fourier transform of
$\ell_{\rm{R}}(x)$ and the $\tilde{\ell}(x)$ represents an estimate of
$\ell(x)$, then,
\begin{equation}
\tilde{\ell}(x) = \ift{\fts{\ell}_{\rm{R}}
\exp\left(i\frac{\sigma}{\sigma_{\rm{G}}}\phi_{\rm{G}}\right)},
\end{equation}
where $\phi_{\rm{G}}$ is phase of the metrology fringes recorded during
observation. A waterfall plot of a set of recovered metrology fringes (or
$\tilde{\ell}$) over time is shown in Fig.~\ref{fig:drp_slmet}(b). The phase
variation of the metrology fringes in Fig.~\ref{fig:drp_slmet}(a) is preserved
in Fig.~\ref{fig:drp_slmet}(b). However, the SNR of the fringes is enhanced
significantly.

The phase delay of the fringes is then computed by inverting the function,
$\tilde{\ell}(x)$. Mathematically,
\begin{equation} \label{eq:drp_x}
\tilde{x} = \frac{N_{\sigma\rm{G}}}{\sigma_{\rm{G}}} + \tilde{\ell}\,^{-1}\left(\tilde{\ell}(x)\right).
\end{equation}
The function $\tilde{\ell}(x)$, which is in the form of,
\begin{equation}
\tilde{\ell}(x) = A\cos\left(2\pi\sigma_{\rm{G}}x + \phi_{\rm{G}}\right),
\end{equation}
is not invertible by definition due to the periodic nature of a cosine function
but with additional computation (e.g.\ the derivative of the function),
$\tilde{x}$ can be recovered for the entire scan range.

The phase delay in the middle of the scan range is defined to be zero thereby
providing a reference point for measuring positions of phase-referenced fringe
packets in the next stage of the pipeline. However, like the measurement with
the WL metrology, the phase measurement of the SL metrology fringes must also be
corrected for phase-wrapping. The term $N_{\sigma\rm{G}}$ in
Eq.~\eqref{eq:drp_x} is added to represent an integer number of cycles that the
fringe phase has wrapped around and is obtained by unwrapping the phase
measurement collected throughout several sets of observations. The lower plot of
Fig.~\ref{fig:drp_opdwx_wrapped} shows the phase of the SL metrology fringes
before unwrapping.

\subsubsection{Estimating $m$}

The phase delay of the MUSCA fringes can be estimated by combining phase delay
measurements from PAVO and the two metrology systems discussed in
Sec.~\ref{sec:met_wl} and \ref{sec:met_sl}, which is given as,
\begin{equation} \label{eq:drp_m}
\begin{split}
\tilde{m} = \tilde{z} - \tilde{w} + \tilde{x}.
\end{split}\end{equation}
The values of $\tilde{z}$ and $\tilde{w}$ are linearly interpolated
to simulate the same sampling rate as $\tilde{x}$ before Eq.~\ref{eq:drp_m}
is computed. In order to avoid confusion, the values of $\tilde{m}$ do not
represent the photon counts in the MUSCA interferograms. Instead, they represent
OPD at each step in one mirror scan. If the stellar fringes are to be plotted as
a graph, the photon counts are on the ordinate and values of $\tilde{m}$ are on
the abscissa.

\subsubsection{Coherent integration}

\begin{figure}
\begin{center}
\subfloat[]{\includegraphics[width=0.4\textwidth]{\imgdir/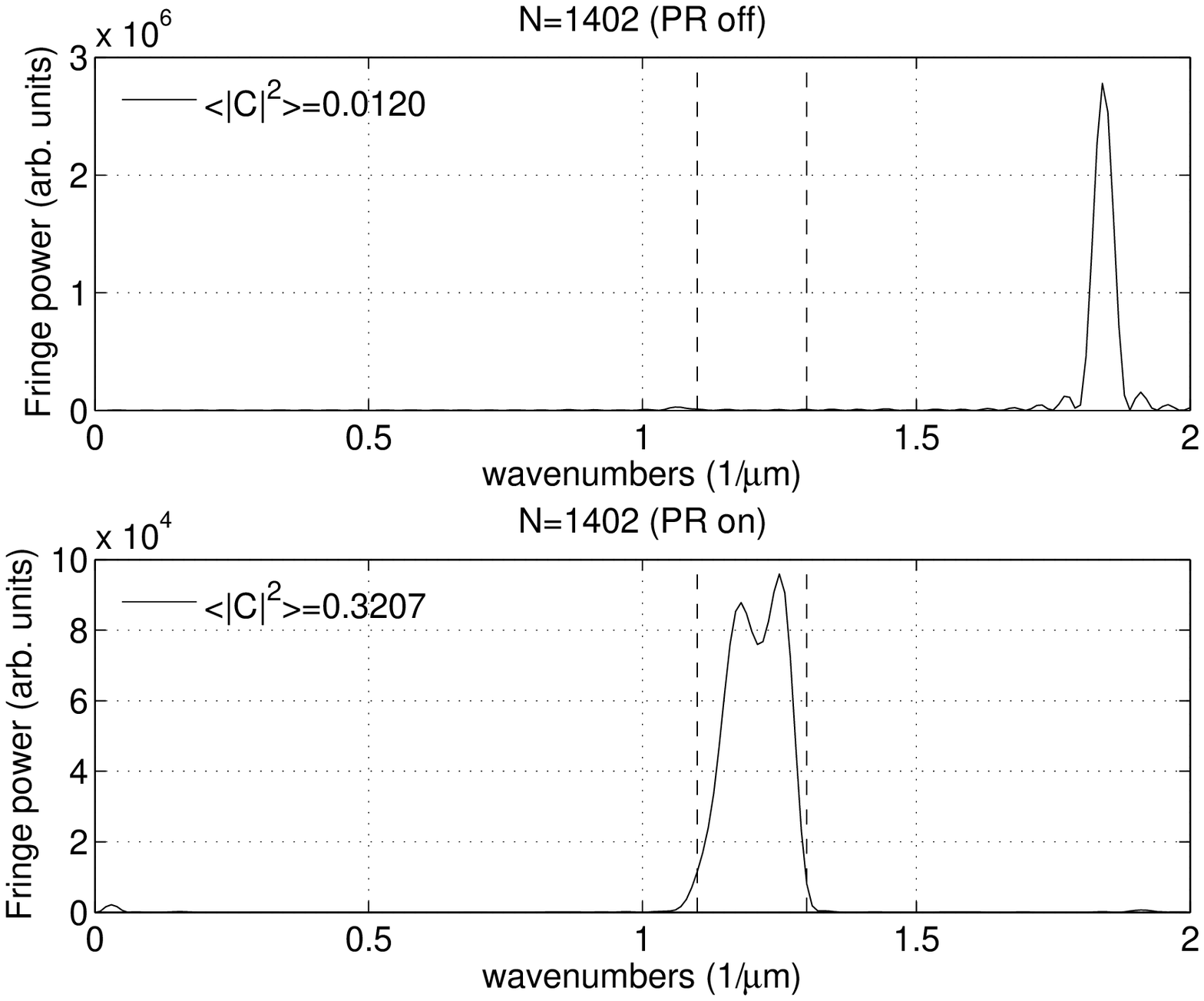}}
\subfloat[]{\includegraphics[width=0.42\textwidth]{\imgdir/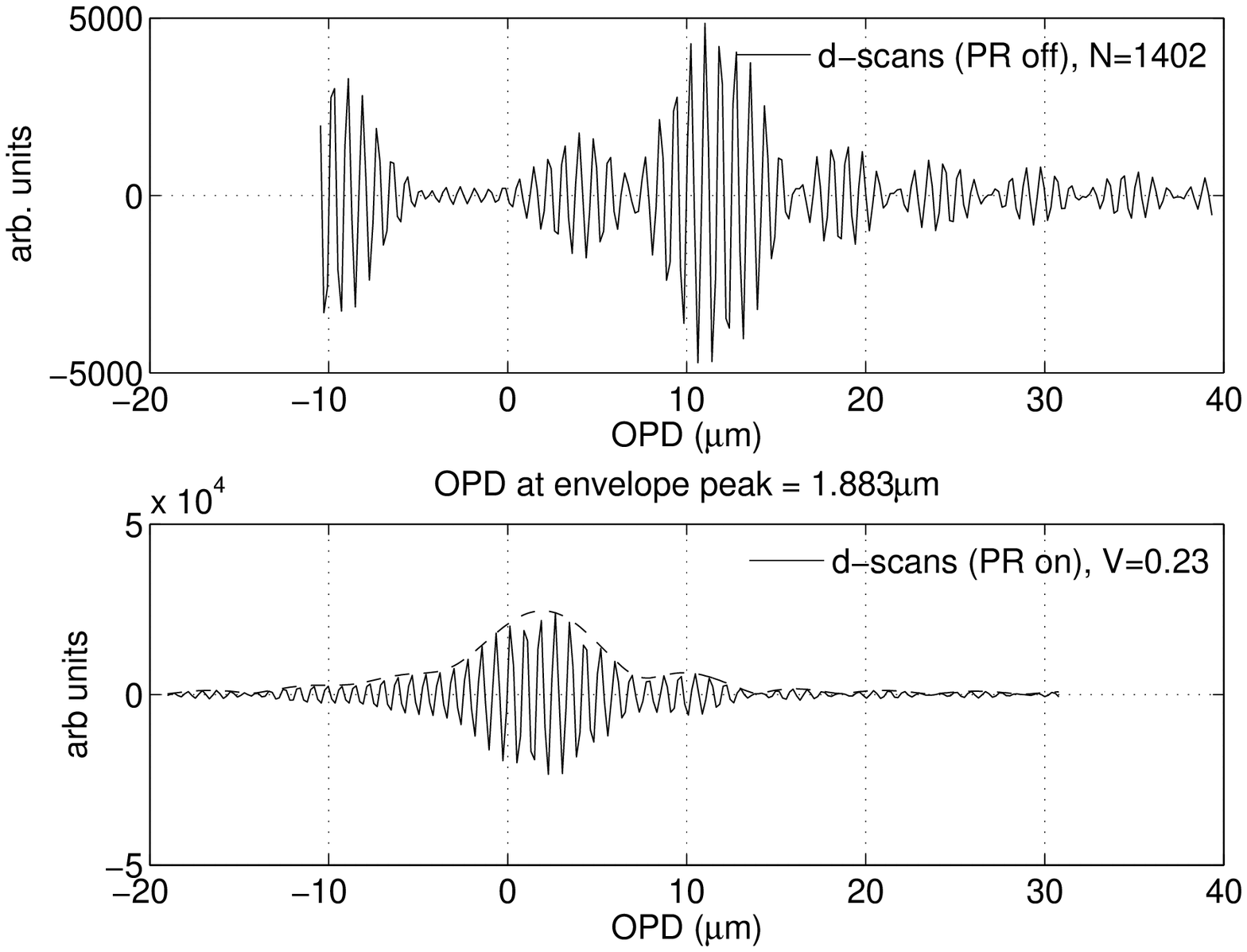}}
\end{center}
\caption{Plots in (a) shows the incoherently (top) and coherently (bottom)
integrated Fourier transform of a self phase-referenced fringe packet of
$\delta$ Orionis Aa. The peak in the top plot of (a) belongs to the SL metrology
laser because the metrology fringes are coherently integrated without
phase-referencing. However, when phase-referencing is engaged the fringes are
incoherently integrated and hence the absence of the metrology laser peak in the
bottom plot. The signal within the two vertical dashed lines in the bottom plot
of (a) belongs to the star. Plots in (b) shows the inverse Fourier transform of
(a) but only with signal between the dashed lines.}
\label{fig:drp_fp1_sample}
\end{figure}

\begin{figure}
\begin{center}
\subfloat[]{\includegraphics[width=0.4\textwidth]{\imgdir/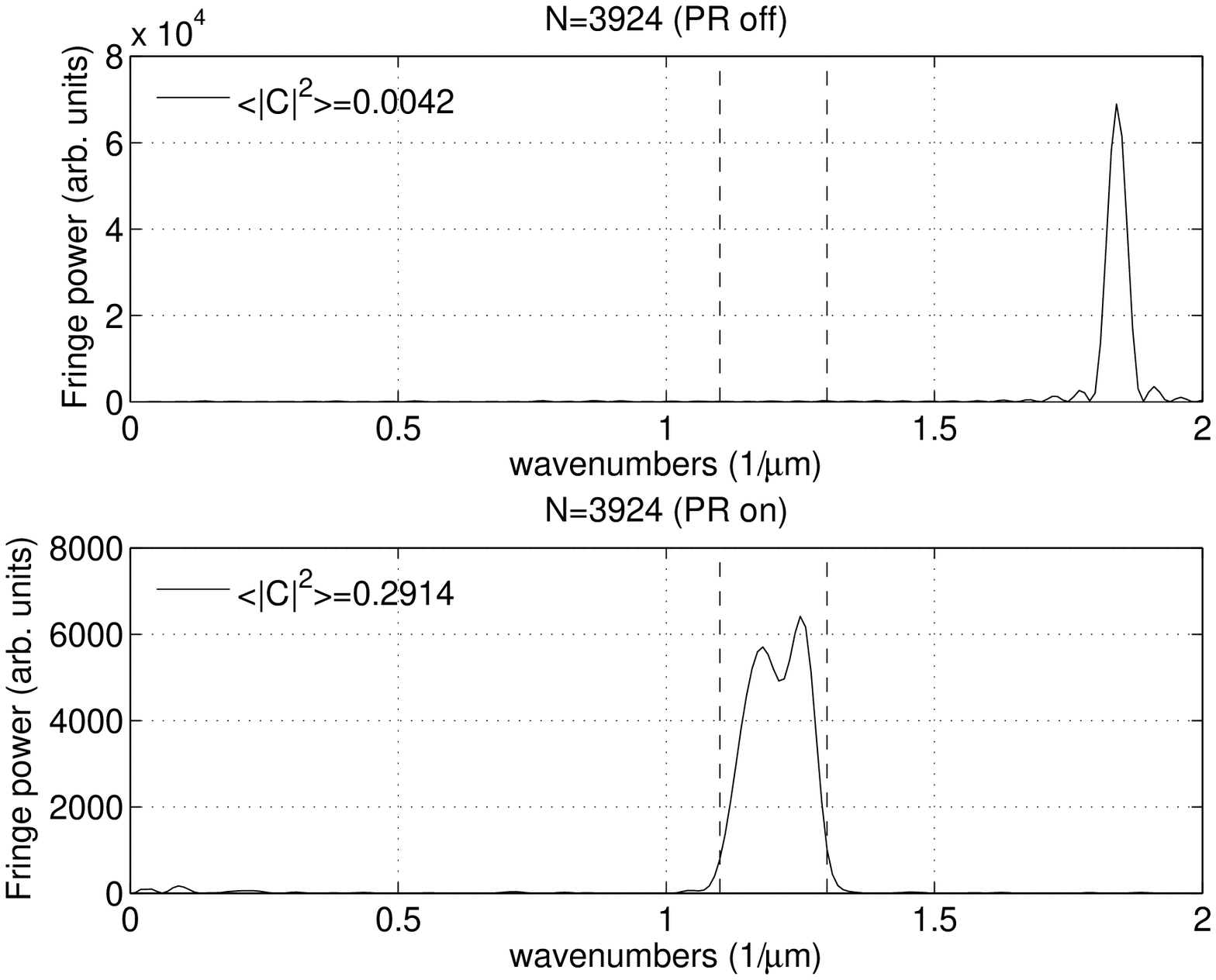}}
\subfloat[]{\includegraphics[width=0.41\textwidth]{\imgdir/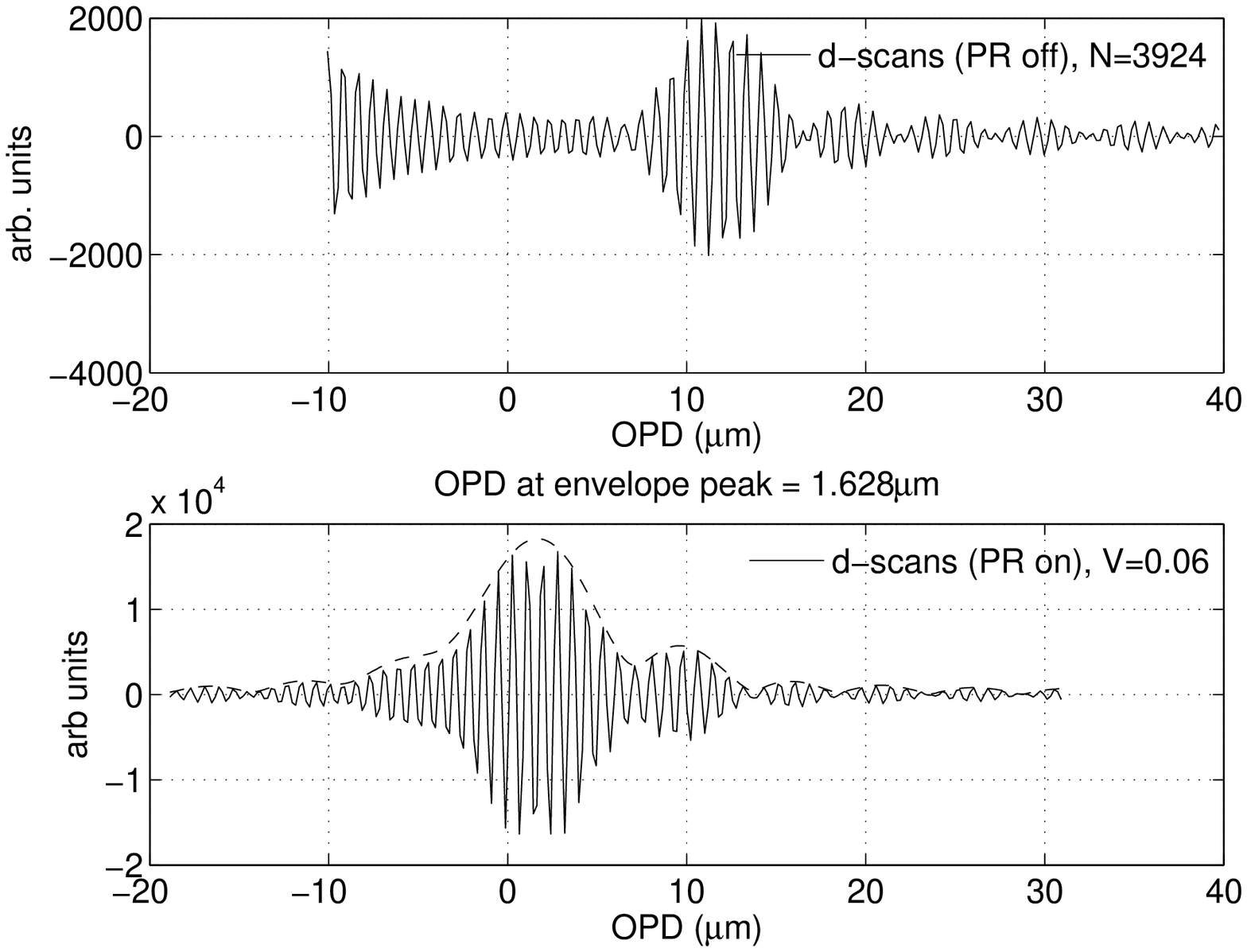}}
\end{center}
\caption{Similar to Fig.~\ref{fig:drp_fp1_sample}. Plots in (a) show the
incoherently (top) and coherently (bottom) integrated Fourier transform of a
dual phase-referenced fringe packet of $\delta$ Orionis Ab, which is separated
by $\sim$0.3$''$ from the reference star $\delta$ Orionis Aa. Plots in (b) show
the inverse Fourier transform of (a) but only with signal between the dashed
lines.}
\label{fig:drp_fp2_sample}
\end{figure}

In order to reduce the uncertainty in the phase delay estimation of $\tilde{m}$,
a number of scans can be coherently integrated together. This means the photon
counts are summed up according to their OPDs. However, only the photon counts
with reliable estimates of $m$ (has $\tilde{z}$ of $\eta \ge 5$) are integrated.
Due to contribution from the other phase delays, $\tilde{m}$ does not have
regular sampling. Therefore, it is easier to integrate the fringes in the
Fourier domain by first computing the Fourier transform of the fringes using a
discrete Fourier transform (DFT) algorithm which is capable of computing the
Fourier transform of a non-uniformly sampled signal \citep{Scargle:1989}. The
computed Fourier transforms have regular frequency spacing in the Fourier
domain. The algorithm also allows the frequency spacing in the Fourier domain to
be appropriately specified. Furthermore, by integrating the fringes in the
Fourier domain, additional phase error due to interpolation in the time domain
can be avoided. Fig.~\ref{fig:drp_fp1_sample} and \ref{fig:drp_fp2_sample} show
the result of coherently integrated phase-referenced fringe packets. They are
also the output of this first stage of the pipeline.

The phase-referenced fringe packet in Fig.~\ref{fig:drp_fp1_sample} is a result
of observing a target in self phase-referencing mode. This means both PAVO and
MUSCA were observing the same star. The phase-referenced fringe packet in
Fig.~\ref{fig:drp_fp2_sample} is a result of observing a target in dual-star
phase-referencing mode. In this mode, PAVO observed one component of a binary
star while MUSCA observed the other. Due to their relative position in the sky,
the
separation between fringe packets changes with time as the pair of stars
traverse the night sky. Therefore a correction term, $\Delta z$, must be added
to Eq.~\eqref{eq:drp_m} in order to accurately estimate the phase delay of the
MUSCA fringes, otherwise the integration will yield fringes of very low or zero
visibility because they are incoherently summed. Eq.~\eqref{eq:drp_m} for
dual-star phase-referencing becomes,
\begin{equation}
\tilde{m} = \left(\tilde{z} + \Delta{z}\right) - \tilde{w} + \tilde{x}.
\end{equation}
In its simplest form, $\Delta z$ is a function of drift velocity and elapsed
time (with respect to a chosen reference time, usually chosen to be at the start
of the dual-star observation). However, the pipeline uses an astronomical model,
which is based on the position angle and the on-sky separation of two stars, to
compute $\Delta z$. The same model is also used for narrow-angle astrometry and
is discussed in Sec.~\ref{sec:results}. Optimizing the amplitude of the
coherently integrated fringes by tweaking the position angle parameter in the
model does not immediately yield the astrometry of the two stars because
$\Delta z$ is also affected by the group delay drift effect, which is also
discussed in Sec.~\ref{sec:results}.

\subsubsection{Phase error}

The coherence of fringes across multiple scans can be assessed using a metric
which is defined as,
\begin{equation} \begin{split}
\langle |C|^2 \rangle
  &= \frac{\int_0^{\infty}\; |\langle\fts{I}(\sigma)\rangle|^2\; d\sigma}
  {\int_0^{\infty}\; \langle|\fts{I}(\sigma)|^2\rangle\; d\sigma}, \\
\end{split}\end{equation}
where $\fts{I}$ is the Fourier transform of the phase-referenced fringes and the
notation $\langle\rangle$ denotes an average over multiple scans. The metric
measures the weighted average of the phase variation across the Fourier
transform of the fringes. Similar to the Strehl ratio for measuring the
performance of an adaptive optics system, this metric has a value between 0 and
1. If the fringe integration is perfectly coherent over all scans, then the
metric will have a value of 1. On the other hand, if the fringes are completely
out of phase then the metric will have a value of 0. Now, if the phase variation
is random and has a normal distribution, then,
\begin{equation} \label{eq:c2_npdf}
\begin{split}
\langle |C|^2 \rangle
  &\approx e^{-\sigma_\varphi^2/2}, \\
\end{split}\end{equation}
where $\sigma_\varphi$ in this context is the standard deviation of the phase
variation. The average phase variation can then be estimated from the coherence
metric.

\subsection{Stage II}

\begin{figure}
\begin{center}
\includegraphics[width=0.45\columnwidth]{\imgdir/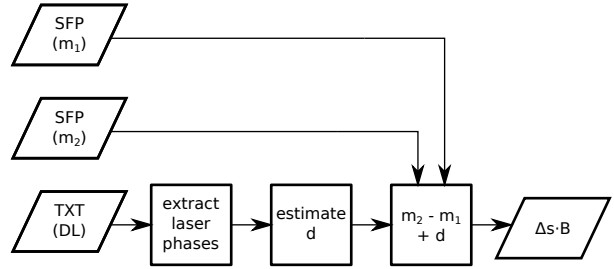}
\end{center}
\caption{The logical flow of Stage II of the PAVO+MUSCA pipeline. The main
inputs comprise the output of Stage I and a set of MUSCA interferograms
containing only the DL metrology fringes. The output is the true separation of
the given pair of fringe packets.}
\label{fig:drp_fpmeas_pipeline}
\end{figure}

The main goal of the second stage of the pipeline is to measure the separation
of two fringe packets at high precision. This stage is executed when the
previous stage has produced two or more phase-referenced fringe packets. In this
stage, the apparent and true separation of a pair of phase-referenced fringe
packets are computed.  Fig.~\ref{fig:drp_fpmeas_pipeline} illustrates the
logical flow of this stage of the pipeline, which is written entirely in MATLAB.

\subsubsection{Estimating $d$} \label{sec:drp_opdd}

The dual-laser metrology is used to measure a change in the position of the DDL
in MUSCA. The DDL is static when stellar fringes are being recorded and it is
only moved to alternate between different fringe packets. Therefore, $\tilde{d}$
is a one-off measurement of the DDL displacement between successive observations
for which metrology fringes are recorded just before and after the DDL is moved.
The estimated change in optical path is given as,
\begin{equation} \label{eq:drp_estd}
\tilde{d} = \frac
  {(\Delta\varphi_2 - \Delta\varphi_1)/2\pi + \Delta N}
  {\sigma_{\rm{R}} - \sigma_{\rm{G}}},
\end{equation}
where the numerator is the difference between the phases of the red and green
laser fringes at two different delay line positions indicated by the subscript 1
and 2, while $\sigma_{\rm{R}}$ and $\sigma_{\rm{G}}$ in the denominator
are the wavenumbers of the two lasers in vacuum. The details of data reduction
for this metrology are discussed in a separate paper \citep{Kok:2013}. Note that
the term $\tilde{d}$ in Eq.~\eqref{eq:drp_estd} is equal to
$n_{\text{M}}\mathtt{d}$, where $n_{\rm{M}}$ is the refractive index of air at
the mean MUSCA wavenumber $\sigma_{\rm{M}}$ and $\mathtt{d}$ is the displacement
of the DDL defined in Eq.~(2) in \citep{Kok:2013}.

\begin{figure}
\begin{center}
\subfloat[]{\includegraphics[width=0.5\textwidth]{\imgdir/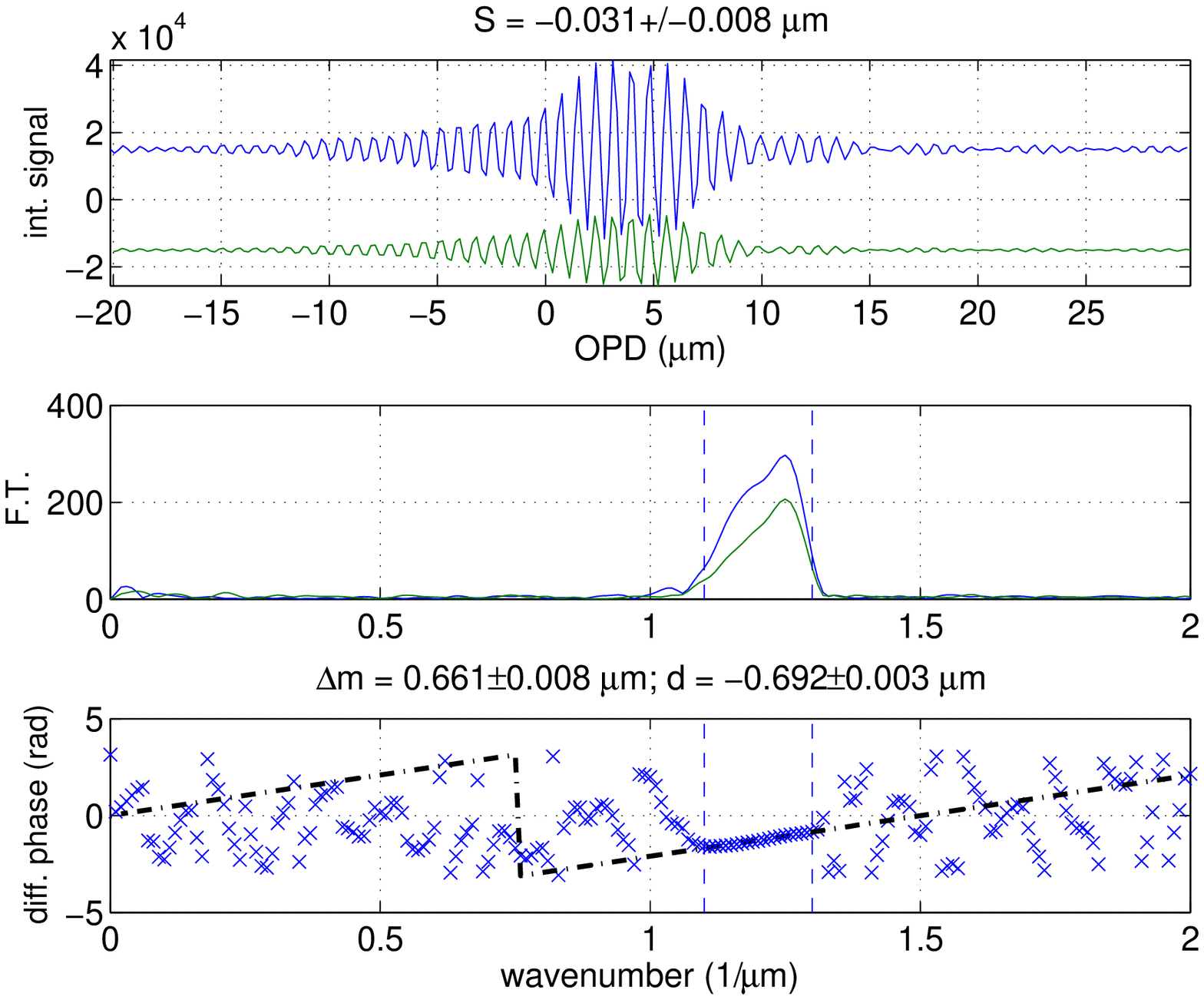}}
\subfloat[]{\includegraphics[width=0.5\textwidth]{\imgdir/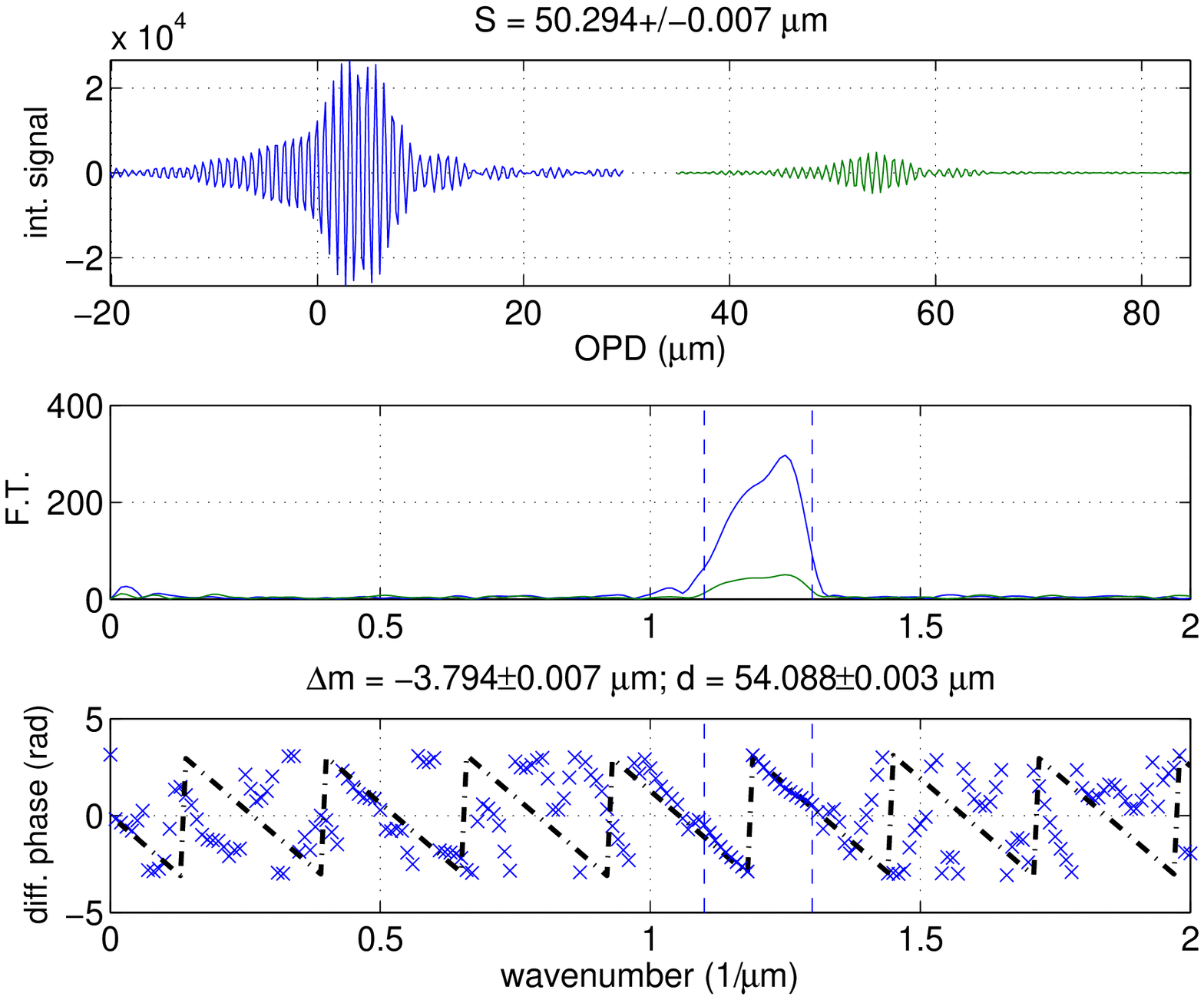}}\\
\subfloat[]{\includegraphics[width=0.5\textwidth]{\imgdir/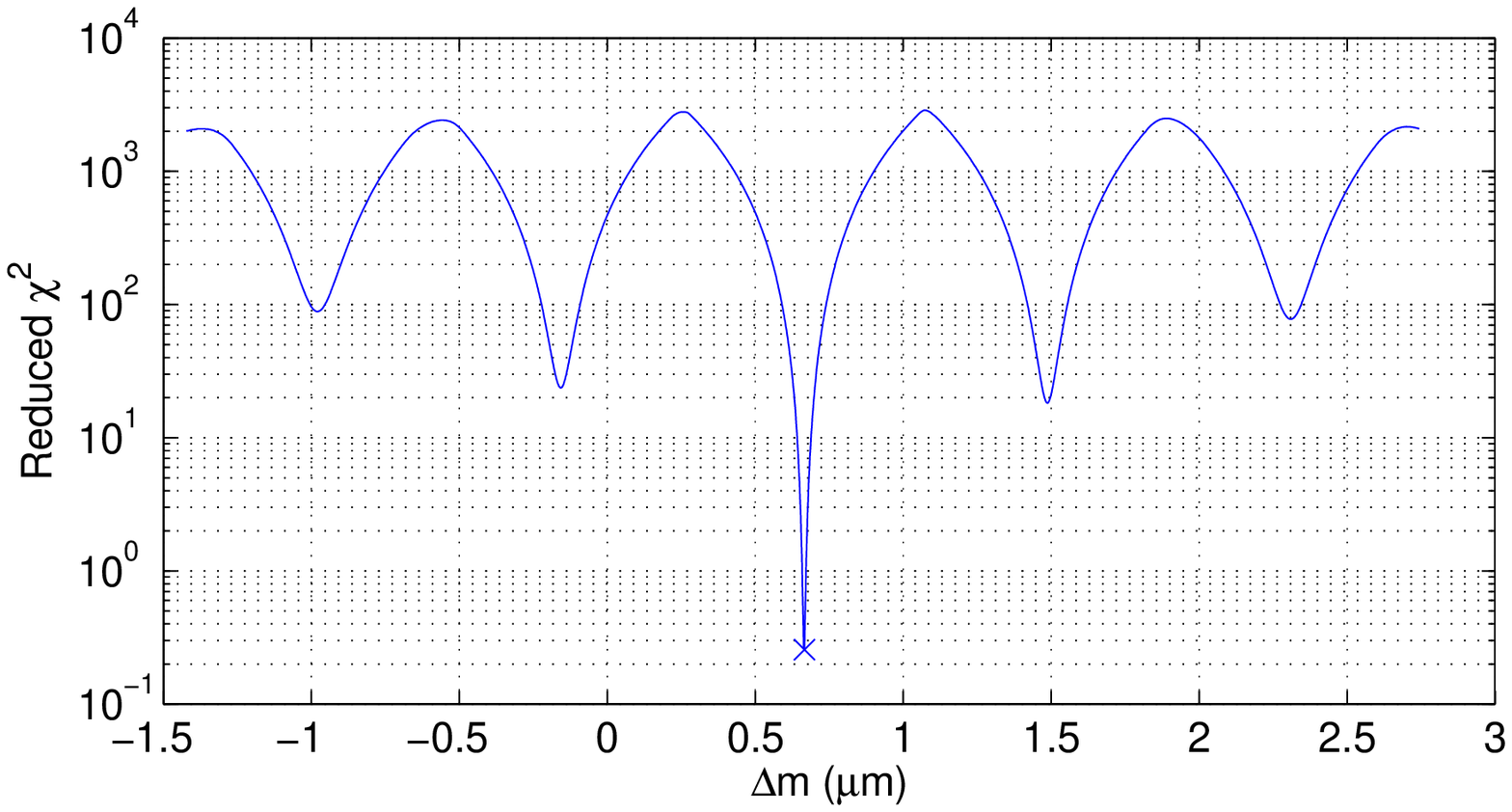}}
\subfloat[]{\includegraphics[width=0.5\textwidth]{\imgdir/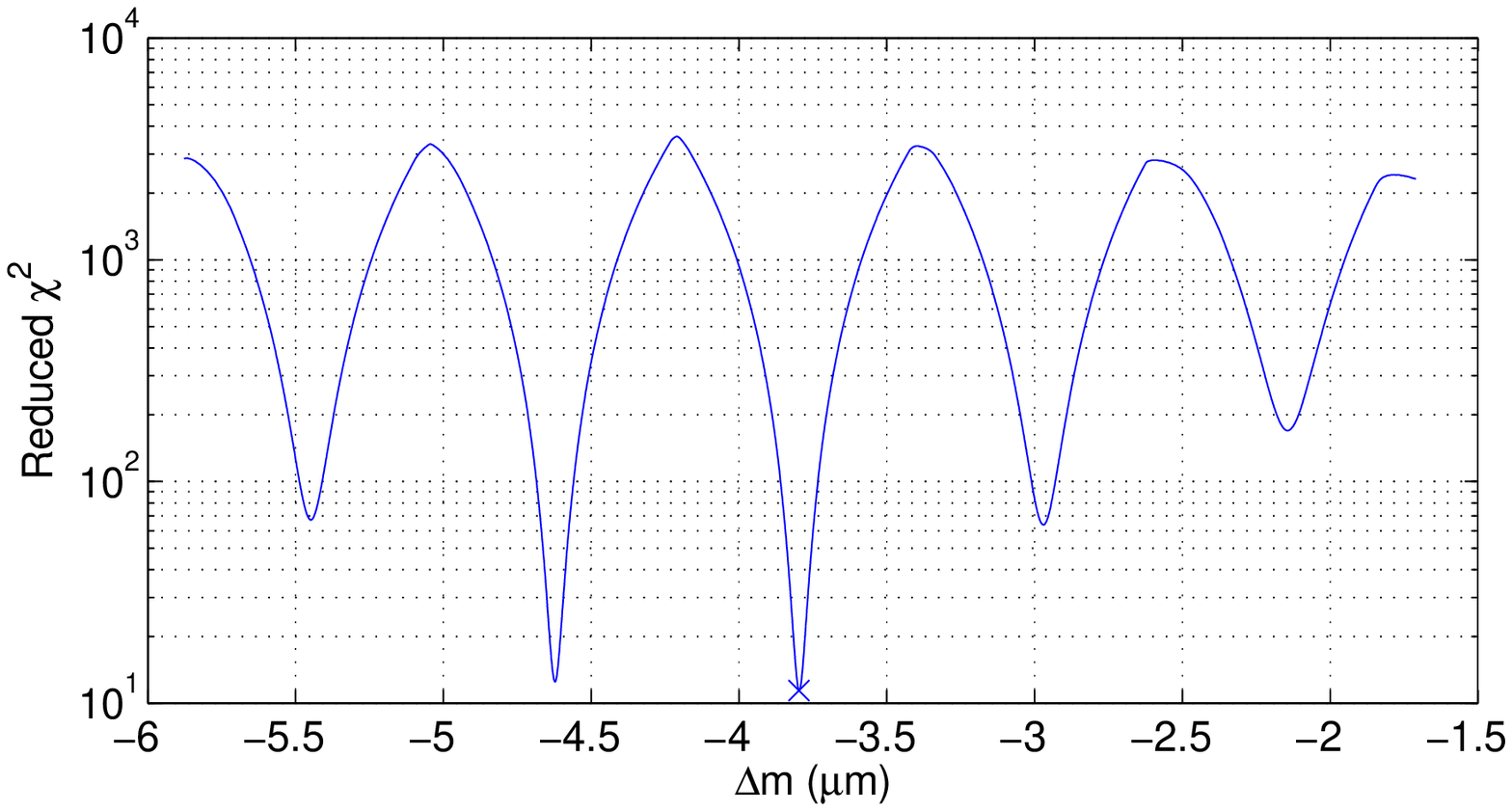}}\\
\end{center}
\caption{The top plot in (a) shows the phase-referenced fringe packet of
$\delta$ Orionis Aa observed at two different times while (b) shows the fringe
packet separation of $\delta$ Orionis Aa and Ab. The level of the interference
signals in the top plot of (a) have been arbitrarily adjusted for clarity. The
middle plots show the modulus amplitude of the Fourier transform of each fringe
packet in the top plot. The bottom plots show the phase difference of the
Fourier transform of the fringe packets (middle plots). The middle and bottom
plots have the same $x$-axis scale. The portion of the phase signal that
determines the apparent separation of the fringe packets, $\Delta m$, lies
between the two vertical dashed lines. The dash-dotted lines show the
least-squares fit to the phases in that region. The plots in (c) and (d) are the
goodness of fit, $\chi^2$, of possible values of $\Delta m$ to the phases in (a)
and (b) respectively. It is usually but not always the case that the value
having the global minimum $\chi^2$ (indicated by the symbol $\times$) is
associated with the accepted value for $\Delta m$.}
\label{fig:drp_fpmeas}
\end{figure}

\subsubsection{Estimating the true separation, $S$} \label{sec:drp_S}

The apparent separation of the two fringe packets, $\Delta m$, is estimated from
the phases of the product of the Fourier transform of one fringe packet and the
conjugate of the Fourier transform of the other fringe packet. Examples of the
modulus amplitude and phase of the Fourier transform products are shown in the
middle and lower plots of Fig.~\ref{fig:drp_fpmeas}. The phase delay at the mean
wavenumber is a measurement of the apparent separation of the fringe packets,
$\Delta\tilde{m}$, and is estimated from a linear least squares fit. Examples of
the goodness of fit, $\chi^2$, of possible values of $\Delta m$ are also plotted
in Fig.~\ref{fig:drp_fpmeas}. The value that has the global minimum $\chi^2$ is
usually but not always the case accepted as the value for $\Delta\tilde{m}$. The
presence of severe differential dispersion (between two fringe packets) and an
unresolved fringe packet embedded within one of the two fringe packets used in
the analysis may cause the result of the fit to be ambiguous by one wavelength.
Example of cases due to the latter are discussed in Sec.~\ref{sec:discussion}.
The true separation of a fringe packet pair, $S$, is then given as,
\begin{equation}
\tilde{S} = \Delta\tilde{m} + \tilde{d},
\end{equation}
where $\Delta\tilde{m} = \tilde{m}_2 - \tilde{m}_1$ is the estimated apparent
separation of the fringe packets. The true separations of the fringe packets in
the examples in Fig.~\ref{fig:drp_fpmeas} are indicated in the top most plots.

\section{Narrow-angle astrometry} \label{sec:results}
A complete set of MUSCA astrometric observations consists of a series of fringe
packet acquisitions. Fringes of a reference (primary component of a binary)
star, then a target (secondary component of a binary) star and finally back to
the reference star are acquired in sequential order. This is done so that the
systematic error in phase delay of the fringe packet caused by the group and
phase delay drift can be corrected.

\subsection{Group delay drift}

\begin{figure}
\centering
\subfloat[]{\includegraphics[width=0.45\textwidth]{\imgdir/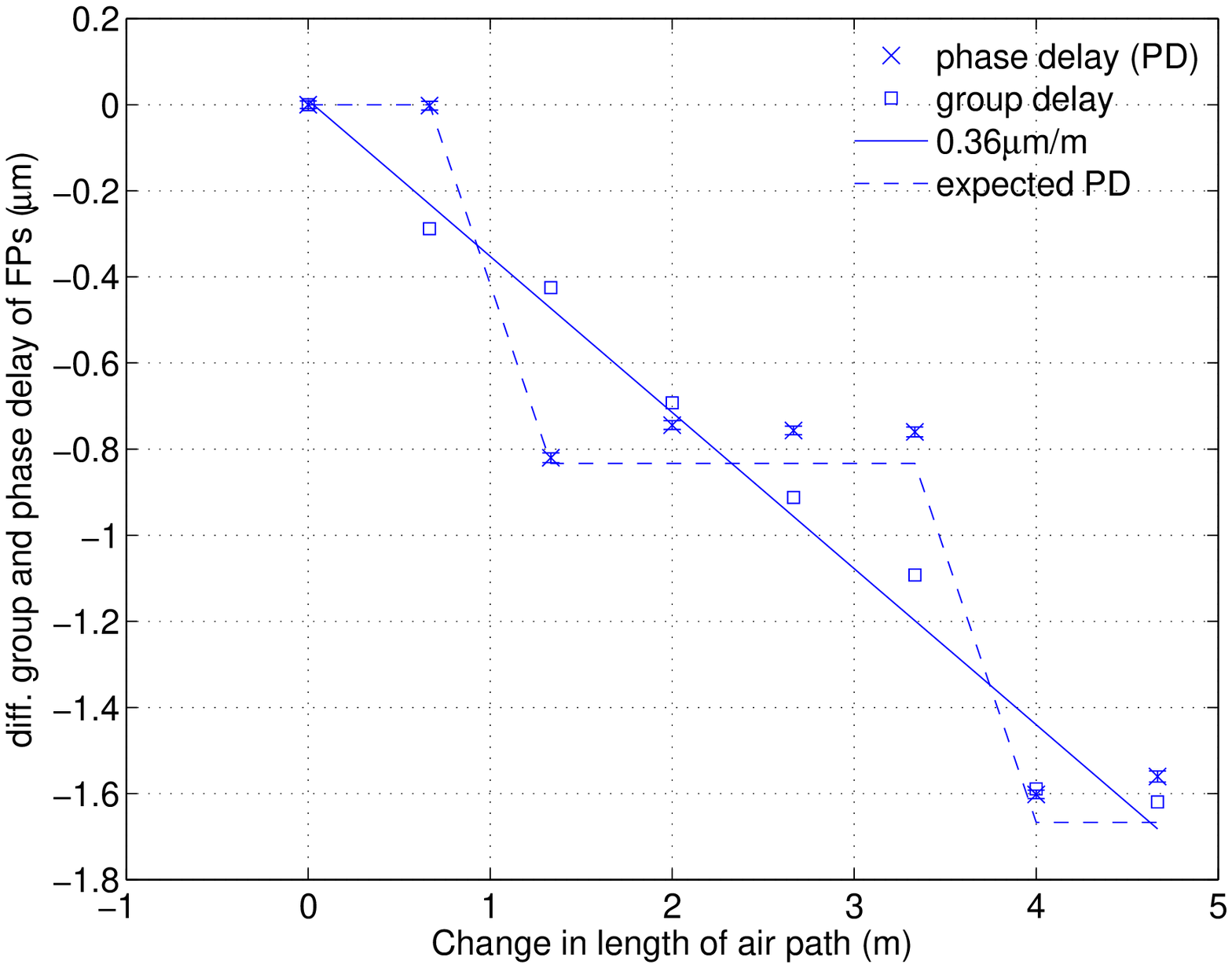}}
\hspace{1em}
\subfloat[]{\includegraphics[width=0.45\textwidth]{\imgdir/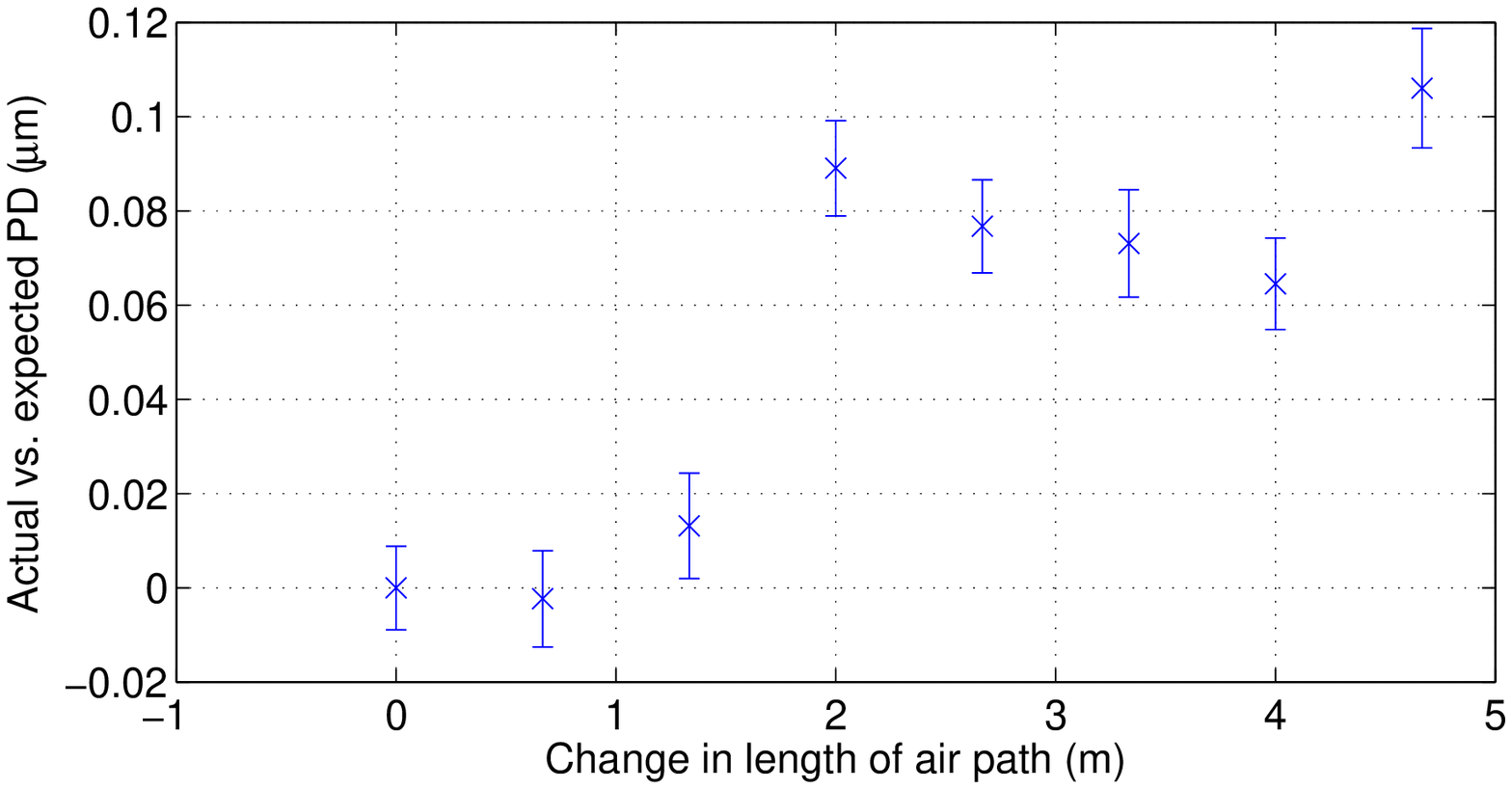}}
\caption[]{(a) Relative position (group and phase delay) of a fringe packet (FP) showing the effect of group delay drift. (b) The actual versus the expected phase delay (PD) of the FP showing the effect of phase delay drift. Both plots are derived from the PAVO and MUSCA simulators without atmospheric phase noise.}
\label{fig:drp_dopd_sample}
\end{figure}

The group delay drift occurs because optical media used to equalize the optical
path in SUSI (air and glass) are dispersive and the mean operating wavelengths
of PAVO and MUSCA are not the same. The rate of the drift is proportional to the length of the optical media along
the common optical path of the interfering starlight and the difference of group
delays at PAVO and MUSCA operating wavelength is about 0.36$\mu$m per 1m of air
path. This estimate, which used extrapolated numbers from \citet{Tango:1990},
takes into account the proportionality of optical path length in both air and
glass used in SUSI. Fig.~\ref{fig:drp_dopd_sample} plots the differential group
and phase delay (at mean wavenumber) of a fringe packet as extracted from the
phases of its Fourier transform. The computation is similar to the one discussed
in the previous section for computing $\Delta\tilde{m}$ but the reference fringe
packet in this case is a fringe packet from the same source but constructed with
a different amount of air path. Essentially the group delay in the plot shows
the relative position of a fringe packet of source when the amount of air path
is varied. The fringe packet in the figure is simulated with the PAVO and MUSCA
simulator. Although the group delay drifts according to the linear relation with
the air path length, the differential phase delay of the fringe packet varies in
discrete steps of about one MUSCA wavelength. This is expected because the
measurement is phase sensitive and does not change unless the differential group
delay changes by more than half a MUSCA wavelength from its previous value.

Due to this group delay drift effect, the separation between the primary and the
secondary fringe packets must be corrected to yield a value that depends only on
the projected separation of the binary star. Suppose $\xi_{S,P}$ represents the
differential group delay between the primary and the secondary fringe packets
which is interpolated from the slope of the phases of the Fourier transform of
the cross-correlation of the two primary fringe packets (e.g.\ the lower plots
of Fig.~\ref{fig:drp_fpmeas}(a)-(b)) of a given astrometric observation, then
the amount of adjustment to the fringe packet separation is,
\begin{equation}
\begin{split}
\Delta\tilde{S} = \xi_{S,P} + \frac{1}{2\pi\sigma_{\text{M}}}
\arg\left(e^{-i2\pi\sigma_{\text{M}}\xi_{S,P} + i\phi_{S,P}}\right),
\end{split}\end{equation}
where $\phi_{S,P}$ is a correction to the phase of the fringe packet due to a
similar phase delay drift effect which is discussed in the next section. In most
cases, $\Delta\tilde{S}$ is small ($\phi_{S,P}/2\pi\sigma_{\rm{M}}$) because the amount
of air path change within the time span of one bracketed astrometric observation
is usually shorter than 3m and therefore $\xi_{S,P}$ is usually less than half
of one MUSCA wavelength.

\subsection{Phase delay drift} \label{sec:drp_pddrift}

The phase delay of the MUSCA fringes also drifts systematically in proportion to
the length of the optical media along the common optical path of PAVO and MUSCA.
This phenomenon occurs because the effect of higher order dispersion is not
taken into account when estimating the phase delay at the mean MUSCA wavelength.
The amount of systematic error is given as
$\psi(\sigma_{\text{M}}-\bar{\sigma})$ where the function
$\psi(\sigma-\bar{\sigma})$ was first introduced in Eq.~\eqref{eq:drp_opdz} and
is explained in detail by \citet{Tango:1990}. The rate of drift is very small
but can become significant when the change in the uncompensated dispersive media
path length is large ($\mathtt{\sim}$5--10m) or the longitudinal dispersion is
not well compensated. A calculation based on refractive indices and dispersion
coefficients extrapolated from \citeauthor{Tango:1990}'s
\citeyearpar{Tango:1990} to MUSCA's wavelength suggests that the air and glass
combination in SUSI gives rise to phase delay drift of about 0.035$\mu$m for
every 1m air path of change. Fig.~\ref{fig:drp_dopd_sample}(b) shows the effect
of the phase delay drifts of a simulated fringe packet. Since the difference in
air path is unlikely to change by more than 10m within a time interval need to
make a bracketed astrometric observation, this error is unlikely to cause an
additional phase delay jump as seen in Fig.~\ref{fig:drp_dopd_sample}(a). The
amount of phase delay adjustment, $\phi_{S,P}$, to be applied to the fringe
separation is empirically interpolated from the measured phase delay difference
between two primary fringe packets within a bracketed observation.

\subsection{Calibration} \label{sec:obs_calibrators}

Single unresolved stars are good targets for instrument calibration. They are
used to test the stability of the reference phase of the fringes in MUSCA and
consequently the intrinsic precision of the instrument and its data reduction
pipeline. The intrinsic precision of the data reduction pipeline was evaluated
with simulated data \citep{Kok:2013a} and the results plotted to illustrate
group and phase delay shift are shown in previous sections. The reference fringe
phase in MUSCA can change due to instrumental and astrophysical reasons. The
main instrumental cause is a change in the position of the differential delay
line (DDL) in MUSCA while the main astrophysical cause is a change in the photo
center of the star under observation. Single unresolved stars are suitable
because they appear as point sources and therefore do not exhibit any photo
center shift.

Calibration tests were carried out in two stages. In the first stage, an
internal white light source was used. The goal of the test is to assess the
ability of the DDL to slew away and then slew back to (or close to) its original
position as well as the ability of the dual-laser (DL) metrology to measure the
displacement of the DDL. The reason an internal white light source is used at
this stage is to eliminate any astrophysical effect and to isolate the cause of
phase shifts to just the DDL and the metrology. The white light source is the
tungsten bulb which is also used for internal fringe searching. The optical
setup for this test is similar to the setup for the DL metrology (a Mach-Zehnder
interferometer). The non-common paths between the white light source and the DL
metrology are negligible because these two sources both originate from the same
pin hole and are injected into the optical path by the same optical components
(see Fig.~\ref{fig:opt_pavo_current}(b)). 

\begin{figure}
\centering
\subfloat[]{\includegraphics[width=0.45\textwidth]{\imgdir/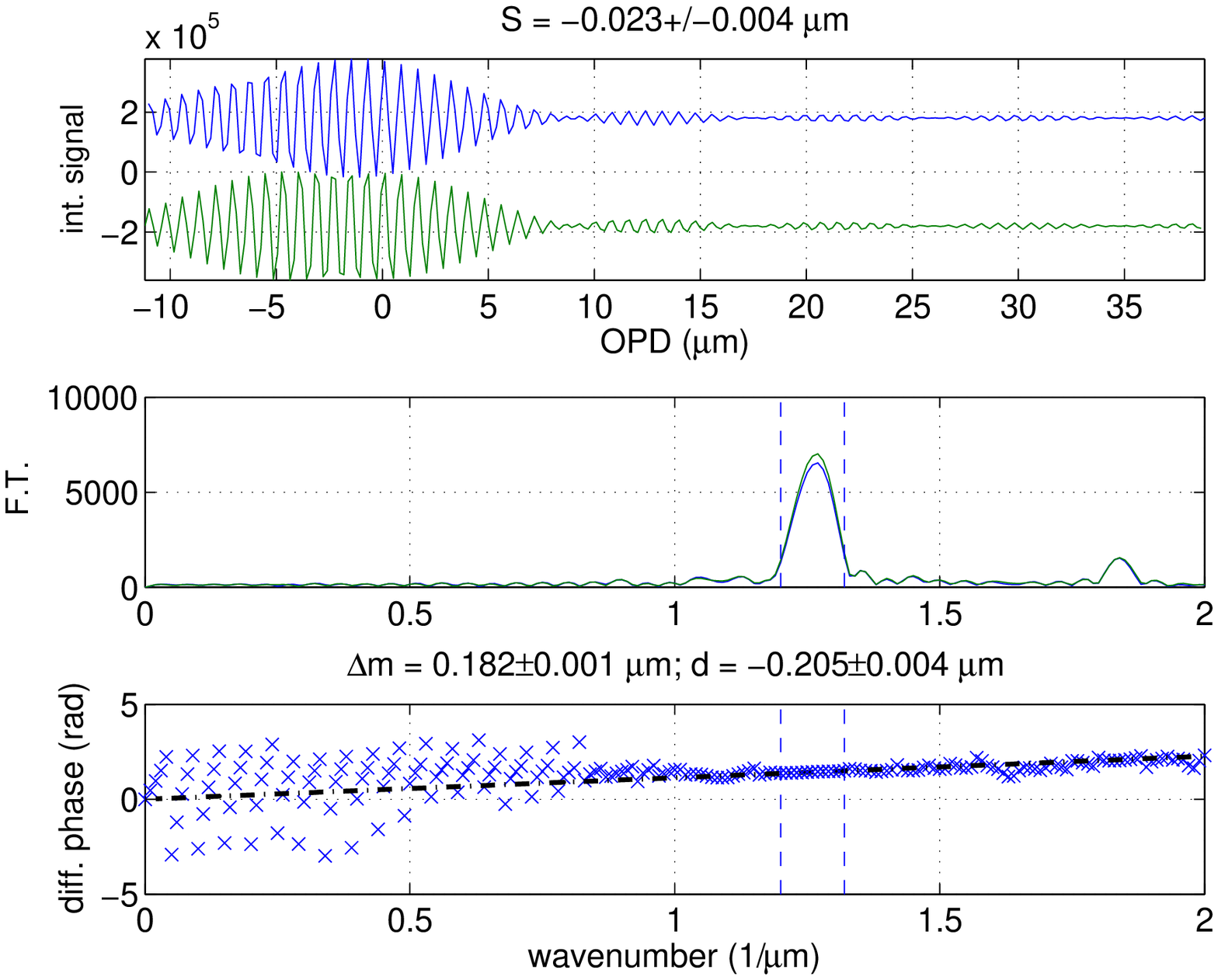}}
\hspace{1em}
\subfloat[]{\includegraphics[width=0.45\textwidth]{\imgdir/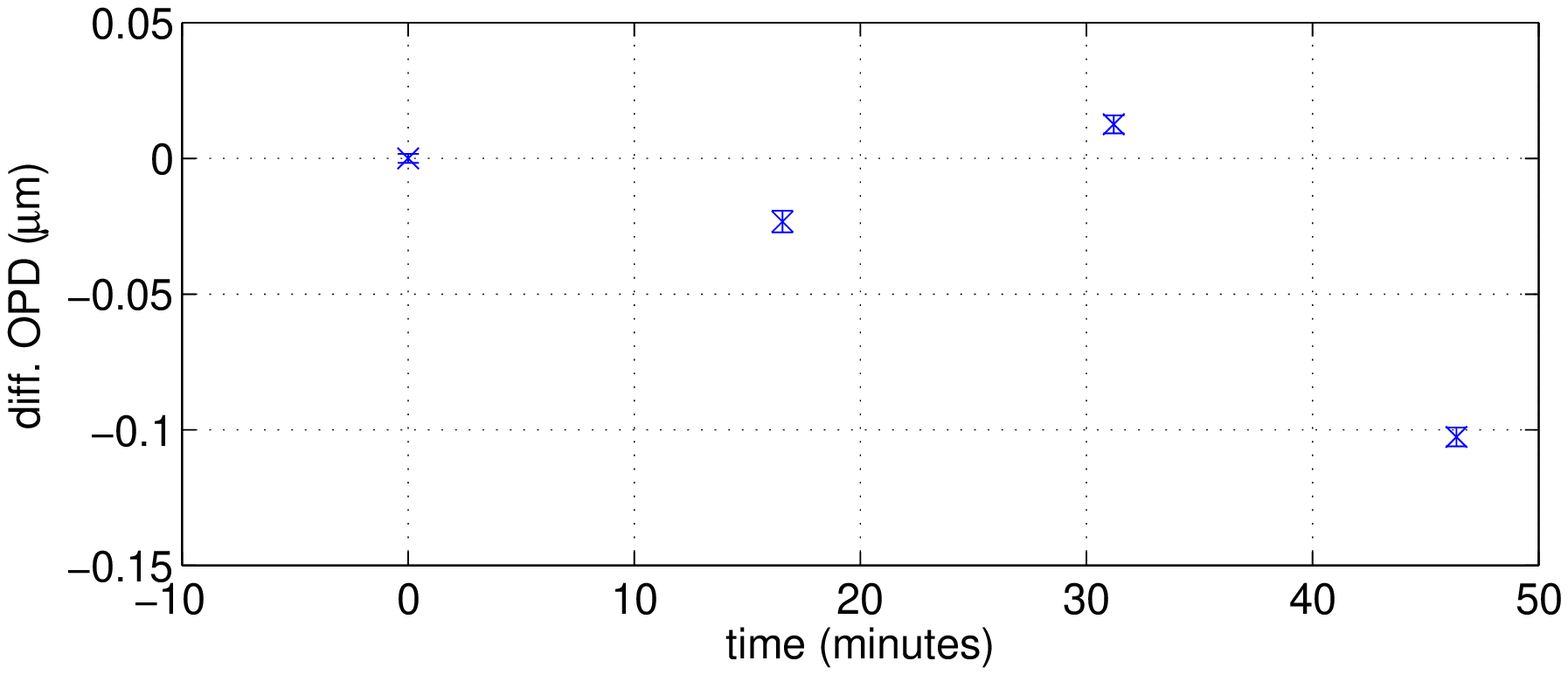}}
\caption[]{Comparison between the phases of a pair of fringe packets. In (a),
the top panel shows the two fringe packets which have their mean level adjusted
so that their phase alignment can be clearly seen while the middle and bottom
panels show the modulus amplitude and the differential phase of the Fourier
transform of the two fringe packets respectively. The displacement of the DDL in
MUSCA as measured by the DL metrology is indicated above the bottom panel. In
(b), the fringe separations of several pairs of fringe packets are shown. The
light source of the fringe packets in this figure is the internal laboratory
tungsten bulb.} \label{fig:obs_wl_130718}
\end{figure}

An example of the fringe packets of the internal white light source recorded
after every subsequent DDL slew away and back are shown in the top panel of
Fig.~\ref{fig:obs_wl_130718}(a). The middle and bottom panels of the figure
respectively show the modulus amplitude and phase of the Fourier transform of
the cross-correlation of the two fringe packets as described in
Sec.~\ref{sec:drp_S}. The differences in OPD or the separations (symbol $S$)
between pairs of fringe packets (measured in $\mu$m) from several consecutive
iterations of this test are shown in Fig.~\ref{fig:obs_wl_130718}(b). The first
fringe packet is used as the reference and therefore is plotted with a fringe
packet separation of zero. Similar but slightly different measurements, the
phase differences between pairs of fringe packets (measured in radians) are
listed in Table~\ref{tab:obs_cal_diffphase}. This measurement is not applicable
to the first data point in Fig.~\ref{fig:obs_wl_130718}(b), hence the table has
one data point less than the figure. The values in the table are equivalent to
2$\pi$ times the $1/\sigma_{\rm{M}}$ modulo of the points in the plot in
Fig.~\ref{fig:obs_wl_130718}(b). $\sigma_{\rm{M}}$ is the mean wavenumber of the
fringes which in this case, $\sigma_{\rm{M}}\approx1.25\mu$m$^{-1}$ but for the
actual stellar fringes $\sigma_{\rm{M}}=1.2\mu$m$^{-1}$.

\begin{wstable}
\caption{Phase stability of the fringes of calibrators in MUSCA}
\begin{tabular}{@{}l l r@{}}
\toprule
Light  & Related & \multicolumn{1}{l}{Differential} \\
source & figures & \multicolumn{1}{l}{phase (rad)} \\
\colrule
\multirow{3}{*}{Internal WL}
& \multirow{3}{*}{Fig.~\ref{fig:obs_wl_130718}(b)\tnote{$\dagger$}}
  & -0.18$\pm$0.03 \\
& &  0.10$\pm$0.03 \\
& & -0.81$\pm$0.03 \\
\colrule
\multirow{6}{*}{Achernar}
& \multirow{4}{*}{Fig.~\ref{fig:obs_singlestars_1}(a)}
  & -0.02$\pm$0.10 \\
& &  0.01$\pm$0.10 \\
& &  0.02$\pm$0.09 \\
& & -0.07$\pm$0.10 \\
\rule{0pt}{3ex}
& \multirow{2}{*}{Fig.~\ref{fig:obs_singlestars_2}(a)-(b)\tnote{$\dagger$}}
  & -1.44$\pm$0.06 \\
& &  0.50$\pm$0.06 \\
\colrule
\multirow{2}{*}{Bellatrix}
& \multirow{2}{*}{Fig.~\ref{fig:obs_singlestars_1}(b)}
  & -0.18$\pm$0.08 \\
& & -0.08$\pm$0.08 \\
\colrule
\multirow{3}{*}{$\beta$~Lupi}
& \multirow{2}{*}{Fig.~\ref{fig:obs_singlestars_1}(c)}
  &  0.04$\pm$0.12 \\
& &  0.17$\pm$0.13 \\
\rule{0pt}{3ex}
& Fig.~\ref{fig:obs_singlestars_2}(c)\tnote{$\dagger$}
  & -0.69$\pm$0.16 \\
\colrule
\multirow{2}{*}{$\phi$~Sagittarii}
& \multirow{2}{*}{Fig.~\ref{fig:obs_singlestars_1}(d)}
  & -0.10$\pm$0.11 \\
& & -0.07$\pm$0.10 \\
\botrule
\end{tabular}
\begin{tablenotes}
\item[$\dagger$] with DDL displacement
\end{tablenotes}
\label{tab:obs_cal_diffphase}
\end{wstable}

The values in Table~\ref{tab:obs_cal_diffphase} provide a very useful metric to
assess how well two fringe packets are aligned in terms of their phases.
Ideally, a perfect alignment gives a zero phase difference while a less than
perfect alignment gives a non-zero phase difference between $-\pi$ and $\pi$. A
large phase difference indicates the presence of a large systematic error in the
instrument which has a direct effect on its narrow-angle astrometric precision.
Since the optical path of the interferometer was not changed during this test,
the phase misalignment observed here is not due to the phase delay effect
mentioned in Sec.~\ref{sec:drp_pddrift}. The range of phase differences
measured with the internal white light source is between 0.1 and 0.8 radian or
$\sim$10--100nm. The lower end of this range is good and is the level of
precision required for high precision astrometry. However, the upper end of this
range is poor.

\begin{wstable}
\caption{Successful observations of calibrators}
\begin{tabular}{@{}c l c l l l@{}}
\toprule
HR & Star & Baseline\tnote{$\dagger$} & Dates\tnote{$\ddagger$} &
\# of MUSCA    & Integration \\
   &      &                           &                         &
scans\tnote{d} & time\tnote{d}\hspace{0.5em} (mins) \\
\colrule
 472 & Achernar			& N3-S2 & 121120, 121122 & 1100--7000 & 2--15   \\
1790 & Bellatrix		& N4-S2 & 130107         & $\sim$500  & $\sim$1 \\
5571 & $\beta$~Lupi		& N3-S3 & 130717, 130720 & 200--400   & 0.4--0.9 \\
7039 & $\phi$~Sagittarii	& N4-S2 & 130726         & $\sim$300  & $\sim$0.6\\
\botrule
\end{tabular}
\begin{tablenotes}
\item[$\dagger$] N3-S2=20m, N3-S3=40m, N4-S2=60m
\item[$\ddagger$] in YYMMDD format
\item[d] for one data point 
\end{tablenotes}
\label{tab:obs_calibrators}
\end{wstable}

In the second stage, a known unresolved single star (for a given baseline) was
used. The goal in this stage is to assess the ability of the instrument as a
whole (all SUSI subsystems, PAVO and MUSCA) to maintain phase stability during a
dual-star phase-referencing observation. The optical setup is not a Mach-Zehnder
interferometer as in the first stage but the actual MUSCA setup described in
Sec.~\ref{sec:musca} for astrometric observation. Several known single
stars have been used as calibrators but only a selected few are listed in
Table~\ref{tab:obs_calibrators} together with records on nights they are
observed. These selected observations either have long enough observation time
or were observed more than once in one night which allows their phase stability
to be analyzed.

\begin{figure}
\centering
\subfloat[]{\includegraphics[width=0.5\textwidth]{\imgdir/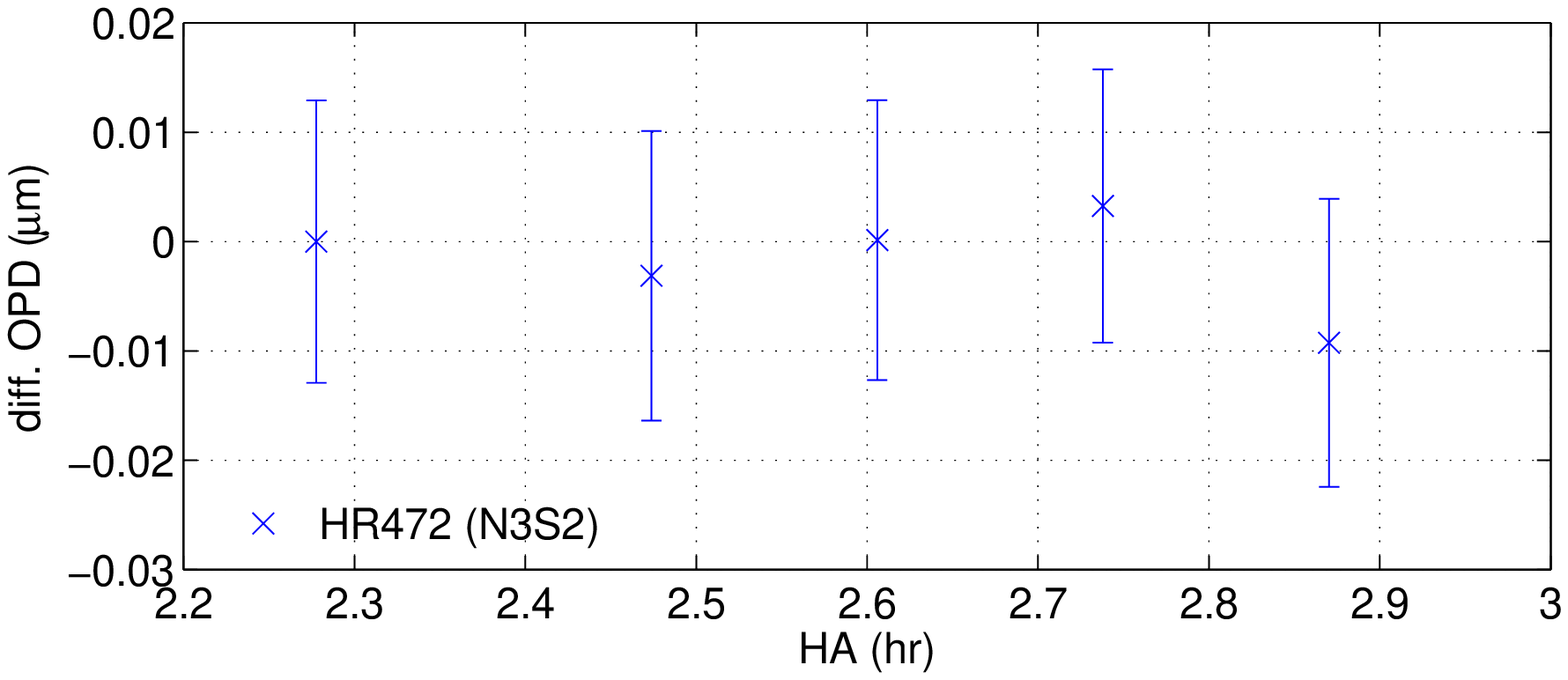}}
\subfloat[]{\includegraphics[width=0.5\textwidth]{\imgdir/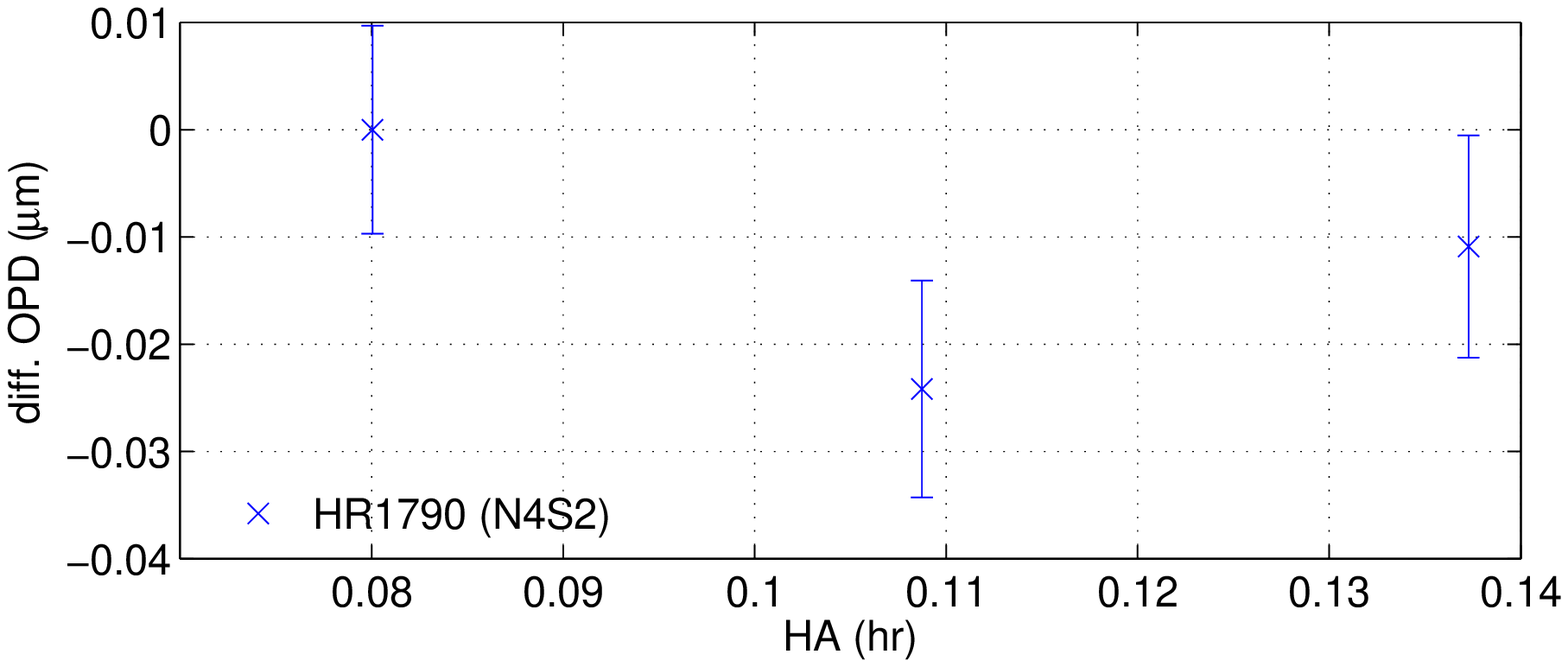}}\\
\subfloat[]{\includegraphics[width=0.5\textwidth]{\imgdir/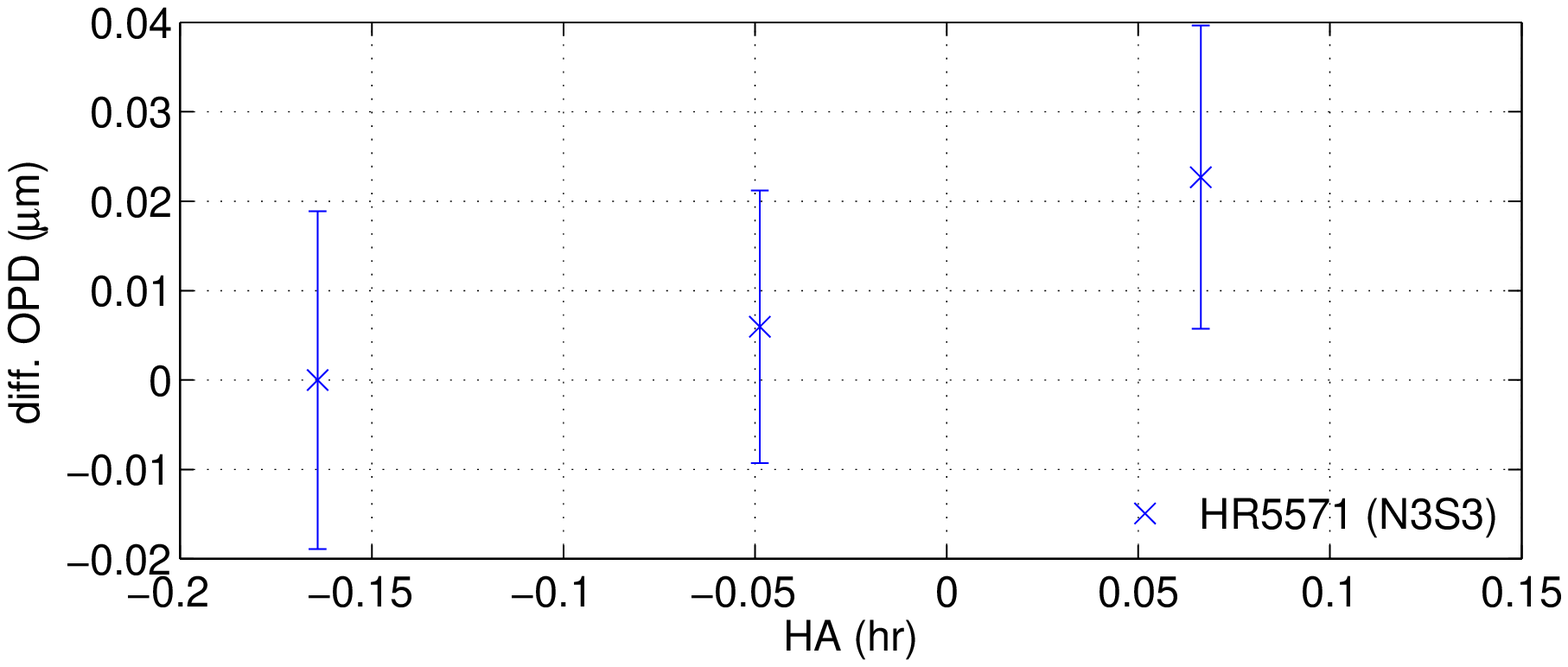}}
\subfloat[]{\includegraphics[width=0.5\textwidth]{\imgdir/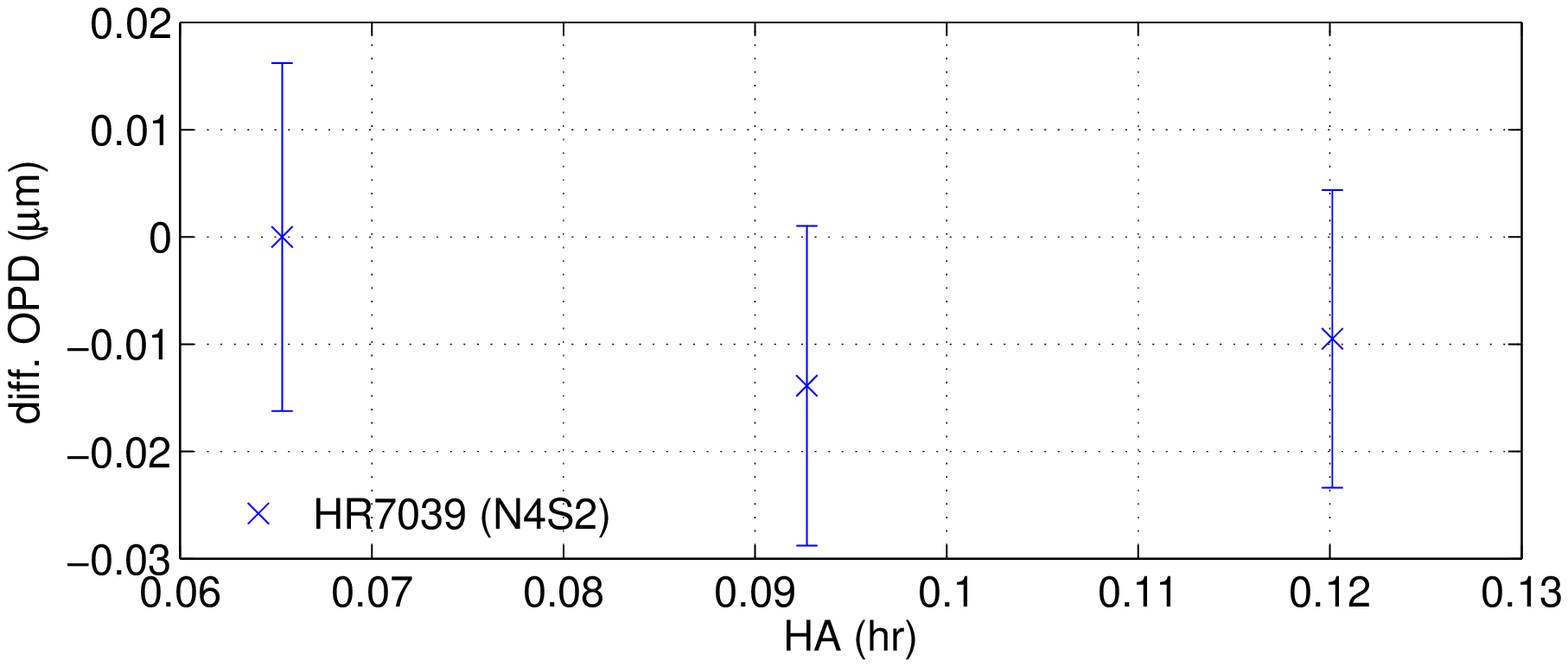}}\\
\caption[]{Similar to Fig.~\ref{fig:obs_wl_130718}(b) but single stars
(indicated in the legend of each plot) replaced the tungsten bulb as light
sources. No DDL displacement took place.}
\label{fig:obs_singlestars_1}
\end{figure}

\begin{figure}
\centering
\subfloat[Bellatrix]{\includegraphics[width=0.5\textwidth]{\imgdir/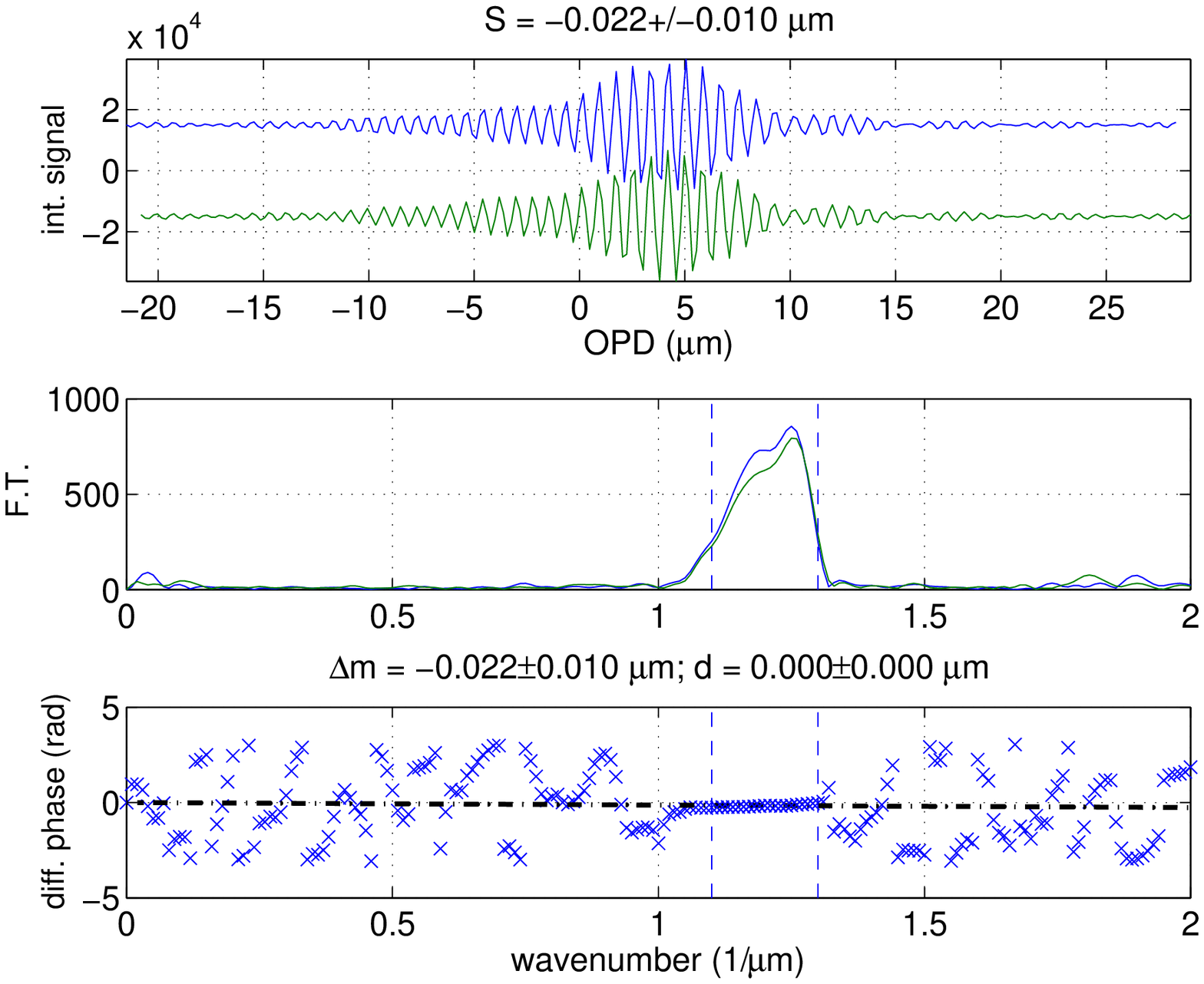}}
\subfloat[$\phi$~Sagittarii]{\includegraphics[width=0.5\textwidth]{\imgdir/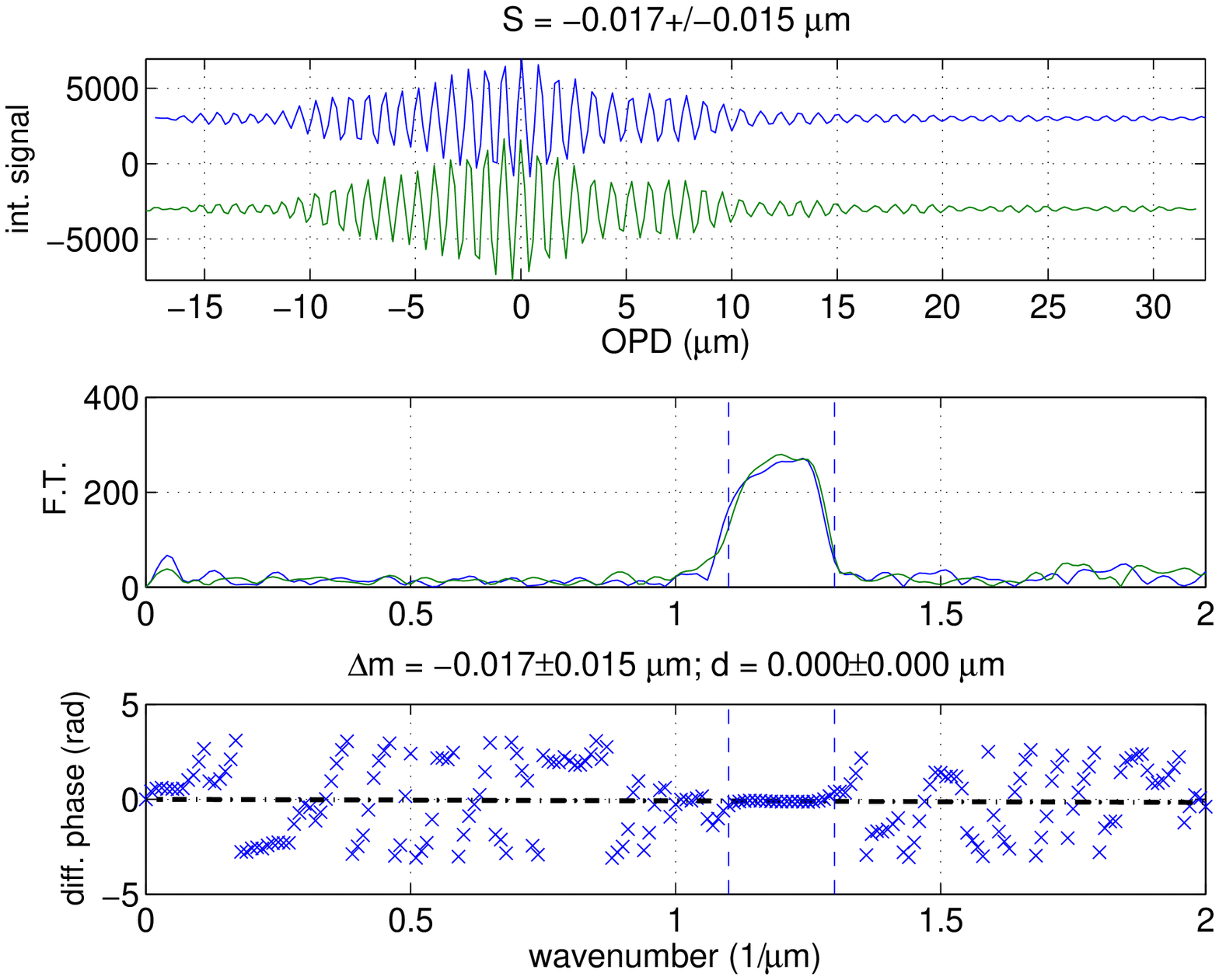}} \\
\caption[]{Similar to Fig.~\ref{fig:obs_wl_130718}(a) but single stars replaced
the tungsten bulb as light sources. No DDL displacement took place.}
\label{fig:obs_singlestars_widefp}
\end{figure}

Similar to Fig.~\ref{fig:obs_wl_130718}(b), the plots in
Fig.~\ref{fig:obs_singlestars_1} shows the phase sensitive relative position of
fringe packets (with respect to the first data point) of Achernar, Bellatrix,
$\beta$~Lupi and $\phi$~Sagittarii. The fringe packets shown were recorded at
different times but on the same nights for each star. The DDL in MUSCA was not
moved during these observations. The results here show the intrinsic precision
of the instrument and the data reduction pipeline without the effect of the DDL
and the DL metrology. Fringe packets of Bellatrix and $\phi$~Sagittarii recorded
at different times are plotted in Fig.~\ref{fig:obs_singlestars_widefp} as
examples. The results are discussed at the end of this section.

\begin{figure}
\centering
\subfloat[Achernar]{\includegraphics[width=0.5\textwidth]{\imgdir/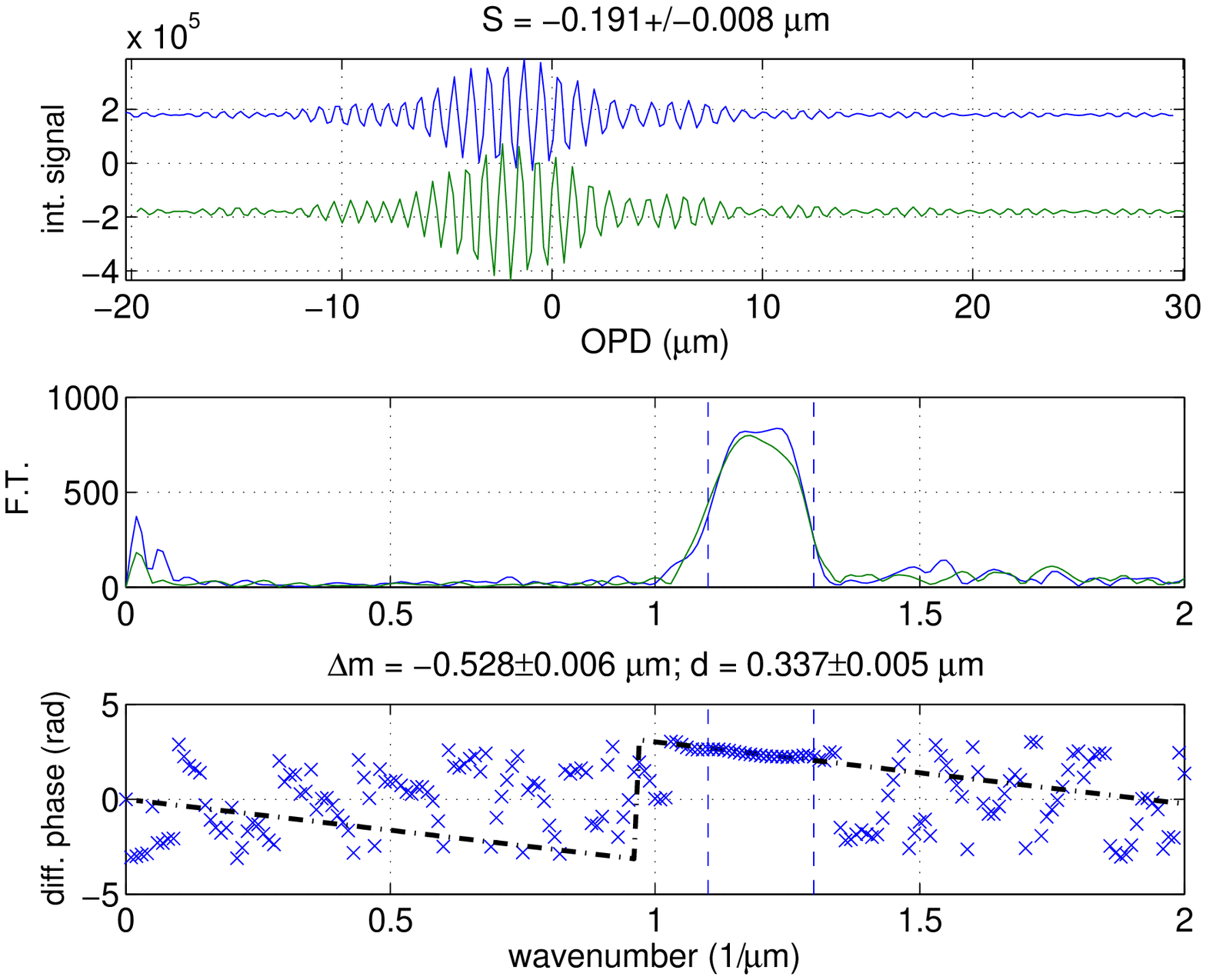}}
\subfloat[Achernar]{\includegraphics[width=0.5\textwidth]{\imgdir/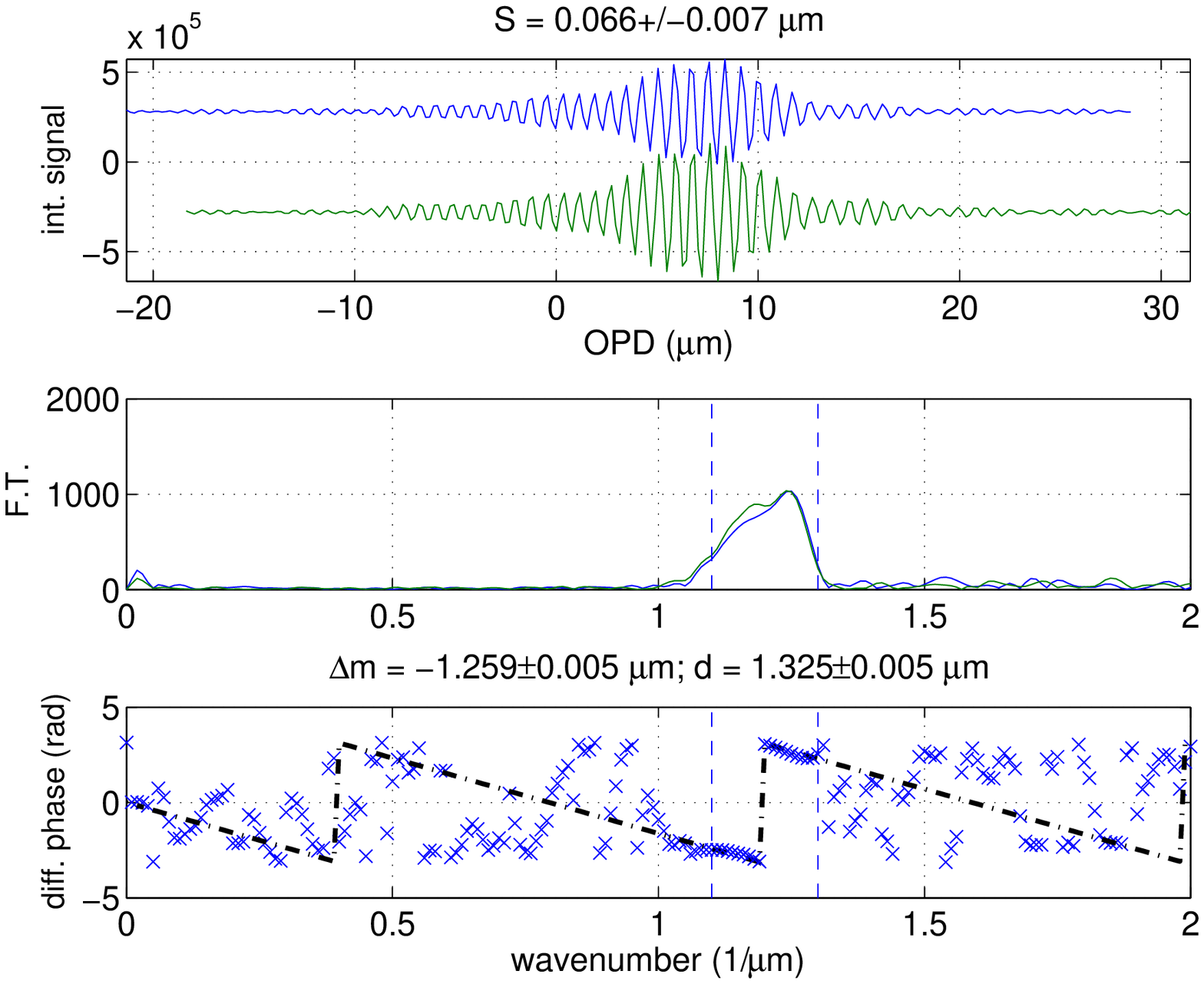}}\\
\subfloat[$\beta$~Lupi]{\includegraphics[width=0.5\textwidth]{\imgdir/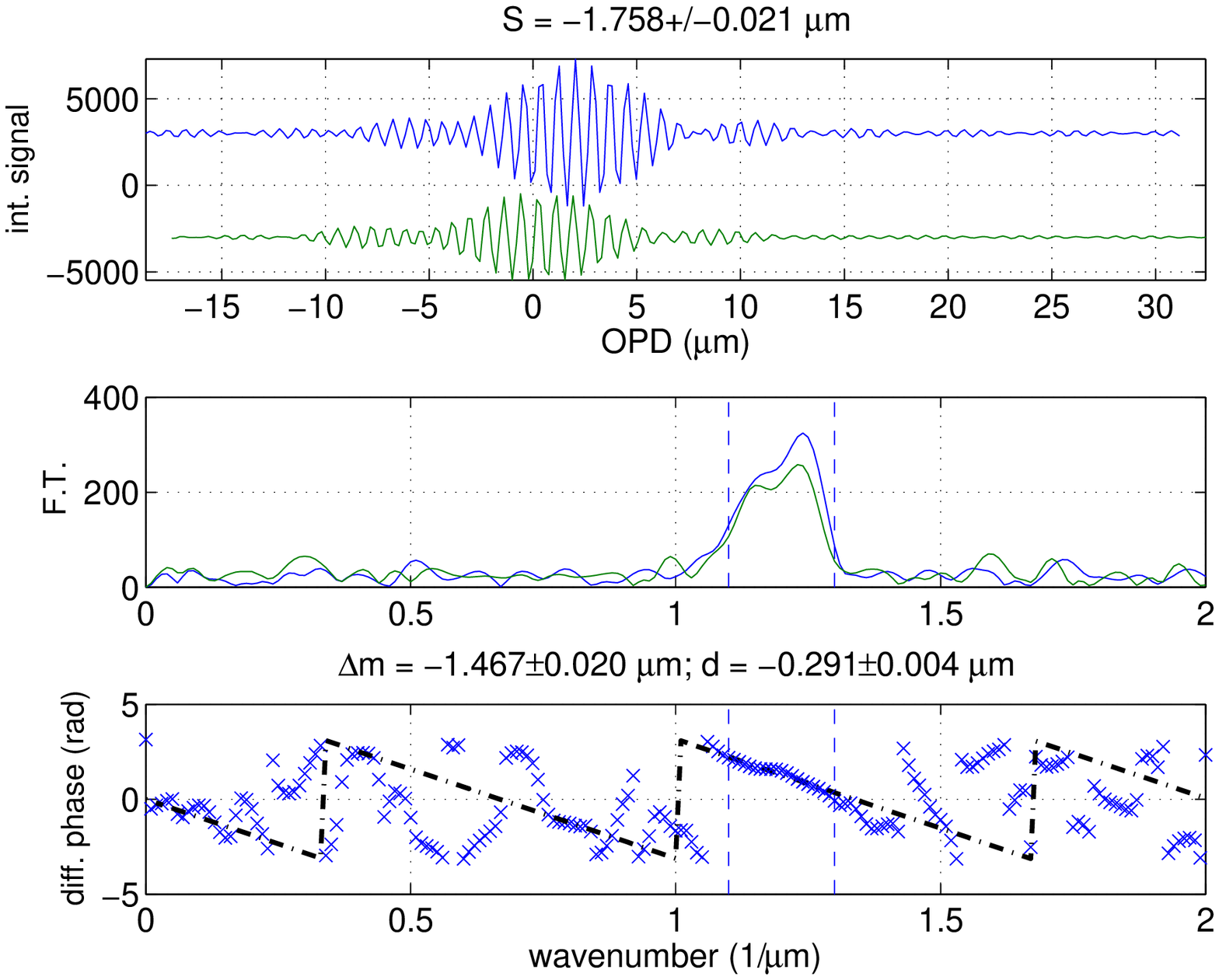}}
\caption[]{Similar to Fig.~\ref{fig:obs_wl_130718}(a) but single stars replaced
the tungsten bulb as light sources. DDL in MUSCA was displaced after each fringe
packet is recorded. The amount of displacement as measured by the DL metrology
is indicated above the bottom panel of each sub-figure.}
\label{fig:obs_singlestars_2}
\end{figure}

With the DDL in MUSCA moved away and back after one fringe packet is recorded,
the separations of pairs of fringe packets are shown in
Fig.~\ref{fig:obs_singlestars_2}. The phase differences of the fringe packets
are listed in Table~\ref{tab:obs_cal_diffphase}. The phase misalignments
reported in the table are due to the intrinsic misalignment observed in the
previous section as well as the phase delay drift effect because the stellar
fringe packets were observed at different times and hence with different air
path length. The fringe separation of the pair of $\beta$~Lupi fringe packets in
Fig.~\ref{fig:obs_singlestars_2}(c) is more than one MUSCA wavelength because of
a change in the amount of air path in the main delay line as fringe packets were
observed at different times in the night and the longitudinal dispersion
compensator (LDC) was adjusted in between observations. Despite that, the phase
alignment between the fringe packets is still less than 1/8th of the mean MUSCA
wavelength ($<$0.8 radian).

The results from the series of calibration tests discussed in this section
indicate that the intrinsic precision of the instrument and the data reduction
pipeline is better than $\sim$20nm when the DDL in MUSCA is not displaced
between observations but degrades to $\sim$100nm when the DDL is displaced. This
lower precision is not adequate for mircoarcsecond precision astrometry. 

\begin{figure}
\centering
\subfloat[N pupils (before)]{\includegraphics[width=0.47\textwidth]{\imgdir/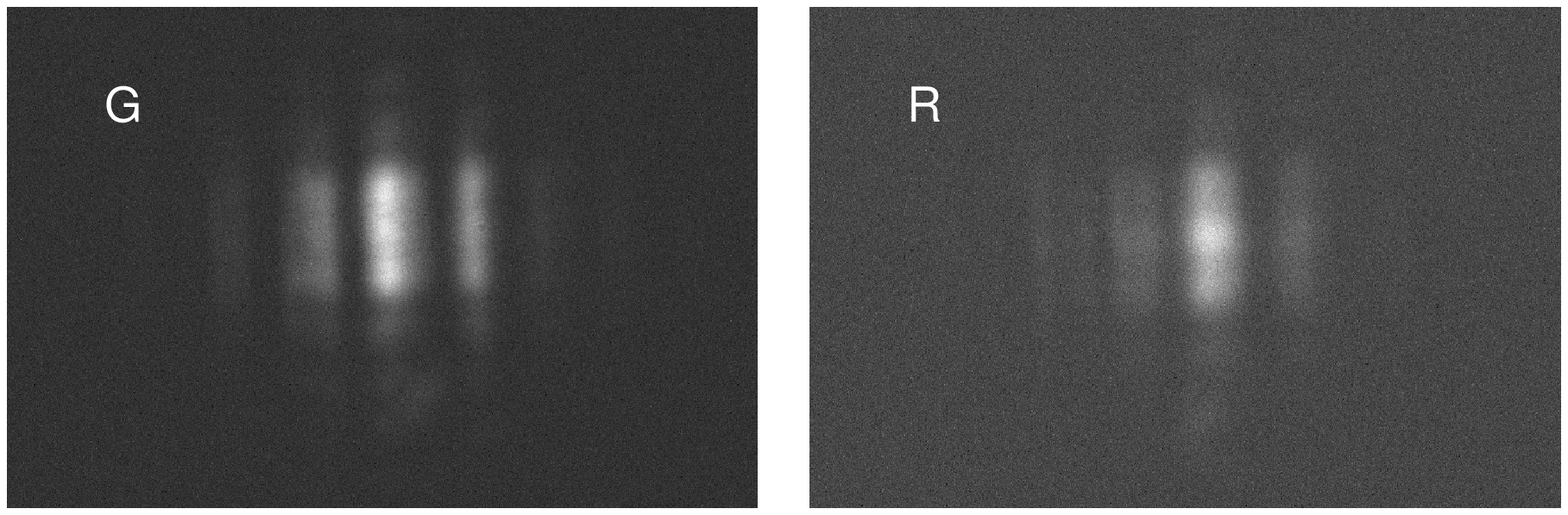}}\hspace{0.5em}
\subfloat[N pupils (after)]{\includegraphics[width=0.47\textwidth]{\imgdir/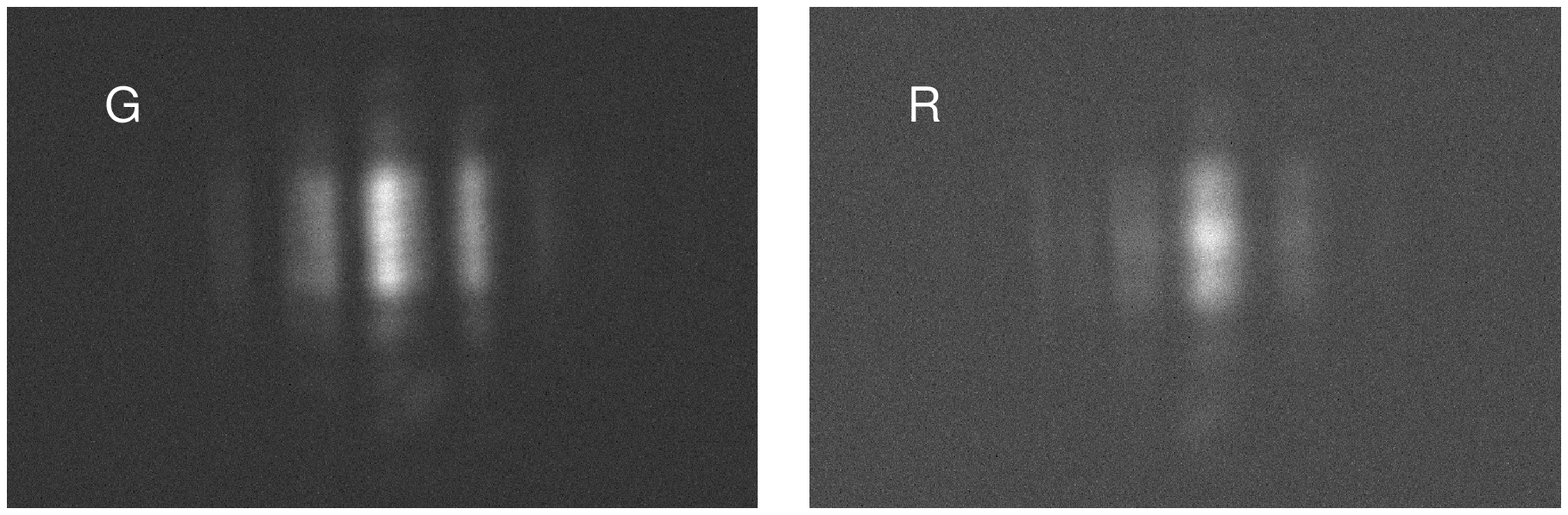}}\\
\subfloat[S pupils (before)]{\includegraphics[width=0.47\textwidth]{\imgdir/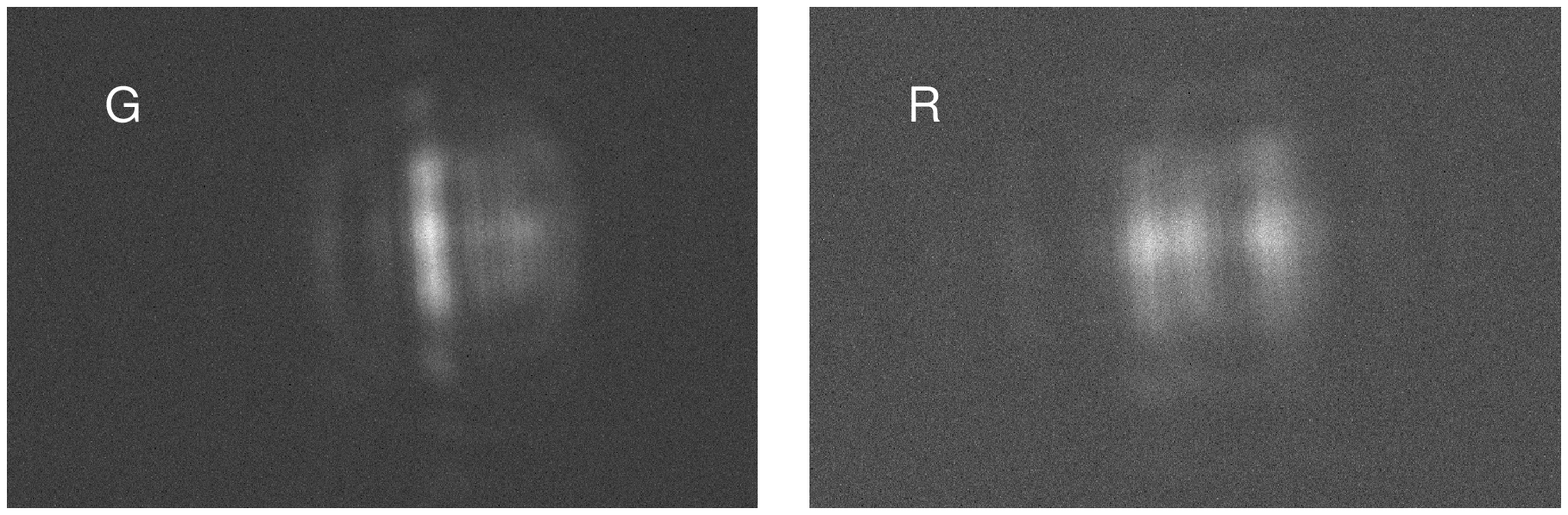}}\hspace{0.5em}
\subfloat[S pupils (after)]{\includegraphics[width=0.47\textwidth]{\imgdir/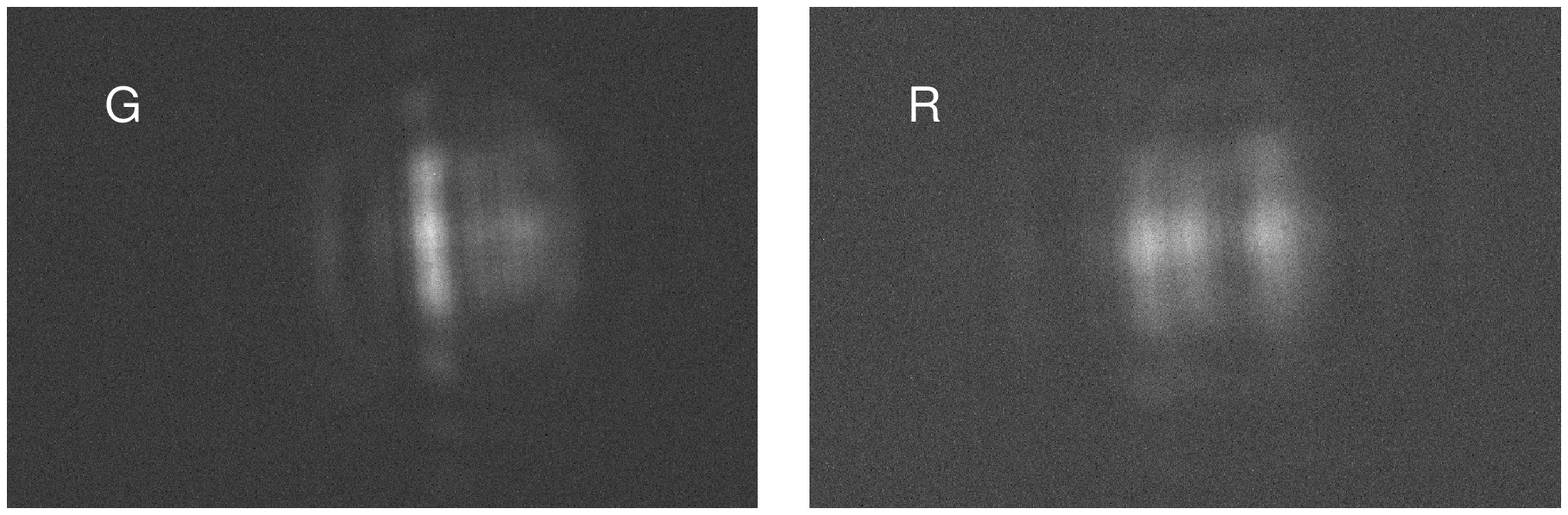}}\\
\caption[]{Images of the green (G) and red (R) laser pupils as seen with the
MUSCA alignment CCD camera. The spatially modulated fringes like structures seen
across the pupils are caused by overlapped and displaced pupils from spurious
reflections off the back side of the M16 dichroic filters. The structures are
not identical between the North (N) and the South (S) pupils because the 6mm
parallel glass windows the dichroic filters were fabricated on most probably
have different reflectivity profile (exact cause was not investigated). More
importantly the structure within the pupils remains unchanged after the DDL is
displaced and then slew back to its original position. This result is expected
because the spurious reflections occur before the laser beams propagate into
MUSCA and through the DDL.}
\label{fig:obs_laserpups}
\end{figure}

The correlation between the precision of measurement and the displacement of the
DDL suggests that the accuracy of the DL metrology could be a problem despite
having a precision of $\sim$5nm or better. However further investigation
revealed that the suggestion is not definitive. Firstly, the non-common path
error arising from the difference in optical path between the metrology lasers
and the white light source (tungsten bulb) is negligible because both light
sources originate from the same pin hole and their fringes are recorded with the
same photo-detectors. Secondly, the position of the linear stage on which the
DDL is mounted varies due to changes in internal (electronics) and external
temperature but this variation is measured by the SL metrology system and
corrected for in all the tests. The low duty cycle ($<$5\%) motion pattern of
the DDL (the ratio between the time spent slewing the stage to idle) during the
above tests and observations ensured no excessive heat is generated by the motor
driving currents in the electronics. Lastly, there were no
obvious changes in the structure of the metrology laser pupils that could have
altered the phase of the laser fringes. Images of the laser pupils as seen with
the MUSCA alignment CCD camera are shown in Fig.~\ref{fig:obs_laserpups}. The
structures in the pupils originate from spurious reflections off the back side
of the M16 dichroic filters, which are fabricated on 6mm thick parallel glass
windows (wedge angle of $\lesssim$30$''$). The pupils of the green and red
lasers are not identical and are labeled in the figure as ``G'' and ``R''
respectively. Neither are the pupils of the North and South beams identical. The
exact cause of the observed asymmetries is uncertain but are probably due to the
fabrication process. These structures are neither observed with the internal WL
source, the IR LEDs at the siderostats nor
starlight. The structures within the laser pupils are not expected to change
because the spurious reflections occur before the laser beams enter into MUSCA
(see Fig.~\ref{fig:opt_pavo_current}(a) and \ref{fig:opt_musca}(a)) and
Fig.~\ref{fig:obs_laserpups} clearly shows this. Due to the presence of such
structures, the probability of unsuspectingly altering the phase of the laser
fringes is not zero. The spatially modulated fringes across the laser pupils are
sensitive to the differential tilt of the original and the displaced laser beam.
Changes in the fringe pattern can alter the total intensity recorded by the
MUSCA APDs at a given scan step and eventually the phase of the metrology
fringes.

\subsection{$\delta$~Orionis~Aa-Ab}

MUSCA achieved ``first light'' on an evening of January 2011 but its full suite
of hardware and data reduction pipeline for performing dual-star
phase-referencing and binary astrometry was not complete until May/June 2013.
The usage of and the results from the pipeline are discussed in this section
using observation data of $\delta$~Orionis~Aa-Ab.

$\delta$~Orionis~Aa-Ab (HR1852; Del Ori) is a
close binary star system which has an expected on-sky separation of 0.3$''$, a
position angle of 131$^\circ$ and an orbital period of 201
years \citep{Mason:2009}. The primary component is itself a known eclipsing
binary ($\delta$~Orionis~Aa1-Aa2) which has an orbital period of 5.7 days
\citep{Harvey:1987}. The primary component is a spectral type O star but it is
suspected to have yet another tertiary component of the same spectral type
\citep{Zasche:2009,Mayer:2010}.

$\delta$~Orionis~Aa-Ab was observed with MUSCA in 2013 as a target for testing
the dual-star phase-referencing mode of the instrument as well as the data
reduction pipeline. Table~\ref{tab:obs_delori} lists the total number of
observations attempted on this ternary system. An observation session is
considered successful when there are at least 3 sets of bracketed observations
of the secondary fringe packet and a bracketed observation consists of a well
phase-referenced primary fringe packet and then a well phase-referenced
secondary packet and finally another well phase-referenced primary fringe
packet. Since the time span for phase-referencing one fringe packet typically
last about 10 minutes, the success rate of an astrometric observation of a
binary system with MUSCA is not high because not only is it highly dependent on
good atmospheric seeing, but also the duration of the good seeing. Out of 7
consecutive attempts, 2 successful observations were made.

\begin{wstable}
\caption{Observations of $\delta$~Orionis~Aa-Ab}
\begin{tabular}{@{}c c l c l l l@{}}
\toprule
Date\tnote{$\ddagger$} & Baseline\tnote{$\dagger$} & Outcome & Range of & 
Calibrators & \# of MUSCA     & Integration \\
                       &                           &         & HAs (hr) & 
            & scans\tnote{c} & time\tnote{c}\hspace{0.5em} (mins) \\
\colrule
130105 & N4-S2 & Fail    & -- & -- & -- & -- \\
130106 & N1-S2 & Fail    & -- & -- & -- & -- \\
\multirow{2}{*}{130107} & 
\multirow{2}{*}{N4-S2} & 
\multirow{2}{*}{Success} & 
\multirow{2}{*}{0.5 -- 2.0} & 
\multirow{2}{*}{Bellatrix} & Aa: 500--2000  & Aa: 1--4 \\
& & & & &                    Ab: 1000--2000 & Aa: 2--4 \\
130108 & N4-S2 & Fail    & -- & -- & -- & -- \\
130110 & N4-S2 & Fail    & -- & -- & -- & -- \\
\multirow{2}{*}{130111} & 
\multirow{2}{*}{N4-S2} & 
\multirow{2}{*}{Success} & 
\multirow{2}{*}{0.0 -- 2.5} & 
\multirow{2}{*}{Bellatrix} & Aa: 500--1500  & Aa: 1--3 \\
& & & & &                    Ab: 1000--4000 & Ab: 2--9 \\
130115 & N4-S2 & Fail    & -- & -- & -- & -- \\
\botrule
\end{tabular}
\begin{tablenotes}
\item[$\dagger$] N1-S2=10m, N4-S2=60m
\item[$\ddagger$] in YYMMDD format
\item[c] label indicates the applicable fringe packet of the close binary component
\end{tablenotes}
\label{tab:obs_delori}
\end{wstable}

\begin{figure}
\centering
\subfloat[]{\includegraphics[width=0.45\textwidth]{\imgdir/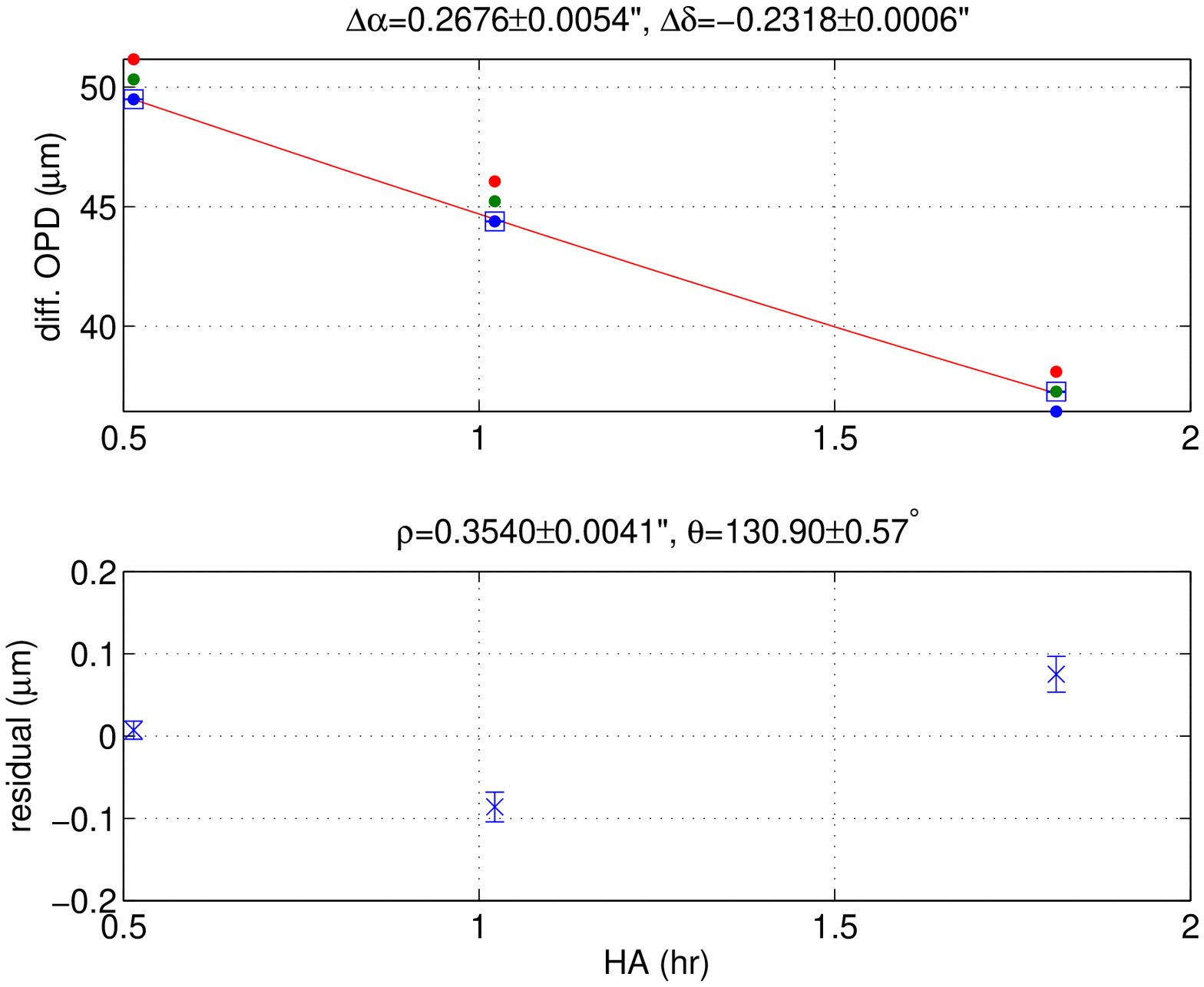}}
\hspace{1em}
\subfloat[]{\includegraphics[width=0.45\textwidth]{\imgdir/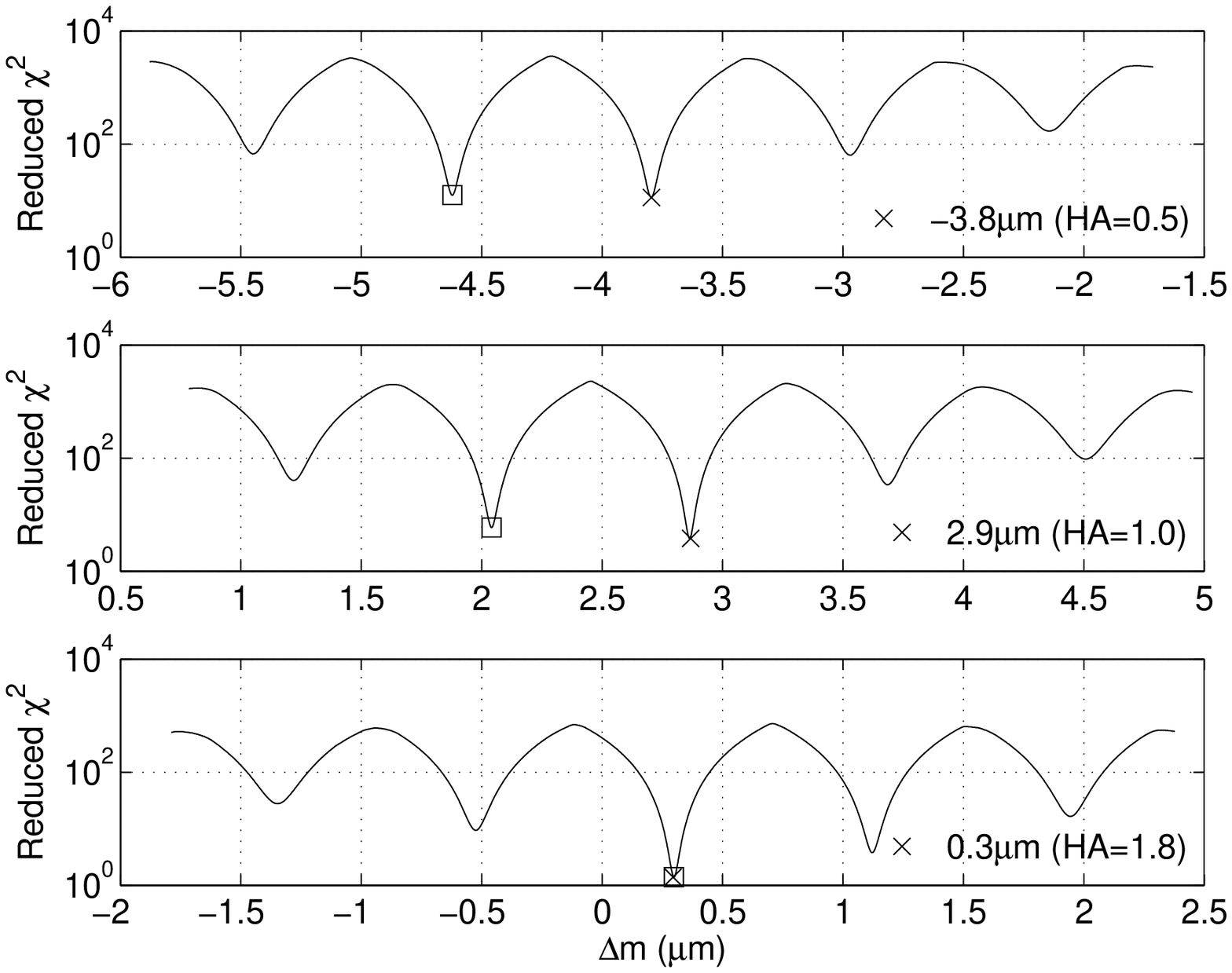}}
\caption[]{Fringe packet separation of $\delta$~Orionis~Aa-Ab at different hour
angles measured by MUSCA with a 60m baseline (N4-S2) on 7 Jan 2013. Plots in (b)
show the reduced $\chi^2$ values of probable apparent separation of the fringe
packets. The symbol $\times$ represents the global minimum and the symbol $\Box$
represents the most probable value based on the model-fitting in (a).}
\label{fig:obs_delori_musca_07}
\end{figure}

\begin{figure}
\centering
\subfloat[]{\includegraphics[width=0.45\textwidth]{\imgdir/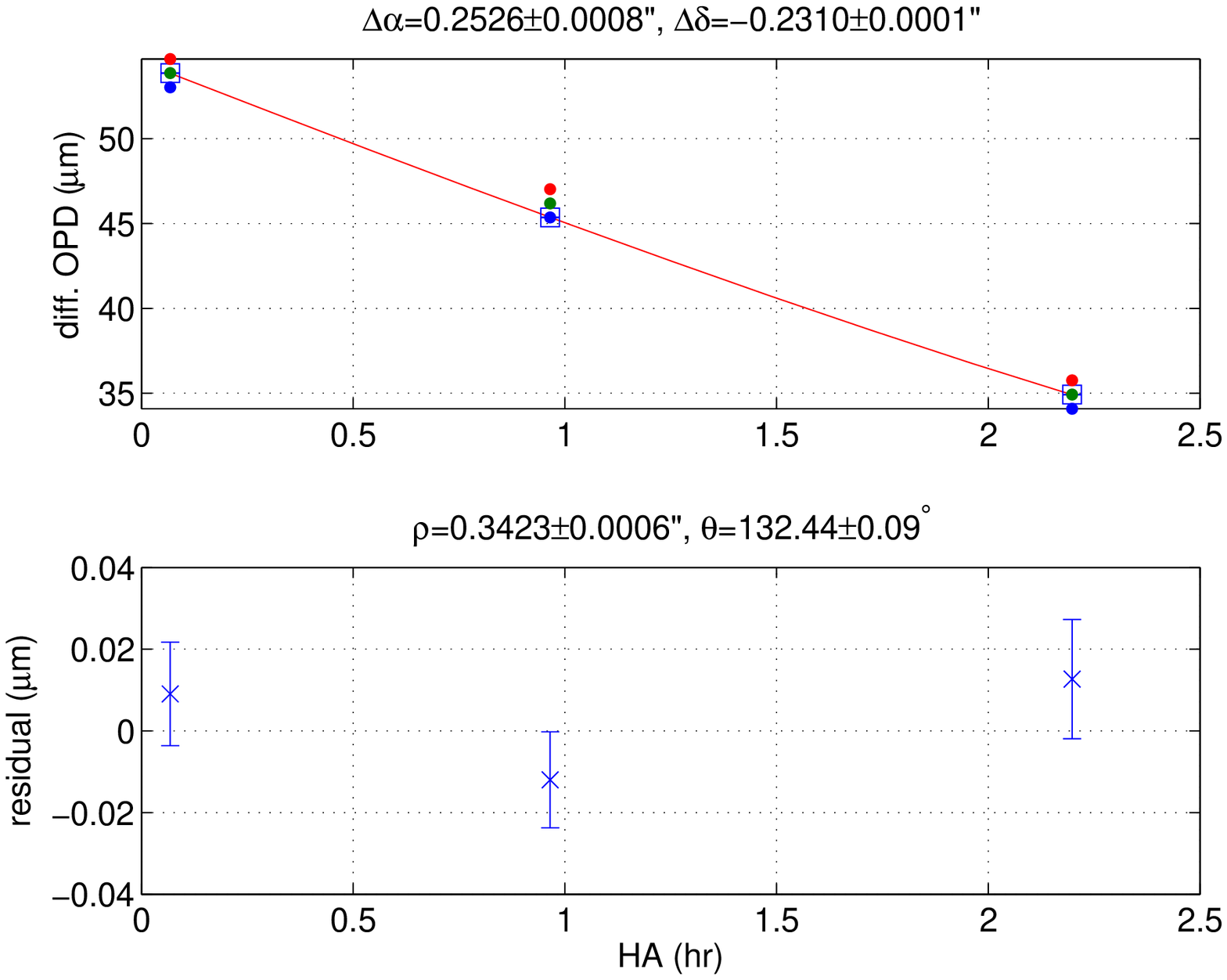}}
\vspace{1em}
\subfloat[]{\includegraphics[width=0.45\textwidth]{\imgdir/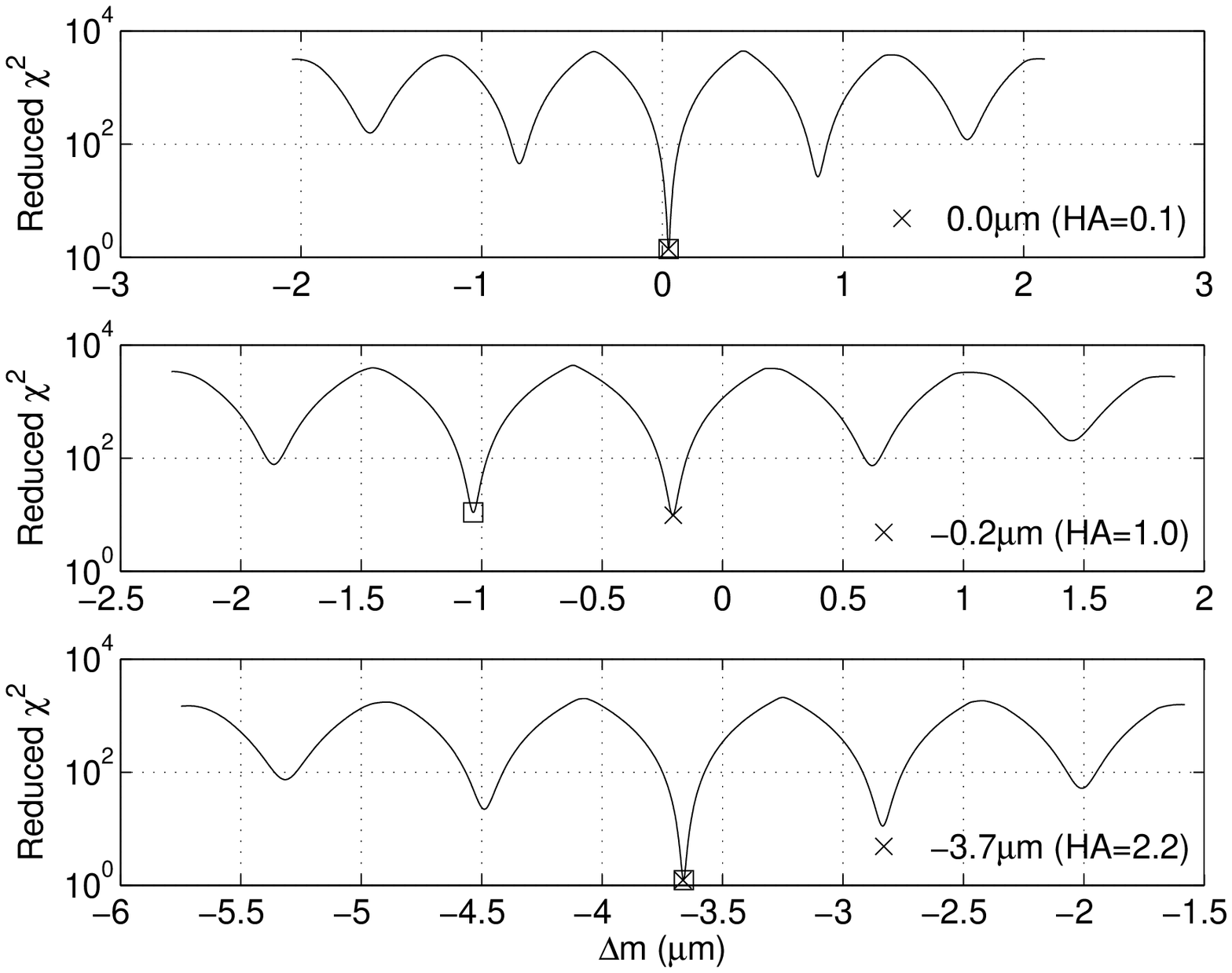}}
\caption[]{Similar to Fig.~\ref{fig:obs_delori_musca_07} but with data taken on
11 Jan 2013.}
\label{fig:obs_delori_musca_11}
\end{figure}

Fig.~\ref{fig:obs_delori_musca_07}(a) and \ref{fig:obs_delori_musca_11}(a) show
the separations between the primary and the secondary fringe packets of
$\delta$~Orionis~Aa-Ab plotted over different observation times. For each
measurement of the fringe packet separation, 3 possible values are plotted. The
first is the fringe separation computed from the phase measurement and the
remaining two are the computed fringe separation plus and minus one MUSCA
wavelength. The reduced $\chi^2$ (shown in Fig.~\ref{fig:obs_delori_musca_07}(b)
and \ref{fig:obs_delori_musca_11}(b)) computed during the least square fitting
of the fringe separation to the measured differential phases of the fringe
packets is used to choose one of the 3 possible values that produces the
smallest residual when the fringe separation data are fitted with a binary star
model. This is done because the fringe separation obtained from the best least
square fit may not be the \emph{correct} value as it is ambiguous by one MUSCA
wavelength due to an astrophysical effect which is discussed in
Sec.~\ref{sec:summary}. The selected data points are indicated in the plots with
square symbols. In addition to correcting the one MUSCA wavelength ambiguity,
the fringe separation measurements are also empirically compensated for errors
arising from the phase delay drift effect and the accuracy of the DL metrology
using the phase misalignment measured with the primary fringe packets.
Table~\ref{tab:obs_delori_diffphase} shows the phase misalignment between the
first and the second primary fringe packet of several bracketed observations.
The magnitude of the phase misalignment is within expectation because the DDL is
displaced between each measurements. If the fringe packets in the figure are
recorded without perturbing the position of the DDL, the phase alignment (not
shown in the table) between the fringe packets is measured to be better than
1/12th of a MUSCA wavelength.

\begin{wstable}
\caption{Phase stability of the fringes of $\delta$~Orionis~Aa in MUSCA}
\begin{tabular}{@{}c c r@{}}
\toprule
Date & Hour Angle & \multicolumn{1}{l}{Differential} \\
     & (Hr)       & \multicolumn{1}{l}{phase (rad)} \\
\colrule
\multirow{3}{*}{130107}
     & 0.5 & -0.21$\pm$0.06 \\
     & 1.0 & -0.33$\pm$0.11 \\
     & 1.8 &  0.10$\pm$0.11 \\
\colrule
\multirow{2}{*}{130111}
     & 0.1 & -0.51$\pm$0.07 \\
     & 1.0 & -1.48$\pm$0.07 \\
     & 2.2 & -1.28$\pm$0.08 \\
\botrule
\end{tabular}
\label{tab:obs_delori_diffphase}
\end{wstable}

The corrected measurements of the projected separation of the primary and
secondary fringe packets, $\tilde{S}+\Delta\tilde{S}$, are used to estimate the
position angle, $\theta$, and on-sky separation, $\rho$, of the binary through
model-fitting. The model used for parameter extraction is given as,
\begin{equation} \label{eq:drp_sepmod1}
\begin{split}
S
&= \Delta\vec{s}\cdot\vec{B} \\
&= \begin{bmatrix}
-\cos\delta_2\sin\rm{HA}_2 + \cos\delta_1\sin\rm{HA}_1 \\
\cos\phi_{\text{LAT}}\left(\sin\delta_2-\sin\delta_1\right) - 
  \sin\phi_{\text{LAT}}\left(\cos\delta_2\cos\rm{HA}_2 - \cos\delta_1\cos\rm{HA}_1\right) \\
\sin\phi_{\text{LAT}}\left(\sin\delta_2-\sin\delta_1\right) +
  \cos\phi_{\text{LAT}}\left(\cos\delta_2\cos\rm{HA}_2 - \cos\delta_1\cos\rm{HA}_1\right)
\end{bmatrix}_{xyz}
\cdot
\begin{bmatrix}
B_x \\
B_y \\
B_z
\end{bmatrix}_{xyz},
\end{split}\end{equation}
where $S$ is the projected fringe packet separation of two stars, $\vec{B}$ is
the vector of the chosen baseline and $\Delta\vec{s} = \hat{s}_2-\hat{s}_1$ is
the difference between pointing vectors of the target and the reference stars.
Both vectors are resolved into their components in a rectangular coordinate
system, taking the ground as the frame of reference. The $x$-axis points to the
East, $y$-axis points to the North and $z$-axis points to zenith. The celestial
coordinates, $(\rm{HA}_1,\delta_1)$ and $(\rm{HA}_2,\delta_2)$, define the
location of the primary and secondary stars in the sky where $\delta_2 =
\delta_1+\rho\cos\theta$ and $\rm{HA}_2 = \rm{HA}_1-\rho\sin\theta$. The symbols
$\rho$ and $\theta$ represent the on-sky separation and position angle of the
two stars in a polar coordinate system. Although the uncertainty in the position
of the primary star translates to an uncertainty in projected fringe packet
separation, the value is extremely small and negligible. For example, through
computer simulation, it was found that a standard deviation of 0.5$''$ in both
axes of the coordinate system produces a standard deviation of less than 1nm in
projected fringe packet separation measurements, even at an unfavorable pointing
angle and with a long (160m) baseline. By unfavorable pointing angle, we mean
that the angle subtended by the pointing vector, $\hat{s}$, and the baseline
vector, $\vec{B}$, is very small. This means the parameters extracted from
fitting the model to the measured projected separation are insensitive to the
absolute position of the primary star if the accuracy is better than 0.5$''$.
This value is large compared to the proper motion of stars suitable for MUSCA
observation which are mainly less than 0.25$''$. Finally the symbol
$\phi_{\rm{LAT}}$ in the above equation represents the geographical latitude of
SUSI.

An alternative model described by \citep{Rizzuto:2013} can also be used. However
it may not be suitable for high-precision narrow-angle astrometry of stars
separated by more than 3$''$ and/or using baselines longer than 160m because the
approximation in $\Delta\vec{s}\cdot\vec{B}$ may exceed 10nm. A discussion of
this approximation error is included in Appendix A.

The fitting of the binary star model to the fringe packet separation
measurements from the two observations have residuals of less than $\sim$100nm,
a magnitude which is consistent with the intrinsic misalignment observed in the
internal phase stability test discussed earlier. The residuals of the fits are
plotted in the lower panels of Fig.~\ref{fig:obs_delori_musca_07}(a) and
\ref{fig:obs_delori_musca_11}(a).

The relative astrometry of $\delta$~Orionis~Aa-Ab obtained from the two
observations is summarized in Table~\ref{tab:results_delori}. Since the
position of the vernal equinox, which is the origin of the equatorial coordinate
system, changes over time due to perturbation in the Earth's rotation axis, the
astrometric parameters in the table are also expressed in an equatorial
coordinate system that is based on equinox 2000\footnote{the origin of the
coordinate system is centered on the position of the vernal equinox in the year
2000} for comparison with astrometry performed with other techniques and
published values in the literature. The coordinate transformation involved took
the effect of precession and nutation of the Earth's rotation axis into account.

\begin{wstable}
\caption{Relative astrometry of $\delta$~Orionis~Aa-Ab}
\begin{tabular}{@{}l r r r r@{}}
\toprule
Parameter & \multicolumn{2}{c}{As fitted} & \multicolumn{2}{c}{Eq=J2000} \\
\colrule
Epoch & 2013.0178 & 2013.0287 & 2013.0178 & 2013.0287 \\
$\Delta\alpha_{\rm{Aa,Ab}}$  (mas) &
267.6$\pm$5.4 &
252.6$\pm$0.8 &
267.3$\pm$5.4 &
252.3$\pm$0.8 \\

$\Delta\delta_{\rm{Aa,Ab}}$  (mas) &
-231.8$\pm$0.6 &
-231.0$\pm$0.1 &
-232.1$\pm$0.6 &
-231.3$\pm$0.1 \\

$\rho_{\rm{Aa,Ab}}$    (mas) &
354.0$\pm$4.1 &
342.3$\pm$0.6 &
354.0$\pm$4.1 &
342.3$\pm$0.6 \\

$\theta_{\rm{Aa,Ab}}$  ($^\circ$) &
130.90$\pm$0.57 &
132.44$\pm$0.09 &
130.97$\pm$0.88 &
132.51$\pm$0.14 \\

$\chi^2_{\rm{R}}$ & 35.2 & 2.3 & -- & -- \\
\botrule
\end{tabular}
\label{tab:results_delori}
\end{wstable}

For comparison, measurements of the separation and position angle of
$\delta$~Orionis~Aa-Ab obtained from the Fourth Catalog of Interferometric
Measurements of Binary
Stars\footnote{\url{http://ad.usno.navy.mil/wds/int4.html}} (INT4) and the
Washington Visual Double Star\footnote{\url{http://ad.usno.navy.mil/wds}} (WDS)
catalog are listed in Table~\ref{tab:obs_delori_cmp}. The measurements obtained
in this paper agree only moderately well (to within $\sim$1$^\circ$ in position
angle and $\sim$0.02$''$ in binary separation) with measurements from the WDS
catalog.  An ephemeris of the secondary component computed based on parameters
from the Sixth Catalog of Orbits of Visual Binary
Stars\footnote{\url{http://ad.usno.navy.mil/wds/orb6.html}} (ORB6) is also
included for comparison.

\begin{wstable}
\caption{Comparison of $\delta$~Orionis~Aa-Ab astrometry.}
\begin{tabular}{@{}l l l l l l@{}}
\toprule
Source & Epoch   & $\theta$   & $\delta\theta$ & $\rho$ & $\delta\rho$  \\
       & (+2000) & ($^\circ$) & ($^\circ$)     & ($''$) & (mas)		\\
\colrule
WDS
       & 13.1311 & 131.0  & --   & 0.3273 & 0.2 \\
Ephemeris\tnote{*}
       & 13.1000 & 130    & --   & 0.31   & --  \\
 & & & & & \\
\multirow{2}{*}{This work}
       & 13.0287 & 132.51 & 0.14 & 0.3423 & 0.6 \\
       & 13.0178 & 130.97 & 0.88 & 0.3540 & 4.1 \\
 & & & & & \\
MA2010\tnote{$^\dagger$}
       & 08.0449 & 132.66 & 0.47 & 0.325  & 1.6 \\
\botrule
\end{tabular}
\begin{tablenotes}
\item[*] based on a grade 4 (preliminary) orbit from ORB6
\item[$^\dagger$] \citet{Maiz-Apellaniz:2010}
\end{tablenotes}
\label{tab:obs_delori_cmp}
\end{wstable}

\section{Discussion} \label{sec:discussion}
\begin{figure}
\centering
\subfloat[]{\includegraphics[width=0.5\textwidth]{\imgdir/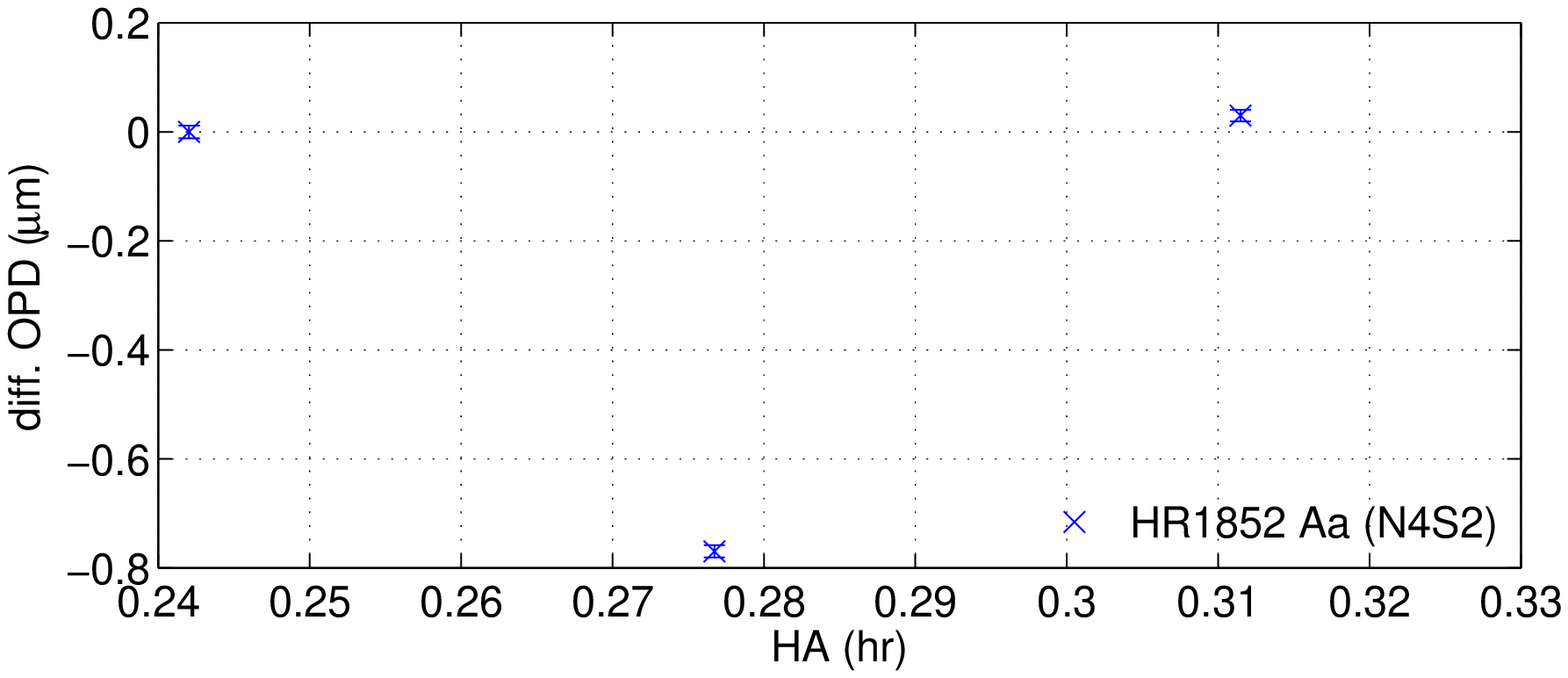}}
\subfloat[]{\includegraphics[width=0.5\textwidth]{\imgdir/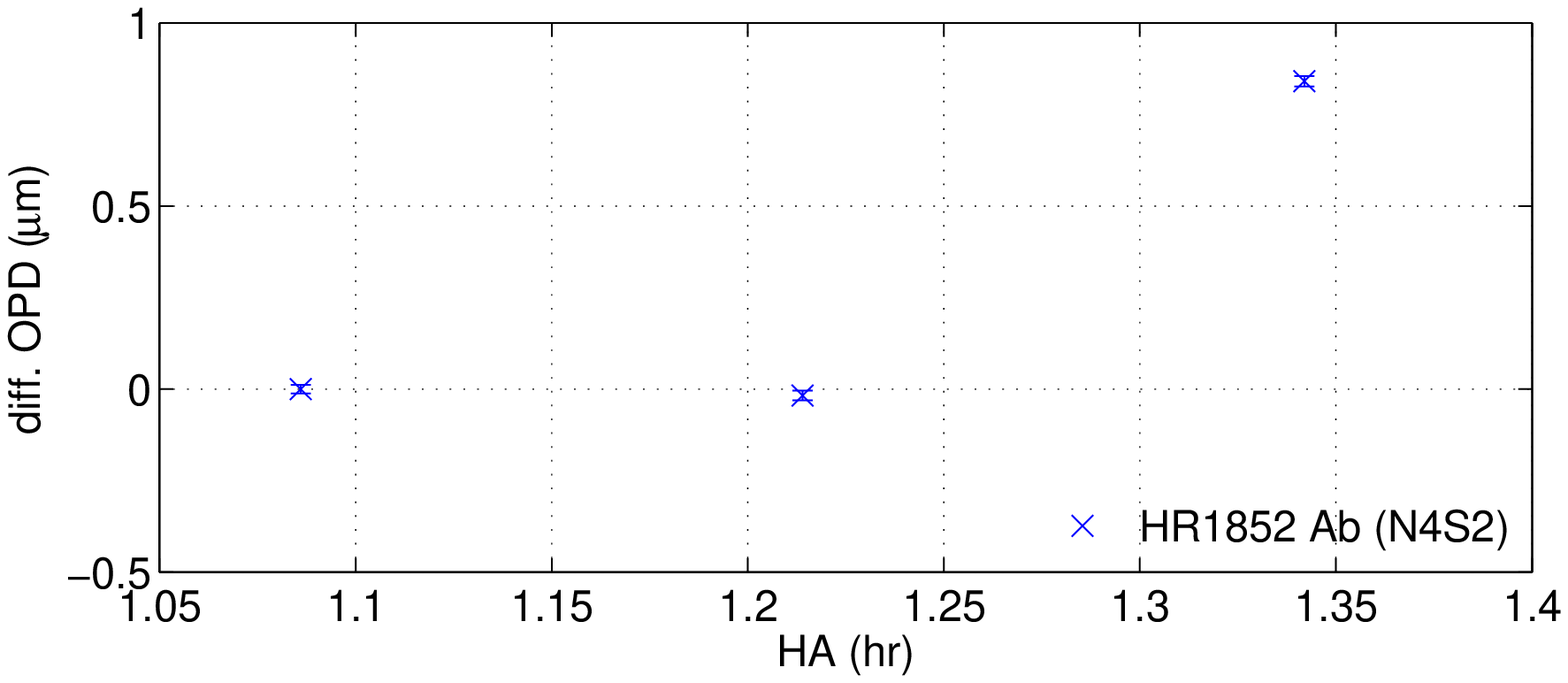}}
\caption[]{Relative fringe packet position of $\delta$~Orionis~Aa1 measured at
different times showing jumps of one MUSCA wavelength occasionally due to the
presence of an unresolved fringe packet of $\delta$~Orionis~Aa2.}
\label{fig:obs_delori_musca1}
\end{figure}

Despite the expected alignment of the fringe phases of better than 1/4th of a
wavelength ($<$1.5 radian), the primary fringe packet registers discrete jumps
of about one MUSCA wavelength in its relative position. Plots in
Fig.~\ref{fig:obs_delori_musca1} show the jumps at hour angles (HA) of
$\sim$0.275 and $\sim$1.35 respectively. The jumps are not related to the
accuracy of the dual laser metrology because fringe packets in the figure are
recorded without any displacement of DDL in between. Furthermore, these jumps
occur within time intervals that are too short for an unresolved secondary
fringe packet to drift by the same amount.

\begin{figure}
\centering
\subfloat[]{\includegraphics[width=0.6\textwidth]{\imgdir/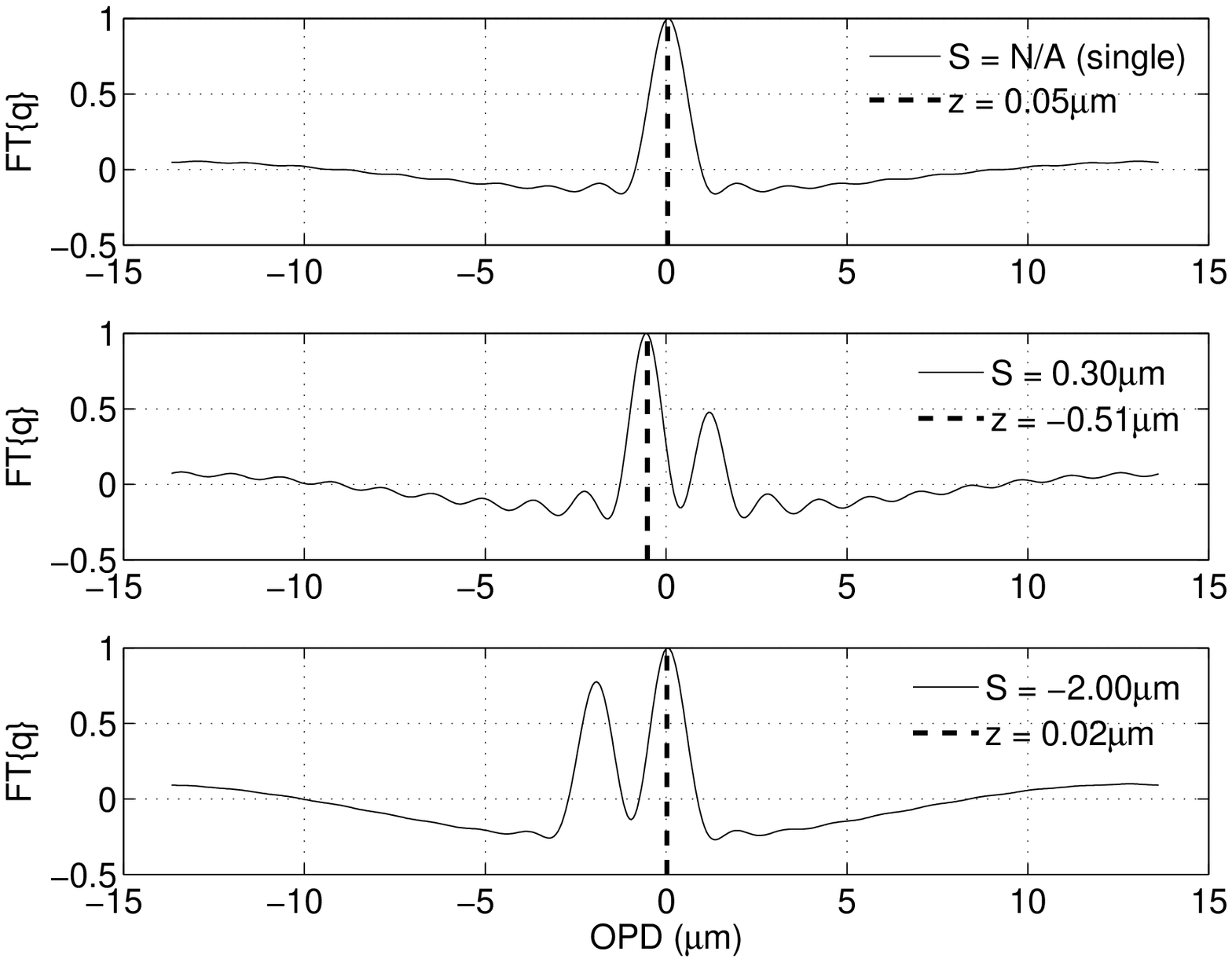}}\\
\subfloat[]{\includegraphics[width=0.5\textwidth]{\imgdir/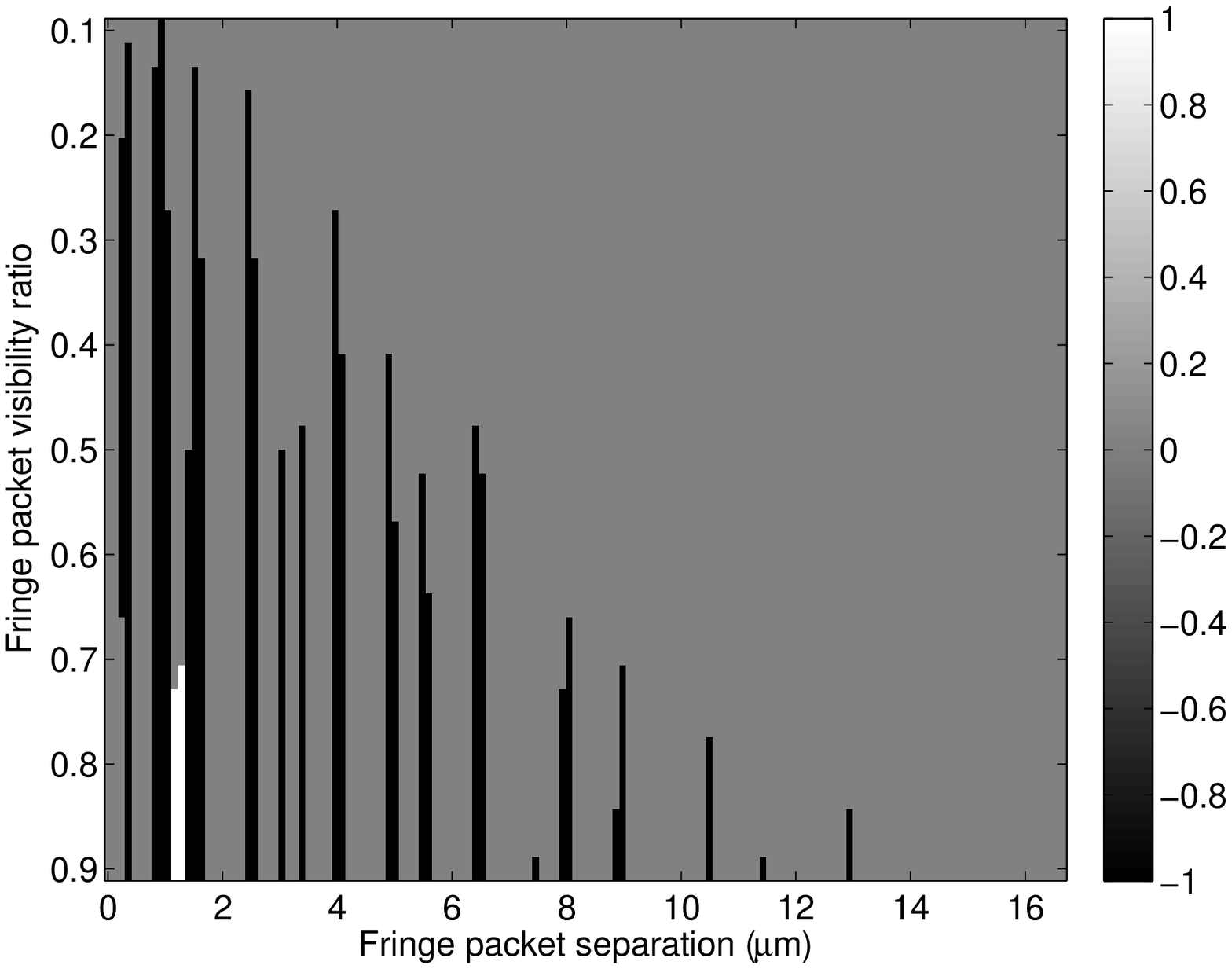}}
\subfloat[]{\includegraphics[width=0.5\textwidth]{\imgdir/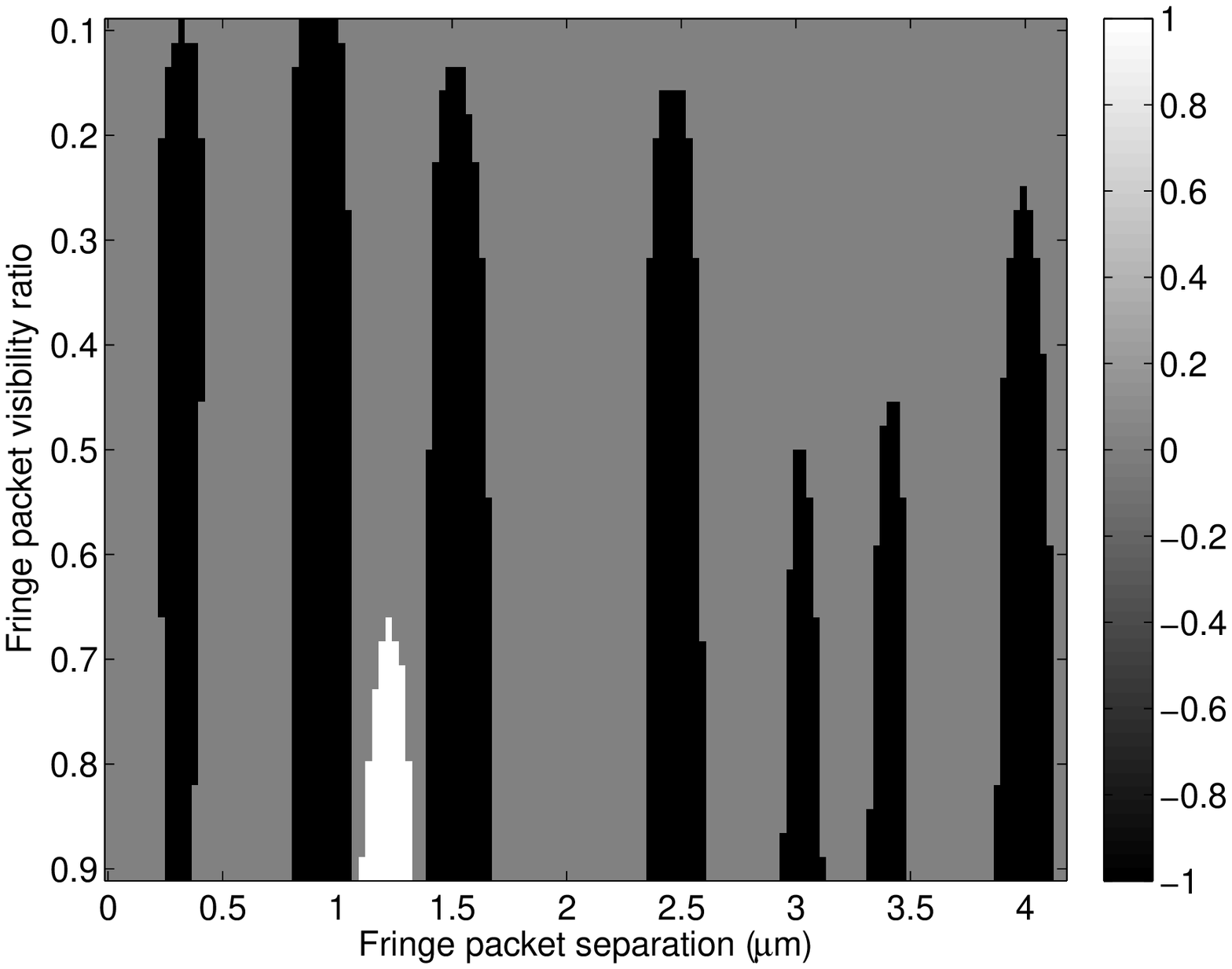}}
\caption[]{The plots in (a) show the Fourier transform of the complex
visibility, a variable computed in the data reduction pipeline to obtain the
phase delay of a set of fringes from the location of the peak in the plots (see
Eq.~\eqref{eq:drp_ftq}). The addition of a second set of fringes at the right
fringe separation can artificially shift the peak, thereby resulting in an
incorrect phase delay computed for the first. Due to this error, the relative
phase delay of the primary fringes between the middle and bottom panels of (a)
will wrongly be computed as one MUSCA wavelength despite them being kept
unchanged. Plots in (b) and (c) show the relative phase delay error in MUSCA
wavelengths for various fringe separations and visibility ratios (primary vs.\
secondary set of fringes).}
\label{fig:obs_binary_pavo_sim}
\end{figure}

The origin of the fringe packet position jump observed can
be traced back to the method the phase delay was computed from the PAVO
interferograms. Phase delays are computed from group delays which in turn are
estimated from the position of the peak of the Fourier transform of the complex
visibility of the fringes across the PAVO bandpass. The detailed description of
this algorithm is given in Sec.~\ref{sec:drp_opdz}. A computer simulated example
plot of the Fourier transform, $\hat{\mata{q}}$ in Eq.~\eqref{eq:drp_ftq}, is
shown in the top panel of Fig.~\ref{fig:obs_binary_pavo_sim}(a). In the example,
the fringes belong to a single star and the phase delay (indicated by the symbol
$z$ in the figure) is zero (within the uncertainty of the algorithm). When
fringes from a secondary star are introduced into the simulation, the phase
delay of the fringes from the primary star is altered depending on the fringe
separation and the visibility of the secondary star fringes. Similar example
plots of $\hat{\mata{q}}$ when PAVO interferograms consist of fringes from two
stars are shown in the middle and bottom panels of
Fig.~\ref{fig:obs_binary_pavo_sim}(a). Different values of fringe separations,
indicated by the symbol $S$ in the legend, are chosen in order to illustrate the
effect of additional fringes on the phase delay of the primary. The ratio of the
visibility of the two sets of fringes in the plots is chosen to be 0.9. The
position of the peak of $\hat{\mata{q}}$ in the middle panel clearly shows the
phase delay deviates from zero. If the phase delay deviates by more than half of
a MUSCA wavelength, then the fringe separation computed between the current and
a reference MUSCA fringe packet will register a jump of one MUSCA wavelength, a
phenomenon observed in Fig.~\ref{fig:obs_delori_musca1}. The magnitude of the
jump for a range of primary-secondary fringe packet separations and visibility
ratios is shown in Fig.~\ref{fig:obs_binary_pavo_sim}(b) and (c). The images in
the figure show various locations where phase delay jumps can occur and the
grayscale of the image shows the magnitude of the jump in MUSCA wavelengths.
Fig.~\ref{fig:obs_binary_pavo_sim}(c) is a zoomed in version of (b). The latter
also shows that the probability of a phase delay jump occurring is zero when the
fringe packet separation is larger than $\sim$15$\mu$m which is half of the
coherence length of one PAVO spectral channel. This is expected because the
fringe packets will no longer overlap each other at that separation.

\begin{figure}
\centering
\subfloat{\includegraphics[width=0.5\textwidth]{\imgdir/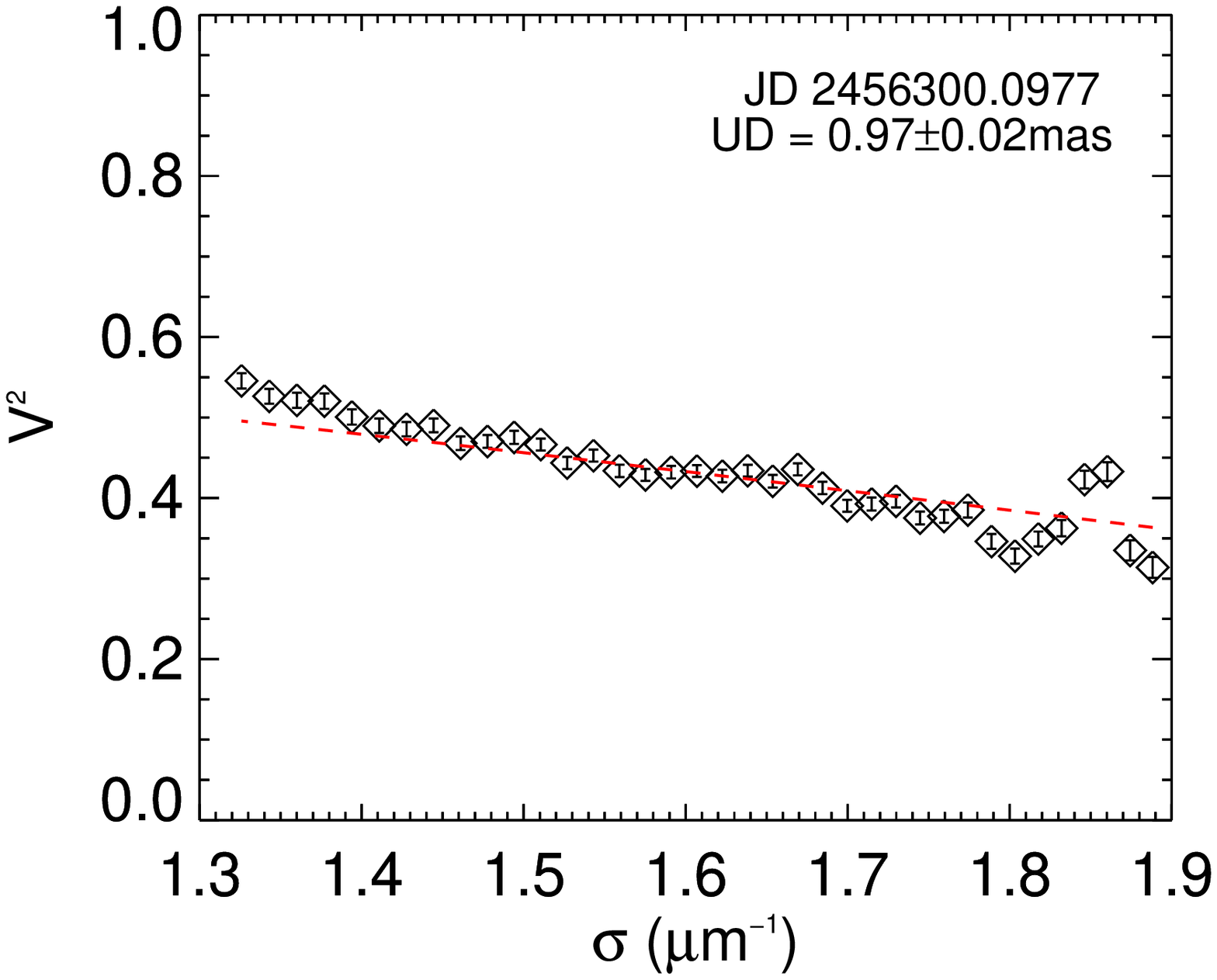}}
\subfloat{\includegraphics[width=0.5\textwidth]{\imgdir/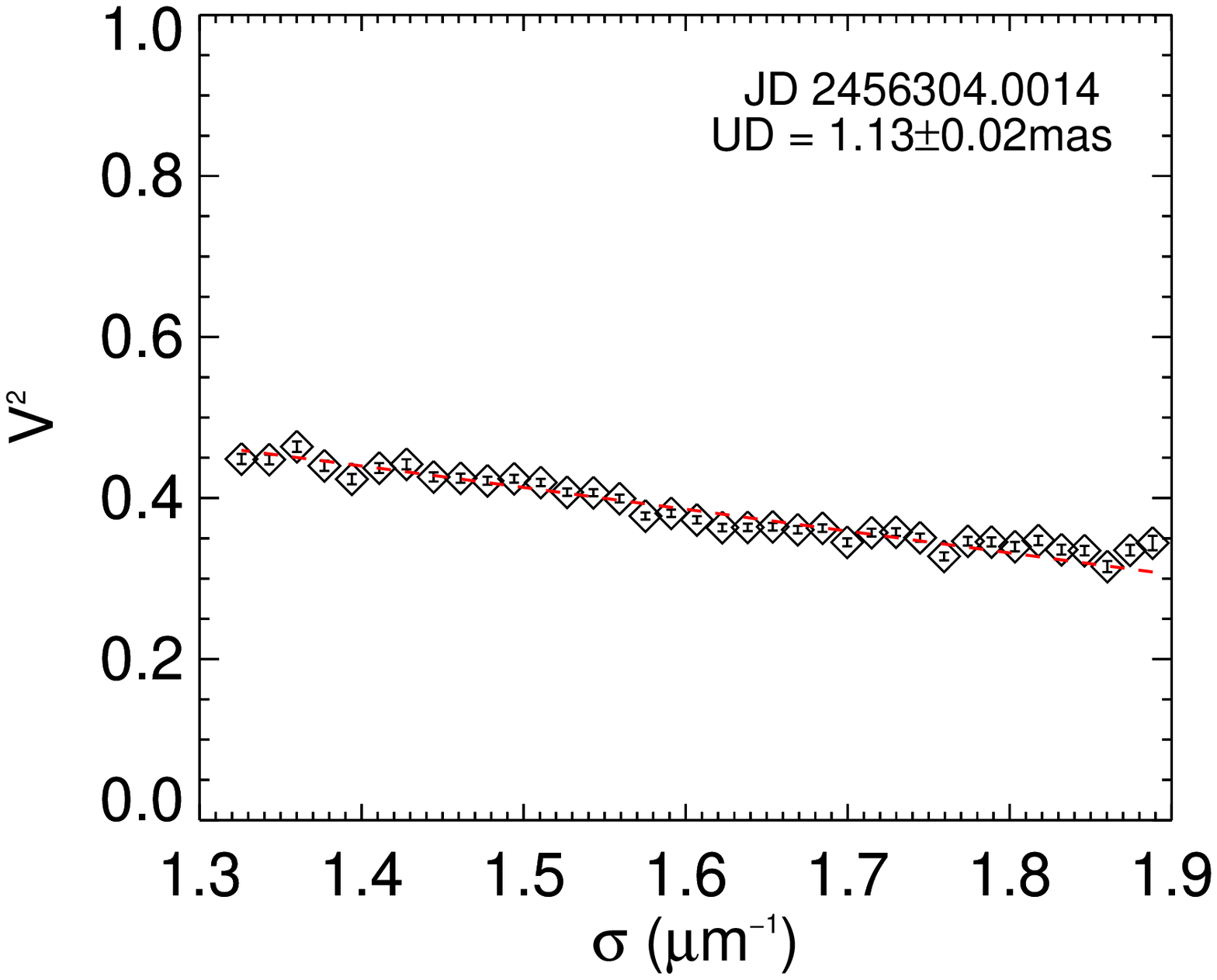}}
\caption[]{Calibrated $V^2$ data of $\delta$~Orionis~Aa obtained with PAVO and a
60m baseline (N4-S2). Uniform disk (UD) models that best fit the data are
plotted as dashed lines and the UD diameters from the fits are indicated in the
plots.}
\label{fig:obs_delori_pavo}
\end{figure}

The PAVO $V^2$ data from the successful observations were analyzed to evaluate
the structure of the primary component. The $V^2$ data showed no amplitude
modulation from the secondary ($\delta$~Orionis~Ab) fringe packet because the
separation of the primary-secondary fringe packet pair ($\sim$35--55$\mu$m) is
larger than the width of a single fringe packet and hence there is little
overlap. However, the presence of incoherent flux from the secondary component
limits the maximum $V^2$ of the primary to $\sim$0.66. Taking this into account,
the data when fitted with a simple uniform disk (UD) model suggests that the
primary star has an average angular diameter (over two observations) of
1.05$\pm$0.02mas. The data and the fitted models are shown in
Fig.~\ref{fig:obs_delori_pavo}. The estimated UD diameter corresponds to
$\sim$48 solar radii (R$_{\odot}$) at $\sim$212pc (computed from parallax
measurement \citep{van-Leeuwen:2007}). This is consistent with the estimated
diameters of the eclipsing components ($\delta$~Orionis~Aa1 and
$\delta$~Orionis~Aa2), which consist of an O-type and a B-type star and have
estimated angular diameters of $\sim$30R$_{\odot}$ and $\sim$10R$_{\odot}$
respectively \citep{Mayer:2010}. Due to such astrophysical structures, the phase
of the primary ($\delta$~Orionis~Aa) fringe packet changes over time as the
asymmetry of the structure rotates with respect to the baseline of the
interferometer. For the sake of estimating the magnitude of the fringe packet
separation of the eclipsing components, the binary separation is assumed to be
of the order of the estimated UD diameter, i.e.\ $\approx$1mas, and the position
angle is assumed to be parallel to the baseline. In this case, the fringe packet
separation observed with a 60m baseline (N4-S2) is $\lesssim$0.3$\mu$m. Given
the sensitivity of the phase delay computation algorithm to a secondary fringe
packet at this separation, it is likely that the jumps in the relative position
of the primary fringe packets are caused by the presence of an unseen fringe
packet of the tertiary component ($\delta$~Orionis~Aa2). In general, the double
fringe packet may not be easily distinguished from the data if the tertiary
fringe packet is partially resolved or buried within the primary fringe packet
of a binary star and especially when the integration time is for long enough for
the tertiary to move by more than one fringe. The systematic error in the
relative positions of the primary fringe packet also affects the estimated
fringe packet separation because the fringes of the secondary
($\delta$~Orionis~Ab) component were phase-referenced to the fringes of the
primary. The reduced $\chi^2$ values from a least square fit analysis carried
out to determine the relative position of fringe packets can be used to estimate
the \emph{correct} position in such circumstances.

The binary separations in the declination axis ($\Delta\delta_{\rm{Aa,Ab}}$)
measured on the two different nights of observation differ by less than 1mas,
which is consistent with the dimension of the angular diameter of the primary
component or the binary separation of $\delta$~Orionis~Aa1-Aa2. Since the two
observations were 3.9 days apart, this consistency is also supported by the
orbital period of the eclipsing binary (5.7 days) which suggests the photo
center of the primary component could be on opposite sides of the uniform disk
on the two nights of observation. On the other hand, the binary separations in
the right ascension axis ($\Delta\alpha_{\rm{Aa,Ab}}$) measured on the two
different nights differ by more than 10mas. This discrepancy cannot be an effect
of the orbital motion of the secondary component (Ab) around the primary (Aa)
because the estimated orbital period for the pair is 201 years
\citep{Mason:2009} and the estimated motion within $\sim$4 days is $\ll$0.1mas.
It is also unlikely to be an effect of narrow-angle baseline misalignment
because in order to produce the astrometric uncertainty ($\sim$10mas over
$\sim$260mas of projected separation) obtained from the data analysis the
East-West component of the narrow-angle baseline on the two nights must differ
by $\sim$30cm, a dimension which is larger than the diameter of the main
siderostat mirror. Therefore, the discrepancy in the right ascension axis
suggests that there are systematic errors on top of the known phase misalignment
which are not accounted for in the data analysis.

At the time of writing, none of the close binary systems observed for
performance testing, either with MUSCA or PAVO, consists of just 2 component
stars. More MUSCA observations, especially with targets that do not have
spectroscopic companions, and data analyses are needed to understand the source
of the systematic errors in MUSCA which may or may not be related to the
differential delay line and the dual-laser metrology. In that aspect, the low
success rate of a MUSCA observation is a challenge that must be overcome
especially for future observation campaigns.

\section{Conclusion} \label{sec:summary}
This paper has documented a successful attempt at high-precision narrow-angle
astrometry by means of optical long baseline interferometry. The astrometric
facility setup at SUSI is now operational and its first generation results have
been demonstrated. The PAVO+MUSCA setup in SUSI is the first dual beam combiner
that is capable of performing dual-star phase-referencing interferometry at
visible wavelengths and it achieves that through a novel post-processing
approach. However, the current best astrometric precision of MUSCA is in the
order of 100$\mu$as in the declination axis while the precision in the right
ascension axis is uncertain as the source of the large discrepancy between two
measurements is currently unknown. Further observations and improvements are
needed to ascertain and optimize the performance of the facility and the
precision of its astrometry. The detection principles outlined in this paper
could also be applied to interferometers at other sites, especially a potential
astrometric interferometer located on the high Antarctic plateau.

\section*{Acknowledgments}
This research was supported by the Australian Research Council's Discovery
Project funding scheme. Y.K. was supported by the University of Sydney
International Scholarship (USydIS) and would like to acknowledge the helpful
discussions with Vicente Maestro. The authors would also like to acknowledge the
use of the electronic bibliography maintained by NASA/ADS system, the Washington
Double Star Catalog maintained by the U.S.~Naval Observatory and the
SIMBAD/VizieR database maintained by CDS, Strasbourg, France.

\appendix{} 

The model in Sec.~\ref{sec:results} was chosen over an alternative model
described by \citep{Rizzuto:2013}. The latter is worth a brief discussion even though it is
not used in this work as it is suitable for less precise narrow-angle
astrometry. The model uses a coordinate system that takes the sky as the frame
of reference. In order to differentiate the frame of reference from that used
for this work, the axes of the coordinate system are labeled $u$, $v$ and $w$.
The $u$-axis points to the celestial East, the $v$-axis points to the celestial
North and the $w$-axis points to towards the sky. The model is given as,
\begin{equation} \label{eq:drp_sepmod2}
\begin{split}
S &= \vec{\rho}\cdot\vec{B}_p \\
&=\begin{bmatrix}
\rho\sin\theta \\
\rho\cos\theta \\
0
\end{bmatrix}_{uvw}
\cdot
\begin{bmatrix}
B_u \\
B_v \\
0
\end{bmatrix}_{uvw},
\end{split}\end{equation}
where $\vec{\rho}$ is a vector defining the separation of two stars in the
two-dimensional $uv$-plane of the sky and $\vec{B}_p$ is the component of
$\vec{B}$ projected onto the same plane. Since $\vec{B}$ is commonly quoted in
the ground ($xyz$) frame because it is more natural to an observer on the
ground, it can be converted into the sky ($uvw$) frame as shown,
\begin{equation}
\begin{split}
\begin{bmatrix}
B_u \\
B_v \\
B_w
\end{bmatrix}_{uvw} = \mata{T}
\begin{bmatrix}
B_x \\
B_y \\
B_z
\end{bmatrix}_{xyz},
\end{split}\end{equation}
where $\mata{T}$ is a matrix for coordinate transformation
\citep{Fomalont:1974}, which is given as,
\begin{equation} \label{eq:tm_skygnd}
\begin{split}
\mata{T}
&= \left[
{\setlength\arraycolsep{0.3em}
\begin{array}{l l l}
  \cos\rm{HA}
  & \sin\phi_{\text{LAT}}\sin\rm{HA}
  & \cos\phi_{\text{LAT}}\sin\rm{HA} \\
  \sin\delta\sin\rm{HA}
  & \cos\phi_{\text{LAT}}\cos\delta+\sin\phi_{\text{LAT}}\sin\delta\cos\rm{HA}
  & \sin\phi_{\text{LAT}}\sin\delta-\cos\phi_{\text{LAT}}\sin\delta\cos\rm{HA} \\
  -\cos\delta\sin\rm{HA}
  & \cos\phi_{\text{LAT}}\sin\delta-\sin\phi_{\text{LAT}}\cos\delta\cos\rm{HA}
  & \sin\phi_{\text{LAT}}\sin\delta+\cos\phi_{\text{LAT}}\cos\delta\cos\rm{HA} \\
\end{array}}\right]. \\
\end{split}
\end{equation}

\begin{figure}
\begin{center}
\includegraphics[width=0.45\textwidth]{\imgdir/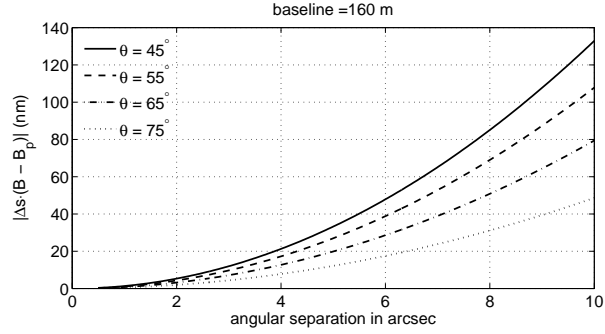}
\end{center}
\caption{The top plot shows the optical path length contributed by
$\Delta\vec{s}\cdot\vec{B}_p-\Delta\vec{s}\cdot\vec{B}$ while the lower plot is
a plot of $\vec{\rho}\cdot\vec{B}-\Delta s\cdot\vec{B}$. The angle $\theta$ in
the legend represents the angle between $\hat{s}_1$ and $\vec{B}$.}
\label{fig:drp_sepmod2_err}
\end{figure}

The RHS of Eq.~\eqref{eq:drp_sepmod2} is an approximation of
$\Delta\vec{s}\cdot\vec{B}$ and therefore the accuracy of the parameters
extracted from the model could be systematically less than parameters extracted
from our model (Eq.~\ref{eq:drp_sepmod1}). The calculated projected separation
using Eq.~\eqref{eq:drp_sepmod2} could be off by as much as 6nm under certain
conditions. This undesirable value translates to more than 10$\mu$as of
astrometric error if the baseline of the interferometer is 100m or less. The
factors affecting the approximation error are given as,
\begin{equation} \label{eq:drp_sepmod2_err}
\begin{split}
\Delta\vec{s}\cdot\vec{B} 
&= \Delta\vec{s}\cdot\vec{B}_p -
  2|\hat{s}_1\cdot\vec{B}|\sin^2\frac{\Delta\theta}{2} \\
&\approx \vec{\rho}\cdot\vec{B}_p -
  2|\hat{s}_1\cdot\vec{B}|\sin^2\frac{\Delta\theta}{2},
\end{split}\end{equation}
where $\Delta\theta$ is the angle subtended by the pointing vectors of the
primary and secondary stars. In the narrow-angle regime, $\Delta\theta \approx
|\Delta\vec{s}|$ and $\Delta\vec{s} \approx \vec{\rho}$. The approximation error
of the latter is negligible even at long baseline ($>$100m). On the contrary,
the plot in Fig.~\ref{fig:drp_sepmod2_err} shows that the approximation error of
$\Delta\vec{s}\cdot\vec{B}_p \approx \Delta\vec{s}\cdot\vec{B}$ is negligible
only for very closely separated stars. In summary, this alternative model is
still (just) suitable for MUSCA, which has a FOV of $<$2$''$, but may not be for
other instruments with a wider FOV.

\bibliographystyle{ws-jai}

\end{document}